%% file: main.tex
\documentclass[12pt,american]{article}
\usepackage{multirow}
\usepackage{afterpage}
\usepackage{float}
\usepackage{etoolbox}
\AtBeginEnvironment{tabular}{\normalsize}
\usepackage[skip=0.1\baselineskip,justification=centering]{caption}
\usepackage[a4paper,
            left=1in,
            right=1in,
            top=1in,
            bottom=1in]{geometry}
\usepackage[english]{babel}
\usepackage[utf8x]{inputenc}
\usepackage{amsmath}
\allowdisplaybreaks[2]  
\usepackage{amssymb} 
\usepackage[retainorgcmds]{IEEEtrantools}
\usepackage[flushleft]{threeparttable}
\usepackage{rotating,booktabs}
\usepackage{graphicx}
\usepackage{tabularx}
\usepackage{kpfonts}    
\usepackage{microtype} 
\usepackage{booktabs}   
\usepackage{bm}         
\usepackage{listings}   
\usepackage{verbatim}   
\usepackage{color}  
\usepackage{hyperref}
\hypersetup{
	colorlinks = true,
	linkcolor = blue,
	citecolor = blue,
	runcolor = blue,
	urlcolor = blue
}
\usepackage{url}
\usepackage{eurosym}
\usepackage[toc,page,header]{appendix}
    \usepackage[]{minitoc}
    \noptcrule 

\usepackage{rotating}
\usepackage{lscape}
\usepackage{subcaption}
\usepackage{mwe}
\usepackage{natbib}
\usepackage{multibib}
\newcites{sec}{References in the Online Appendix}
\usepackage{setspace,caption}
\usepackage{lipsum}
\usepackage{amsthm}
\theoremstyle{plain}

\newtheorem*{proposition*}{Proposition}
\newtheorem*{proof*}{Proof}
\setcitestyle{round,citesep={,}}
\usepackage{adjustbox}
\usepackage{placeins}

\usepackage{pdflscape}

\usepackage{dsfont}

\newcolumntype{H}{>{\setbox0=\hbox\bgroup}c<{\egroup}@{}}

\newcounter{hcount}

\usepackage{soul}

\renewenvironment{abstract}
 {\small
  \begin{center}
  \bfseries \abstractname\vspace{-.5em}\vspace{0pt}
  \end{center}
  \list{}{
    \setlength{\leftmargin}{0.25cm}%
    \setlength{\rightmargin}{\leftmargin}%
  }%
  \item\relax}
 {\endlist}

\makeatletter
\let\oldsection\section
\renewcommand{\section}{\vspace{-\parskip}\oldsection}

\let\oldsubsection\subsection
\renewcommand{\subsection}{\vspace{-\parskip}\oldsubsection}

\let\oldparagraph\paragraph
\renewcommand{\paragraph}{\vspace{-\parskip}\oldparagraph}
\makeatother

\begin{document}
	
\doparttoc 
\faketableofcontents 

\title{{\Huge{The Shared Costs of Pursuing \\ Shareholder Values}\thanks{We thank Pat Akey, Ghazala Azmat, Heski Bar-Isaac, Giorgia Barboni, Roland B\'enabou, Claire C\'elerier, Cl\'ement de Chaisemartin, Vicente Cu\~nat, Alex Edmans, Thierry Foucault, Nickolay Gantchev, Juanita Gonzalez-Uribe, St\'ephane Guibaud, Luigi Guiso, Sergei Guriev, Junnan He, Emeric Henry, Yael Hochberg, Panagiotis Koutroumpis, Nicolas Inostroza, Fausto Panunzi, Naciye Sekerci, David Sraer, David Thesmar, Jie Yang, Noam Yuchtman, and Riccardo Zago for helpful discussions and comments, and participants at seminars and conferences at Boca Conference, Bocconi, CREST, Erasmus Corporate Governance Conference, European Finance Association, FED Board, Lyon Gate, RCF ECGI, Sciences Po, Singapore Management University, and SIOE. We are grateful to Jeffrey Sonnenfeld and his team for responding to our requests for data clarifications. Mohammad Atif Haidry, Runpei He, Yujia Huang, Takayoshi Tokai, and Yuxuan Xu provided excellent research assistance. Fioretti and Saint-Jean thank the EIEF and Rotman for their hospitality during their respective visits. This study received funding from a Sciences Po-McCourt Institute grant. The views expressed in this paper are those of the authors and do not necessarily reflect the views and policies of the Board of Governors or the Federal Reserve System. All errors and omissions are ours.
}
}}
\author{\textbf{Michele Fioretti}\thanks{Universit\`a Bocconi, Department of Economics, IGIER, and CEPR. e-mail: \href{fioretti.m@unibocconi.it}{fioretti.m@unibocconi.it} }
	\and
\textbf{Victor Saint-Jean}\thanks{ESSEC Business School, Department of Finance. e-mail: \href{saintjean@essec.edu}{saintjean@essec.edu} } 
	\and
\textbf{Simon C. Smith}\thanks{Federal Reserve Board. e-mail: \href{simon.c.smith@frb.gov}{simon.c.smith@frb.gov} }}
	
\date{February 11, 2026}

\maketitle\vspace{-1cm}
\begin{abstract}\singlespacing \vspace{-0.5cm}
We study how shareholder values shape firms’ costly prosocial actions and who bears their costs. We develop a model in which some shareholders are publicly associated with a firm (e.g., founders or other prominent individual blockholders). When the firm takes a visible action under intense media scrutiny, these shareholders can plausibly claim credit and gain reputation, while diversified institutional investors cannot. The key empirical challenge is that influence is rarely observed: many consequential decisions are not subject to shareholder proposals or votes. We therefore use predetermined annual general meeting (AGM) timing combined with large, sudden crises---COVID-19 and the invasion of Ukraine---to generate quasi-experimental variation in attention and attribution, and to study highly visible, high-cost actions that were not legally required at onset. Firms with prominent individual blockholders are more likely to donate or exit when their AGM falls at crisis onset, while firms with large diversified institutional owners are less likely to do so. Consistent with our mechanism, online searches rise for prominent individuals after firm actions but not for institutions. Using an intent-to-treat triple-difference design on the 1{,}000 largest U.S.-listed firms, we find that exposed firms reduce investment, productivity, and profitability by 1--3\% for up to two years, highlighting the shared costs of pursuing the values of a visible minority.


\vfill
\end{abstract}



\thispagestyle{empty}
\newpage

\onehalfspacing

\setcounter{page}{1}

\section{Introduction}
Through private meetings, public letters, and sustained informal engagement, the descendants of John D. Rockefeller urged the oil company that became ExxonMobil to confront climate change and invest in cleaner energy \citep{kaiser2016rockefeller}. Their campaign relied on two ingredients: visibility in the public sphere and access to management. The episode illustrates a broader possibility: shareholders may induce firms to undertake costly corporate social responsibility (CSR), even when the financial benefits are unclear and the costs are borne by other investors.

Shareholders may push for such actions for both \textit{financial reasons} and reasons rooted in \textit{investor values} \citep{starks2023presidential}. Some are motivated by ethical convictions or by reputational rewards from being publicly associated with doing good \citep{broccardo2020exit,fioretti2022caring,bonnefon2025moral}. Others support CSR for instrumental reasons, such as mitigating reputational risk or catering to consumer demand \citep{lins2017social,hoepner2024esg}. Yet causal evidence on when and why shareholders push firms toward costly CSR in settings where financial returns are unclear remains limited. A central challenge is that influence is rarely observed \citep{dimson2015active}: we typically observe ownership and outcomes, but not the pressure shareholders apply or the conditions under which it is effective. We also know relatively little about how shareholder influence affects firms’ real decisions and performance, and how these effects are distributed across shareholders.

We focus on a reputational channel for shareholder influence. A large literature studies social preferences, image concerns, and the role of reputation in shaping economic behavior \citep{andreoni2006philanthropy, benabou2006incentives}. Related work emphasizes that reputation and status can also shape the incentives and effectiveness of shareholder activism \citep{johnson2021reputation, wiersema2020activist}. We emphasize a distinct margin: whether a shareholder can credibly claim credit for a firm’s visible actions, and thus capture private image benefits from it \citep{benabou2010individual}. When attribution is strong---as for prominent individuals whose identity is publicly linked to the firm (e.g., a founder or prominent individual)---costly prosocial actions can generate reputational ``rents'' that do not accrue to other investors. When attribution is weak---as for more anonymous financial shareholders---those private benefits are limited, weakening incentives to support costly CSR absent clear financial returns. This wedge implies an \textit{intra}-shareholder conflict: visible shareholders may press the firm toward prosocial actions that enhance their personal image while shifting costs onto investors who do not share in the reputational gains.

The key empirical challenge is that shareholder influence is typically unobserved: many consequential decisions are not governed by proposals or votes, so we rarely see when shareholders intervene. We test these predictions by exploiting predictable visibility spikes around annual general meetings (AGMs), moments when large shareholders are more publicly associated with firm decisions. We pair these visibility windows with salient external shocks that sharply increase public demand for corporate action, so the reputational payoff to being seen pushing a prosocial response rises discontinuously, and compare firms with different pre-existing shareholder composition. We find that shareholder influence is strongest among highly visible blockholders and when the prosocial action is relatively less costly, consistent with a trade-off between visibility gains and value losses even in large corporations with formal governance. Finally, we show that these influence-induced actions are followed by weaker operating and market outcomes for up to two years, including a 1–2\% decline in earnings per share driven by lower investment.

The findings have two broader implications. First, they suggest that existing governance mechanisms may not aggregate shareholder objectives neutrally when prosocial actions are salient and publicly attributable: attention can amplify the voice of visible minorities, even when costs are borne widely \citep{hart2017companies,bebchuk2020illusory}. Second, they speak to the sustainability transition. If visible investors can capture private image benefits from firm actions, firms’ climate and social policies may reflect the visibility of their shareholders as much as their fundamentals or the preferences of the marginal dollar \citep{flammer2015does,pastor2021sustainable}. Together, the results highlight a channel through which capital-market governance can shape the incidence and efficiency of corporate responses to social and environmental crises.

We formalize these ideas in a simple model of shareholder influence under public
scrutiny. Shareholders differ in whether they can be publicly credited for firm
actions. When scrutiny is high, publicly attributable shareholders obtain private
image benefits from visible prosocial actions, while more anonymous shareholders
do not. The model predicts that (i) increases in scrutiny raise the likelihood of
visible prosocial actions at firms with prominent individual shareholders, and
(ii) the effect is strongest when financial costs are low, reflecting a trade-off
between visibility gains and value losses.

We begin by establishing that AGMs generate sharp, predictable increases in public attention. News coverage and Google search intensity rise around annual general meetings (AGMs) for both firms and their most prominent shareholders. Consistent with visible owners exploiting these attention windows, we show that prosocial communications are strategically timed: in the month preceding the AGM, the share of positive environmental and social (E\&S) press releases rises by about 35\% more, and donation announcements by roughly 50\%, at firms \emph{with} large individual shareholders relative to otherwise similar firms \emph{without} such shareholders. While we show that these results are not driven by seasonality, they provide \textit{prima facie} evidence for a shareholder visibility channel.


Our empirical challenge is that investor pressure is typically unobserved. We address this by utilizing the AGM as an institutionally driven 'attention shock.' In U.S. practice, AGM timing is predetermined and rigid: SEC Rule 14a-8 and advance-notice bylaws require proposals to be filed no later than 90 days before the anniversary of the prior AGM, ensuring the date is fixed independently of contemporaneous firm choices. We exploit the interaction between this fixed timing and the sudden onset of a crisis to move beyond broad sentiment changes and evaluate specific, high-cost corporate actions. By interacting these predetermined dates with pre-crisis ownership, we isolate a quasi-experimental setting to infer how prominent individual shareholders (seeking reputation) and diversified institutional investors (opposing financial costs) pressure managers.

We operationalize this logic by defining treated firms as those with AGMs scheduled
in the first 90 days after crisis onset and interacting this indicator with
pre-crisis shareholder composition. We show that treated and control firms are
similar along observables, including ESG scores and other visible pro-social characteristics. We focus on two sudden crises
that induced costly, highly visible private responses. During COVID, donations
pledged in the first three months amount to about 0.1\% of revenues, comparable to the average
donations made by a S\&P~500 firms in the previous year. During the Russia--Ukraine conflict,
rapid exits impose substantial financial costs \citep{jack2022exit}; for this
setting, we measure exits within the first month after February~24,~2022 to reduce
contamination from subsequent government sanctions.\footnote{Donations data are hand-collected from corporate disclosures and online searches. Results remain consistent when restricted to donations reported in annual accounts, suggesting that firms did not simply shift the timing of donations within the fiscal year \citep{liang2025disaster}. The Russia analysis uses merged data from \cite{sonnenfeld2022business} and \texttt{leave-russia.org}. }

In both settings, firms with prominent individual shareholders are more likely to
take the visible action when their AGM coincides with crisis onset, while firms
with large diversified institutional shareholders are less likely to do so.\footnote{Equity ownership serves as both a measure of public association for individual shareholders and a proxy for management access for institutional investors \citep{hoepner2024esg}. Our results are robust to alternative measures of association, such as the correlation in Google search intensity between shareholders and their firms. Our results are robust to financial motives, consumer pressure, and competition as alternative explanations. }
Consistent with the mechanism, reputational payoffs accrue where attribution is
strongest: searches for prominent individuals rise sharply after donation
announcements, while searches for financial shareholders do not. We also find
evidence of limited pass-through for financial institutions: institutions that
donate directly during COVID reduce donation activity at portfolio firms whose
AGMs occur at crisis onset. The Russia analysis yields analogous patterns in a
high-cost setting, indicating that the mechanism is not specific to charitable
giving.

Heterogeneity analyses confirm the model’s reputational–cost trade-off. When financial slack is high (lower private cost), prominent individual shareholders drive an increase in prosocial actions, while institutional investors remain passive. However, as the cost of action rises—either due to low financial slack or high exposure to the crisis—the individual effect disappears and the institutional effect turns negative. These patterns match the model’s prediction: the same attention shock amplifies individual influence when reputational gains are 'cheap,' but triggers institutional counter-pressure when the financial stakes threaten firm value.

We then quantify consequences for firm outcomes and the incidence of costs across
investors. Using the 1{,}000 largest U.S.-listed firms, we estimate an
intent-to-treat triple-difference design that compares crisis-onset-AGM firms to
other firms, interacting this variation with the presence of prominent individual
shareholders and with pre/post crisis timing. During COVID, exposed firms
experience higher operating expenses with little change in revenues, and
investment and earnings per share each fall by about 2\% in 2020 and 2021. During
the Russia--Ukraine crisis, similarly exposed firms increase asset sales by about
4\% and incur restructuring charges that reduce earnings per share by about 1\% in
2022 and 2023.

Finally, we benchmark magnitudes. A back-of-the-envelope calculation implies that
raising a prominent shareholder’s Google searches by 1\% costs about \$0.025 per
share (ca. \$182{,}000 for the median prominent individual shareholder). This
is comparable to about one month of high-profile public-relations exposure and is
well below the average COVID donation in our dataset, highlighting 
the private value of influence. 


\paragraph{Related literature} Beyond internal governance, our evidence speaks to ongoing debates on
sustainability regulation \citep{christensen2021mandatory}. Most ESG and
sustainable finance frameworks discipline or reward observable outcomes---donations,
exits from controversial markets, headline ratings---rather than the decision
processes that produce them. Yet shareholder engagement often operates through
private channels and can target both short-term actions and long-run strategy
\citep{mccahery2016behind}. As a result, the same observed action can reflect a
broad alignment of investor preferences or the preferences of a small set of
prominent shareholders \citep{hart2017companies}. Treating these cases alike risks
rewarding image rather than shared purpose. Our results therefore point to a
simple distinction that may matter for policy: who initiates prosocial actions,
who captures reputational gains, and who bears the financial cost.

Our first contribution is to identify a reputational channel through which ownership structure shapes corporate decisions. When a prominent individual is publicly linked to a firm, they can credibly claim credit for visible prosocial actions and capture private image benefits. While research on family firms suggests that such socioemotional returns often shape corporate choices \citep{berrone2012socioemotional}, we show that similar identity-based incentives are present even among the large-cap firms in our sample. This adds an owner-level mechanism to the literature linking CSR to public attention \citep[e.g.,][]{benabou2010individual,dyck2019do,akey2021hacking,list2021corporate,conway2024consuming} and shifts the locus of image concerns from managers or stakeholders to shareholders themselves \citep{brendel2025value}. Finally, we distinguish this reputational channel from general media pressure on the firm \citep{dyck2008corporate} and from anonymous ESG preferences \citep{krueger2020importance} by emphasizing the unique role of firm-specific public attribution.


Our second contribution is to show that this mechanism has economically meaningful
consequences in salient crises. Our visibility mechanism helps explain
heterogeneity in firms’ private sanctions and related adjustments in geopolitical
settings \citep[e.g.,][]{korovkin2024supply,nigmatulina2023sanctions,steinbach2023russia,hart2024private}, complementing models of private politics in which firms respond
to social and political pressure \citep[e.g.,][]{baron2001private} by identifying
a source of that pressure in publicly attributable shareholders.

Our third contribution is methodological. Prior work typically identifies
shareholder influence through observable channels such as proposals and votes
\citep[e.g.,][]{gompers2003corporate,cunat2012vote}. Many consequential decisions,
however---including pandemic donations and exits from foreign markets---are not
voted on. We develop a portable framework that uses quasi-random variation in AGM
timing to identify otherwise unobserved shareholder influence and to mitigate
endogeneity concerns in ownership-based designs \citep{edmans2017blockholders}.
Though our empirical settings involve crises, the mechanism is not
crisis-specific: any setting that creates temporary gains across heterogeneous
shareholders can activate similar conflicts.

Finally, our results speak to agency with multiple principals \citep{bernheim1986common}, highlighting a trade-off in shareholder heterogeneity: while diverse objectives can discipline one another, sudden attention shocks can tilt influence toward a visible minority \citep{edmans2011governance}. This perspective reframes governance---from traditional blockholder monitoring to minority protection---by focusing on the diversity of values among large owners rather than simple ownership concentration. Our findings suggest that the benefits of a diverse shareholder base depend on institutional safeguards that prevent corporate policy from becoming one-sided when specific actions are publicly salient but financially costly.\footnote{For classic work on large shareholders and monitoring, see \citet{shleifer1986large,edmans2009blockholder}; for investor protection and control, see \citet{la2002investor}; and for evidence on ownership concentration, see \citet{cronqvist2009large}. The mechanism is consistent with models in which individuals derive private image-based benefits from prosocial actions \citep[e.g.,][]{andreoni1990impure,bar2008seller}. Whether managers respond directly to shareholder pressure or instead preemptively cater to these preferences remains an open question \citep{saint2024exit}.}



The paper is structured as follows. Section \ref{s:conceptual} develops the conceptual framework. Section \ref{s:shareholder_types} discusses the variation generated by AGMs, and Section \ref{s:application} derives the empirical strategy from the theoretical model. In Sections \ref{s:covid} and \ref{s:russia}, we apply this framework to analyze shareholder influence at the onset of the recent pandemic and the invasion of Ukraine, respectively, highlighting the underlying mechanisms and presenting several robustness checks. Section \ref{s:distributional} examines the distributional costs of aligning corporate actions with the preferences of some shareholders and underscores our key contributions. Finally, Section \ref{s:conclusion} summarizes our findings and concludes.

\section{Conceptual Framework} \label{s:conceptual}
This conceptual framework illustrates the scope of our empirical analysis. A firm considers a \textit{costly} but \textit{visible} prosocial action that reduces profits and dividends by $D > 0$ for its two shareholders. Shareholder $A$, publicly associated to the firm, gains $v_A > 0$ in image benefits if the action is adopted, while shareholder $B$, with no public connection, gains nothing ($v_B = 0$). Each shareholder can pressure the manager to pursue their preferred action, creating an \textit{externality} on the other shareholder, with the following probability function:
\begin{equation}\label{eq:prob.don}
\Pr(\text{prosocial action})=s(e_{A},e_{B}),
\end{equation}

\noindent which is increasing in the influence effort of shareholder $A$, $e_A$, and decreasing in that of shareholder $B$, $e_B$. The cost of effort to shareholder $i$, $c(e_i)$, is convex in $i$'s effort level, with $e_i >0$. Shareholder $i$'s utility is 
\begin{equation*}
U_i = (v_{i}-D)\cdot  s(e_A,e_B) - c(e_i).\end{equation*}

\noindent Agent $i$'s  optimal effort sets marginal benefit equal to marginal cost, according to:
\begin{equation}\label{eq:foc}
\begin{aligned}
    (v_{A}-D)\cdot \frac{\partial s(e_{A},e_{B})}{\partial e_A}&=c^\prime(e_A), \, \text{ for shareholder } A, \hspace{0.4cm}  \\
    - D \cdot \frac{\partial s(e_{A},e_{B})}{\partial e_B}&=c^\prime(e_B), \, \text{ for shareholder } B.
\end{aligned}
\end{equation}

\noindent Therefore, $A$ will pressure the firm only in case of a clear association with the firm ($v_A > D$). Importantly, an \textit{exogenous increase} in $v_A$ will raise $A$'s equilibrium effort. In the case of strategic complements, shareholder $B$ will adjust $e_B$ to reduce the probability of a donation.\footnote{Strategic complements naturally arise in this setting with $s(e_A, e_B) = \frac{e_A}{e_A+e_B}$ (with $e_A + e_B > 0$) or the logistic probability, particularly when shareholders have comparable valuations or efforts.} Thus, the two agents compete to influence the firm and will set effort levels to offset each other, considering the dividend loss and their private returns and effort costs.
This model easily extends to a setting with $N$ shareholders, where shareholder $k \in {1, \dots, N}$ is of type $A$ or $B$ and owns $x^k \in (0,1]$ shares, with $\sum_{k=1}^N x^k = 1$. Denoting by $E_i$ the total effort of all type-$i$ shareholders, the utility of shareholder $k$ of type $i$ is 
\begin{equation*}U_i^k = (v_{i}^k - D \cdot x^k) \cdot s(E_A, E_B) - c(e_i^k), \qquad i\in(A,B),
\end{equation*}
where $D$ now denotes the total loss to be shared across shareholders. This maximization problem yields similar first-order conditions as the two-shareholder case, providing us with two new \textit{theoretical predictions}, which we will test in the data.

The first prediction is that large-$B$ shareholders (those with higher $x^k$) will increase their effort more than small-$B$ shareholders when $v_A$ rises, since large-$B$ shareholders stand to lose more as $A$'s effort increases. The second prediction concerns $A$-type shareholders. If the public is less likely to associate minority type-$A$ shareholders with the firm---a hypothesis we confirm empirically in Figures \ref{fig:Google_agm} and \ref{fig:Google_10} in the next sections---then $v_A^k$ may rise more for large-$A$ than for small-$A$ shareholders when their firm is under the spotlight, implying greater effort by the former shareholders.

\paragraph{Implications for the empirical analysis } In the real world, shareholders have different effort costs for influence and dividend expectations. Suppose an exogenous treatment increases $v_A$ at some firms but not others. We expect prosocial actions to be more likely at treated firms with a high proportion of type-$A$ shareholders, particularly type-$A$ blockholders owning large shares. Conversely, prosocial actions should be less likely at treated firms with larger type-$B$ blockholders. The next section explains how we implement this intuition on shareholding data.

Testing these predictions requires addressing three challenges: (i) distinguishing shareholders by their potential reputational gains (type~$A$ vs. type~$B$), (ii) quantifying the costs of prosocial actions ($D$), and (iii) measuring the unobservable effort used to influence corporate policy ($e_A$ and $e_B$). Section~\ref{s:shareholder_types} introduces the ``spotlight'' created by the exogenous timing of AGMs to characterize the visibility of different shareholder classes. Section~\ref{s:application} then exploits this cross-sectional variation in how firms respond to crises to elicit the influence of different shareholder types.

\section{Categorizing Shareholder Types}\label{s:shareholder_types}

This section uses AGM dates as predictable ``spotlight'' episodes to (i) measure which shareholders experience unusually large visibility gains when their connected firm enters the news cycle, and (ii) provide suggestive evidence that firms connected to those shareholders concentrate prosocial communication in the same window. The goal is not to claim a causal channel from shareholder identity to firm messaging here. Rather, we document two empirical patterns that motivate our classification of ``high-visibility'' (type-$A$) shareholders and the strategic-timing interpretation we take to the model and the main identification in later sections.

\paragraph{Data and sample} Our empirical analysis focuses on firms included in the S\&P 500 between 2011 and 2020. We combine four data sources. First, quarterly shareholding data come from LSEG. Second, AGM dates come from \textit{Institutional Shareholder Services} (ISS). Third, we collect monthly Google Trends search intensity for shareholders. Finally, we use RavenPack News Analytics to measure firms’ prosocial communication through donation announcements and environmental and social (E\&S) press releases.

For the shareholder-level visibility analysis, we track all shareholders holding at least 1\% of a firm’s shares in any quarter during the sample period, yielding 1{,}597 distinct shareholders. For each of these shareholders, we collect monthly Google Trends scores, which range from 0 to 100 and are normalized within shareholder so that 100 corresponds to the month with the highest search intensity. For the firm-level prosocial-communication analysis, we merge RavenPack data on donation announcements and E\&S press releases with the same set of firms, AGM dates, and ownership information. Press-release sentiment is computed as the share of positive E\&S press releases among all E\&S press releases issued by the firm. We exclude non-E\&S press releases, which mechanically spike around AGMs.

In both analyses, we define ownership using holdings measured seven months before the AGM. This timing predates both the three-month AGM notice period and the four- to six-month window for submitting shareholder proposals. While ownership may still change for reputational reasons, we interpret results in this section as descriptive and suggestive rather than causal.

\paragraph{Empirical design} We implement two complementary event-study specifications around AGMs. The first operates at the shareholder--firm level and measures how visibility (Google searches) changes around AGMs for individual versus non-individual shareholders. The second operates at the firm level and tests whether prosocial communication is more concentrated around AGMs at firms with large individual shareholders. In both cases, we align firm calendars around the AGM date and set the omitted period to three months before the AGM, consistent with widely followed guidelines that require firms to notify shareholders about an upcoming AGM 90 days in advance (see Section~\ref{s:application}).

We begin by aligning firm calendars so that month $0$ of year $r$ is the AGM month for all firms in that year and estimate:
\begin{equation}\label{eq:Google_agm}
\begin{aligned}
    y_{jtr} &= \sum_{m=-4,\, m\neq -3}^{4} \left(  \theta_m \ \textnormal{AGM}_{f(j),t=m,r} + \gamma_m \ \textnormal{AGM}_{f(j),t=m,r}\times \textnormal{Individual}_{j(f)r} \right)\\
    &\quad + \alpha_{jf(j)r} + \tau_{tr} + \varepsilon_{jtr},
\end{aligned}
\end{equation}
where $y_{jftr}$ is the standardized Google Trends score for shareholder $j$ connected to firm $f(j)$ in month $t$ of year $r$. The indicators $\textnormal{AGM}_{f(j),t=m,r}$ capture event time relative to the AGM month. We interact these with $\textnormal{Individual}_{j(f)r}$, an indicator for whether investor $j$ is an individual (or family) shareholder in firm $f$, measured using shareholding data seven months before the AGM.\footnote{We fix ownership seven months before the AGM, predating both the three-month notice period and the four- to six-month window for submitting shareholder proposals. Still, since ownership may also change for reputational reasons, the results in this section should be viewed as suggestive.} The specification includes shareholder--firm--year fixed effects ($\alpha_{jfr}$), which absorb time-varying ownership composition and other factors at this level (e.g., stake size or an executive/director role), as well as calendar month--year fixed effects ($\tau_{tr}$). Standard errors are clustered at the shareholder--firm--year level.

We then estimate an analogous event-study specification at the firm level using weekly data. Aligning calendars so that week $t=0$ is the AGM week in year $r$, we estimate:
\begin{equation}\label{eq:stacked_did}
    \begin{aligned}
        y_{ftr} &= \sum_{w = -13,\, w\neq -12}^{20} \left( \theta_w \ \textnormal{AGM}_{f,t=w,r} + \gamma_w \ \textnormal{AGM}_{f,t=w,r} \times \text{Individual}_{fr}  \right)  \\
        &\quad+ \alpha_{fr} + \tau_{tr} + \varepsilon_{ftr}.
    \end{aligned}
\end{equation}
where $y_{ftr}$ measures cumulative prosocial communication over the four weeks preceding week $t$ of year $r$. The indicator $\textnormal{Individual}_{fr}$ equals one if firm $f$ in year $r$ has at least one individual shareholder holding more than 5\% of shares as of seven months before the AGM; the omitted category is three months before the AGM ($w=-12$). We include firm-by-year fixed effects ($\alpha_{fr}$) and week-by-year fixed effects ($\tau_{tr}$). Standard errors are clustered at the firm-by-year level.

\begin{figure}[!ht]
    \caption{The Google search gap between individual and non-individual shareholders widens around AGMs, especially for larger holdings}
     \label{fig:Google_agm}
    \centering
        \centering
        \includegraphics[width=0.8\textwidth]{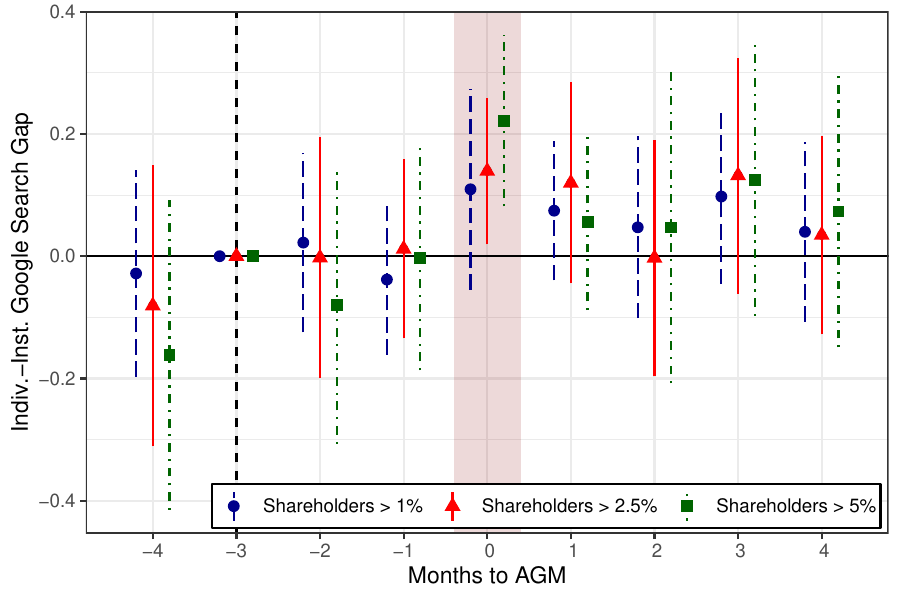} 
\begin{minipage}{1 \textwidth} 
{\footnotesize Note: The figure plots estimates of $\hat{\gamma}_m$ from \eqref{eq:Google_agm}, which regresses a standardized measure of a shareholder's Google Trends score in month $t$ on the interaction between the shareholder is an individual shareholder and month $t$ is $m$ months away from the shareholder's firm's AGM date[cite: 558]. We include shareholder-firm-year fixed effects and month-year fixed effects. Dates are calendarized so that all firms have an AGM on month 0 in each year. We present results from three versions of \eqref{eq:Google_agm}: blue dots, red triangles, and green squares report $\hat{\gamma}_m$ when subsetting the data to shareholders with $>1\%$, $>2.5\%$, and $>5\%$ shares, respectively. Vertical bars denote 95\% confidence intervals (CI). Standard errors are clustered by firm-shareholder-year. The data spans S\&P 500 firms between 2011 and 2020. Shareholding data from LSEG, AGM dates from ISS.\par}
\end{minipage}
 \end{figure}

\paragraph{Results} Figure~\ref{fig:Google_agm} reports estimates of $\hat{\gamma}_m$ from \eqref{eq:Google_agm}. The differential search activity between individual and non-individual shareholders increases sharply around the AGM month and is larger for larger ownership stakes. Estimates exceed 0.1 standard deviations in the AGM month and peak above 0.2 when comparing individual shareholders with more than 5\% ownership to similarly large non-individual shareholders. These patterns indicate that prominent individual shareholders---often founders, public figures, or wealthy families---experience disproportionately large visibility gains when their connected firms enter the AGM spotlight. We therefore classify such prominent individuals/families as type $A$ (high visibility), and other shareholders (e.g., mutual funds, banks, insurance firms) as type $B$.\footnote{Although our main analyses leverage variation in equity shares to measure prominence, we also explore robustness to alternative visibility measures (see Section~\ref{s:robustness}). Throughout, we use ``individual'' to capture individuals or families; the classification is based on shareholder type in LSEG.}

\begin{figure}[!ht]
    \caption{Individual shareholding and strategic timing of prosocial actions} \label{fig:double_es}
    \centering
    \begin{subfigure}{0.5\textwidth}
    \captionsetup{justification=centering}
        \centering
        \includegraphics[width=\textwidth]{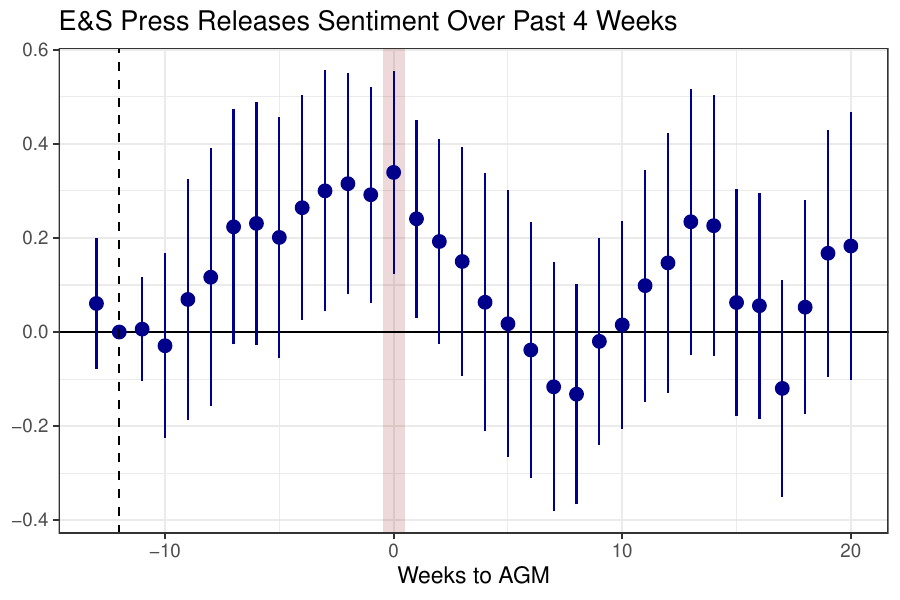} 
        \caption{Share of positive E\&S press  \\ [-1ex] releases in past month}
    \end{subfigure}\hfill
    \begin{subfigure}{0.5\textwidth}
    \captionsetup{justification=centering}
        \centering
        \includegraphics[width=\textwidth]{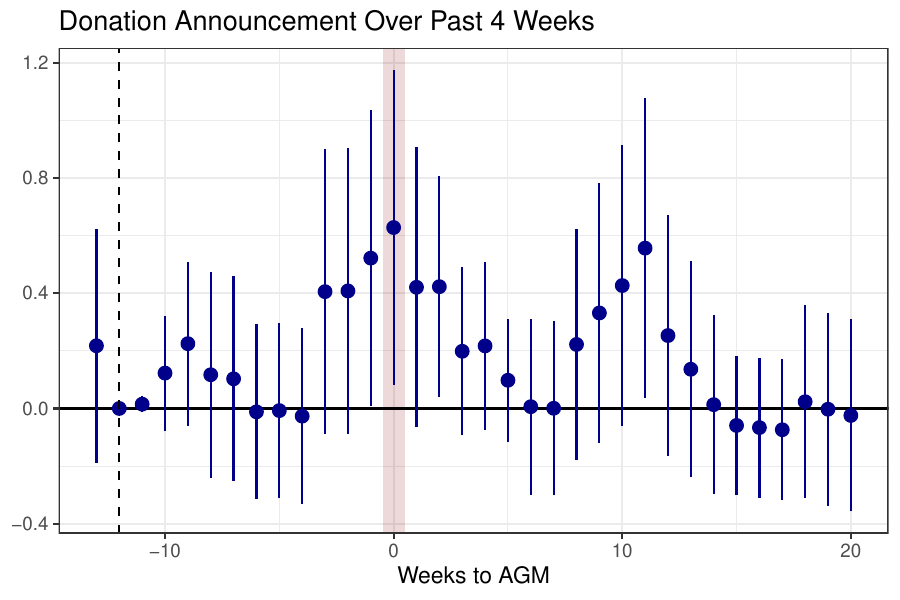} 
        \caption{Indicator for a donation \\ [-1ex] in  past month}
    \end{subfigure}
\begin{minipage}{1 \textwidth} 
{\footnotesize Note: The figure plots estimates of $\hat{\gamma}_w$ from \eqref{eq:stacked_did}, which regresses measures of prosocial communication on the interaction between an indicator for firms with at least one large individual shareholder and week $t$ being $w$ weeks away from the firm's AGM date. We include firm-year fixed effects and week-year fixed effects. Dates are aligned so that all firms have an AGM on week 0. Panel (a) reports results for a standardized measure of the share of positive E\&S press releases, where sentiment is scored using RavenPack’s NLP algorithm. Panel (b) reports results for an indicator of whether a firm published a press release mentioning a donation. We define the treatment group as firms with at least one individual (or family) shareholder holding more than 5\% of shares seven months before the AGM. Vertical bars denote 95\% confidence intervals. Standard errors are clustered at the firm-year level. Shareholding data from LSEG, AGM dates from ISS, press releases and donation announcements from RavenPack.  \par}
\end{minipage}
\end{figure}

Figure~\ref{fig:double_es} reports estimates of $\hat{\gamma}_w$ from \eqref{eq:stacked_did}. Firms with at least one large individual shareholder increase prosocial communication in the run-up to the AGM. In particular, positive E\&S press releases rise in the month preceding the AGM, and donation announcements become more likely over the same period: relative to the reference period three months before the AGM, positive E\&S press releases increase by about 35\%, and donation announcements by about 50\%.\footnote{Later spikes are consistent with other predictable attention events such as earnings calls.} We interpret these patterns as suggestive evidence of strategic timing: when the AGM spotlight is approaching and reputational returns are plausibly higher, firms connected to highly visible shareholders appear to concentrate prosocial announcements in this window. Importantly, these patterns are not, by themselves, causal evidence of shareholder influence; they motivate our use of shareholder visibility as a key dimension in the theoretical model and inform the interpretation of the empirical strategy that follows.

\paragraph{Takeaway} AGMs generate predictable spikes in public attention that disproportionately accrue to prominent individual shareholders. Firms connected to such shareholders appear to time prosocial communication around these same spotlight periods. Together, these results motivate our classification of type-$A$ shareholders and frame why prosocial communication clustered around AGMs is a natural place to look for reputationally driven behavior.

\section{Empirical Framework}\label{s:application}
The results presented thus far are only suggestive, partly because of confounding factors such as share trading, since shareholders can adjust their stakes at opportune times to influence management. In this section, we show how to bring the model in Section~\ref{s:conceptual} to the data to causally infer shareholder influence.  

To sum up our approach, we first proceed by linearizing the system of first-order conditions in \ref{eq:foc} so that the efforts of each agent can be written as functions of $A$'s reputational gains, $v_A$, and damages, $D$. We then plug these functions into Equation~\eqref{eq:prob.don}, which gives the probability of a prosocial action, to derive an empirical specification that can be taken to the data. Finally, we introduce the exogenous variation used for identification, namely the AGM spotlight. Appendix~\ref{apndx:deriv} illustrates all derivations.

\paragraph{Linearizing efforts} 
To move from the theoretical framework to an estimable form, we first show how equilibrium efforts can be approximated as functions of reputational incentives and costs. Starting from the first-order conditions in Equation~\eqref{eq:foc}, we take total differentials and approximate the resulting system with a first-order Taylor expansion around a reference point $(\bar{e}_A, \bar{e}_B, \bar{v}_A, \bar{D})$. This step delivers tractable reduced-form expressions for shareholder effort:
\begin{equation}\label{eq:approx}
e_A \approx \tilde{\alpha}_0 + \tilde{\alpha}_1 v_A + \tilde{\alpha}_2 D, 
\qquad 
e_B \approx \tilde{\delta}_0 + \tilde{\delta}_1 v_A + \tilde{\delta}_2 D.
\end{equation}
Coefficients $(\tilde{\alpha}_0, \tilde{\alpha}_1,\tilde{\delta}_0, \tilde{\delta}_1)$ are equilibrium objects, as detailed in Appendix~\ref{apndx:deriv}.

\paragraph{Empirical specification} 
We next embed these reduced-form efforts into the firm’s decision rule $s(e_A,e_B)$ for prosocial actions in Equation~\eqref{eq:prob.don}. Let $y_f=1$ if firm $f$ behaves prosocially and $y_f=0$ otherwise. According to $s(e_A,e_B)$, the corresponding latent variable $y^*_f$ is specified as
\begin{equation}\label{eq:s-lin}
    y_f^* = \tilde{\theta}_0 + \tilde{\theta}_1 e_{A,f} + \tilde{\theta}_2 e_{B,f} + \tilde{\epsilon}_f,
\end{equation}
where $\tilde{\epsilon}_f$ follows a logistic distribution. Thus, the firm acts prosocially if $y^*_f>0$, and does not otherwise. Substituting \eqref{eq:approx} into \eqref{eq:s-lin} yields
\[
\Pr(y_f=1) = \Lambda(\beta_0 + \beta_1 v_{A,f} + \beta_2 D_f),
\]
where $\Lambda(\cdot)$ is the logistic cumulative distribution function. Under small variations in $v_A$ and $D$ (as induced by AGM shocks) the logistic link is commonly approximated by the following linear probability model:
\begin{equation}\label{eq:base}
y_f = \theta_0 + \theta_1 v_{A,f} + \theta_2 D_f + \epsilon_f.
\end{equation}

\paragraph{Exogenous variation in the model} Consider now two otherwise identical firms, \(f\) and \(f'\), with the same level of damages \(D\) and type-$A$ shareholder ownership, except that only firm \(f\) experiences an exogenous increase in visibility. Then, the expected difference in prosocial actions is
\begin{equation}\label{eq:difference}
    \mathbb{E}[y_f - y_{f^\prime}] = \theta_1 \cdot \Delta v_A + \mathbb{E}[\epsilon_f-\epsilon_{f^\prime}],
\end{equation}
where \(\Delta v_A = v_{A,f} - v_{A,f'}\) is induced by the exogenous change in visibility and is uncorrelated with the difference in $\epsilon$. This is the variation to which we turn next.

\paragraph{Exogenous variation in the data} Our exogenous variation arises from the coincidence of AGM periods with unexpected crises for three reasons.

First, U.S. securities regulation requires firms to hold annual general meetings (AGMs) at which shareholders review performance and vote on management and shareholder proposals. Under SEC Rule 14a-8, shareholders who want their proposals included in the company’s proxy materials must submit them at least 120 calendar days before the anniversary of the release of the prior year’s proxy statement \citep{securities1934securities}.\footnote{A proxy statement (SEC Form DEF 14A) is filed with the SEC ahead of shareholder meetings and discloses information on issues to be voted on, including board elections, executive compensation, and shareholder proposals.} This federal deadline contributes to the stability of AGM dates across years, as shown in Panel (a) of Appendix Figure~\ref{fig:meetings}, where the timing of AGMs for firms covered by ISS varies little from 2012 to 2019.

Second, companies commonly adopt ``advance notice bylaws'' under state corporate law, which govern when shareholders may submit proposals or nominations for consideration at the AGM even if those items are not included in the proxy materials. These bylaws typically require that such proposals be filed no later than 90 days before the anniversary of the previous AGM, making the 90-day mark the final cutoff for bringing new items onto the meeting agenda. Proxy advisory firms such as ISS and Glass Lewis recommend this 90-day deadline as a reasonable market standard, and it has been widely adopted in U.S. corporate practice \citep{iss2022, glasslewis2022}.

Third, we focus on firms' reactions to sudden crises as highly visible actions that news outlets are likely to report in contrast to, say, changes in ESG ratings or other less visible prosocial actions.

We therefore define our treatment group as firms with AGMs scheduled in the first 90 days of a crisis. For these firms, the AGM date was determined in advance of a crisis. The coincidence of this AGM period with the crisis ($\textnormal{AGM}_f=1$) generates an exogenous increase in visibility, $v_{A,f}$, provided that type-$A$ shareholders are present, captured with  $\textnormal{Ownership}_{f}$, as shown in Figure~\ref{fig:Google_agm}. Hence, we rewrite Equation~\eqref{eq:base} as
\begin{equation}\label{eq:lin-prob-model-cross-section}
y_{f} = \underbrace{\beta_0 + \beta_1 \, \textnormal{Ownership}_{f} 
        + \beta_2 \, \textnormal{AGM}_{f} 
        + \beta_{\text{treat}} \, \textnormal{Ownership}_{f} \times \textnormal{AGM}_{f}}_{\theta_0 \, + \, \theta_1 \,  v_{A,f} } + \varepsilon_{f},
\end{equation}
with $\varepsilon_f = \theta_2\, D_f + \epsilon_f$. To the extent that the coincidence of a crisis with the AGM window (AGM$_f$ = 1) is orthogonal to firm fundamentals, governance, or managerial preferences in $\varepsilon_f$, the expected difference in prosocial actions between two otherwise identical firms $f$ and $f^\prime$ that differ only in AGM date is
\[
\mathbb{E}[y_f - y_{f^\prime}] = \beta_2 \, \textnormal{AGM}_{f} + \beta_{\text{treat}} \, \textnormal{Ownership}_{f} \times \textnormal{AGM}_{f} +  \mathbb{E}[\varepsilon_f-\varepsilon_{f^\prime}].
\]
Here, $\beta_2$ absorbs any baseline difference in prosociality between firms with and without AGMs. If prosocial behavior differs only through shareholder composition, then $\hat{\beta}_2=0$, and $\beta_{\text{treat}}$ isolates the causal effect of a visibility shock on shareholder influence, corresponding to the $\theta_1 \Delta v_A$ term in Equation \eqref{eq:difference}. In other words, the treatment shifts $v_{A,f}$ for firm $f$, while $v_{A,f^\prime}$ for control firms remains unchanged.

\paragraph{Identification}
\sloppy Our design allows ownership to be endogenous in levels, so that $\mathrm{Ownership}_f$ may be correlated with firm attributes that affect prosocial behavior. What we require is that the \emph{incremental} effect of ownership on the outcome does not vary systematically with whether the firm's AGM falls in the crisis window, except through the visibility channel induced by the AGM.\footnote{Formally, we assume that AGM timing is predetermined with respect to crisis-period shocks, so that $\mathbb{E}[\varepsilon_f \mid \mathrm{Ownership}_f, \mathrm{AGM}_f] = \mathbb{E}[\varepsilon_f \mid \mathrm{Ownership}_f]$.} This requirement allows $\mathrm{Ownership}_f$ to be correlated with $\varepsilon_f$ in levels, but rules out differential selection on unobservables across early-AGM and late-AGM firms conditional on ownership.\footnote{To corroborate this assumption, we examine firms covered by ISS and MSCI between 2012 and 2020. As shown in Appendix Figure \ref{fig:agm_esg_score}, the distribution of ESG scores does not vary with a firm's AGM month. This suggests that AGM timing is not related to a firm's baseline prosociality, a confounding factor that would otherwise be captured in $\mathbb{E}[\epsilon_f - \epsilon_{f^\prime}]$ in Equation \eqref{eq:difference}.}

\paragraph{Illustration} We illustrate with the COVID-19 crisis. 
Equation~\eqref{eq:lin-prob-model-cross-section} describes a cross-sectional regression of $y_{f}$, an indicator for whether firm $f$ donated to COVID-19, on pre-crisis shareholder composition (measured in December 2019) and an AGM indicator equal to one if the firm had an AGM scheduled within 90 days of January~15, 2020, the date of the first U.S. COVID-19 case \citep{holshue2020first}.\footnote{For the Russian invasion of Ukraine (February 24, 2022), we define treatment on a shorter window to avoid confounding from sanctions introduced shortly thereafter.}

Assume that $y_f=1$ if firm $f$ donates for COVID-19. Here, $\beta_{\text{treat}}$ measures the effect of shareholder visibility on the probability of donating, which is identified as the difference between two differences: (i) the difference in donation rates between firms with both an AGM and individual shareholders, $(\beta_1+\beta_2+\beta_{\text{treat}})$, and firms with an AGM but no individual shareholders, $(\beta_2)$, and (ii) the difference in donation rates between firms with individual shareholders but no AGM, $(\beta_1)$, and those with neither, $(0)$. Thus, the estimator simplifies to $\beta_{\text{treat}}$. A positive $\beta_{\text{treat}}$ therefore implies that visible individual shareholders pushed for donations in response to crisis-induced salience. Replacing $\textnormal{Ownership}_f$ with other types of blockholders (e.g., institutional investors) allows us to test for opposite effects, as predicted in Section~\ref{s:conceptual}.

\paragraph{Access to managers and other confounding factors} 
Because the interaction between crisis onset and pre-determined AGM timing is orthogonal to firm and shareholder characteristics, our design isolates variation in $v_A$ driven by the AGM spotlight. Other factors uncorrelated with AGM timing—such as founder status, shareholder preferences, or corporate governance—cancel out by design.

A separate concern is that changes in ownership around the crisis could bias the link between shareholder composition and firm behavior. We address this by fixing shareholder networks before the crisis, ensuring that post-crisis trading does not contaminate the estimated effect of pre-existing ownership structures.

\paragraph{Distributional consequences} 
To study the distributional effects of shareholder influence on other shareholders, Section \ref{s:distributional} adopts an intention-to-treat approach and adapts Equation \eqref{eq:lin-prob-model-cross-section} into a triple difference-in-differences framework, where the crisis start determines the post-period, and $\textnormal{AGM}_{f}$ and $\textnormal{Ownership}_{f}$ define the treatment. We will examine pre-trends and evaluate results with event studies.

\vspace{3ex}
Next, we estimate $\beta_{\text{treat}}$ in Equation \eqref{eq:lin-prob-model-cross-section} exploiting two difference crisis, COVID-19 (Section~\ref{s:covid}) and the Russian invasion of Ukraine (Section~\ref{s:russia}) to ensure the external validity of the influence mechanism we study.

\section{Shareholder Influence During the Pandemic}\label{s:covid}
This section tests the framework from Section~\ref{s:conceptual} using the COVID-19 pandemic as a laboratory. Section~\ref{s:data} introduces the data, Section~\ref{s:results} presents the main findings that individual shareholders pushed for prosocial actions while financial shareholders opposed them, and Section~\ref{s:mechanism} explores mechanisms and robustness.

\subsection{Data} \label{s:data}
We construct a comprehensive dataset of S\&P 500 firms by integrating financial information from Compustat with ownership data from LSEG (formerly Refinitiv). We augment these records with ESG ratings from MSCI and firm-level branch data from Orbis. To track corporate prosociality during the COVID-19 pandemic, we manually compile a dataset of charitable donations by auditing firm disclosures, press releases, and media reports. Specifically, we searched Google News and corporate websites for donation announcements made between January 15 and April 15, 2020. Appendix Table \ref{tab:covid_contributions} details the source material and provides specific examples of these disclosures.

\input{Tables/summary_statistics_new}

Table \ref{tab:summary-meeting} describes our data. The first panel reports financial and operational characteristics. The second panel describes shareholder composition with a focus on individual and financial investors (defined as banks, mutual funds, and insurers).\footnote{Financial investors, defined analogously, are among the largest institutional investors and account for 87\% of institutional blockholders (shareholders with stakes of at least 5\%). Institutional blockholders also include endowments, pension funds, private equity, and sovereign wealth funds.} The third panel summarizes firms’ prosocial characteristics, including ESG scores, past donations, and donation data collected from media reports, company disclosures, and internet searches. The final panel defines our treatment: firms that held their 2020 AGM before April 15, 2020, based on ISS data. We restrict the sample to 482 U.S. firms with financial and operational data as of December 2019 and not subject to mergers or exit from the index during the entire 2020 year.\footnote{COVID-19 case and death counts are sourced from Johns Hopkins University \citep{dong2020interactive}.}

\paragraph{Balance checks}
Columns 1-3 report distributional statistics (25th, 50th, and 75th percentiles), while Column 4 reports means. Importantly, Columns 5-8 compare treated and control firms, with \emph{t}-tests indicating no significant differences across key observables. Financial characteristics (e.g., market capitalization) and operating metrics (e.g., revenue, EPS, and employment) are balanced across groups. Shareholding structures are also similar: both groups display comparable levels of individual shareholders, although treated firms have slightly less institutional and financial blockholders with a 5\% share (18\% vs. 21\%).\footnote{This does not jeopardize identification because we allow selection in ownership levels and identify the incremental effect from the Ownership $\times$ AGM (crisis-onset) interaction, conditional on controls and fixed effects.} Donation behavior is likewise balanced. First, treated and control firms display similar 2019 donations as a share of revenue (a proxy for firm size), suggesting that AGM timing does not influence a firm's stance toward donations. Second, by April 15, 2020, treated and control firms had donated an average of \$7.36 million for COVID-19 relief---approximately 0.1\% of revenue.\footnote{Dollar figures exclude in-kind contributions (e.g., medical supplies), which are difficult to value.} We find no substantial difference in donation intensity, although treated firms were slightly more likely to donate before April 15 (p-value = 0.09).  Overall, these balance checks support the exogeneity of AGM timing with respect to firm characteristics and pre-crisis donation behavior.

\paragraph{S\&P 500 sample} Focusing on US-listed S\&P 500 firms, the sample size includes 482 firms, which is in line with recent cross-sectional studies leveraging field experiments or surveys. In our setting, most S\&P 500 firms hold AGMs in the spring. As a result, 47 firms in our sample held their AGMs between January 15 and April 15, 2020, representing 10\% of the sample. This imbalance introduces attenuation bias in the estimation of $\beta_{\text{treat}}$ in Equation~\eqref{eq:lin-prob-model-cross-section}, making our estimates conservative.

\subsection{Main Results} \label{s:results}

\subsubsection{Event-Study Evidence} 
Before turning to the estimates of $\beta_{\text{treat}}$ from \eqref{eq:lin-prob-model-cross-section}, we present event-study evidence on COVID-19 donations by AGM timing and shareholder composition to graphically illustrate the identifying variation. We estimate:
\begin{align}\label{eq:agm_event_study}
\textnormal{Donated by April 15}_f
  &= \sum_{w \in \mathcal{W}} \beta_{w}\,\text{AGM}_{w,f} \notag\\
&\quad + \sum_{w \in \mathcal{W}} \beta_{\text{Ind,w}}\,\text{AGM}_{w,f}\times \text{Individual}_f \\
&\quad + \sum_{w \in \mathcal{W}} \beta_{\text{Fin,w}}\,\text{AGM}_{w,f}\times \text{Financial}_f \notag\\
&\quad + \beta_{\text{Ind}}\,\text{Individual}_f + \beta_{\text{Fin}}\,\text{Financial}_f
    + \sigma_{s(f)} + \iota_{i(f)} + \varepsilon_{f}\,, \notag
\end{align}
where $\mathcal{W}$ denotes the set of weeks from January 15 to June 15. The dependent variable equals one if firm $f$ donated by April 15, 2020. This cutoff captures the 90-day window after the first U.S. COVID-19 case and excludes the period when the SEC permitted firms to postpone or reschedule AGMs \citep{securities2020guidance}.\footnote{Among the seven S\&P 500 firms with AGMs between April 7 and April 15, 2020, all had similar dates in 2019, suggesting no endogenous shifts. Appendix Figure~\ref{fig:meetings}, Panel (b), shows that most 2020 timing changes involved firms with May AGMs in 2019.} The omitted category is the week of June 15, so coefficients measure differences relative to firms with AGMs in that week.

Note that this is a cross-sectional regression: the $\{\beta_w\}$ terms capture how the probability of donation varies with AGM timing, while $\{\beta_{\text{Ind},w}\}$ and $\{\beta_{\text{Fin},w}\}$ capture how these effects differ with shareholder composition. We define $\text{Individual}_f=1$ if firm $f$ has at least one individual blockholder, and $\text{Financial}_f=1$ if it has at least one financial institution with a stake of 5\% or more.\footnote{LSEG data records only relatively large shareholders. Our results are robust to using other data sources such as ORBIS \citep{bajgar2020coverage,arndt2023should}.} To avoid endogeneity from crisis-driven trading, we measure shareholding as of December 2019.\footnote{Given our choice of indicator variables, this restriction is inconsequential. For instance, 66\% of individual shareholders with a 5\% share in December 2019 (at the start of our sample) were already holding such a share in January 2010 (ten years before the sample).} Finally, $\sigma_{s(f)}$ and $\iota_{i(f)}$ are state and industry fixed effects.

\begin{figure}[!htbp]
    \centering
       \caption{How AGM date and shareholder composition impact the probability of donating by April 15---Cross-sectional evidence}
       \label{fig:agm_event_study}
    \includegraphics[width=.9\textwidth]{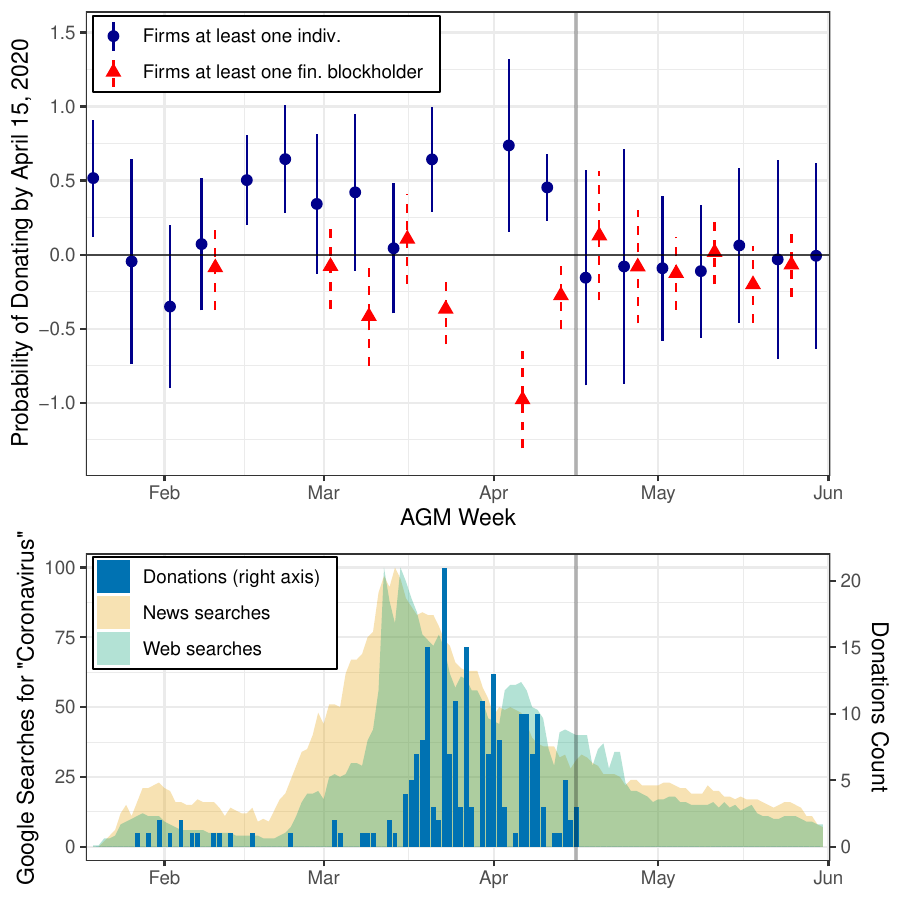}
    \begin{minipage}{\textwidth} 
{\footnotesize Note: The top panel plots estimates of $\{\hat{\beta}_{\text{Ind},w}\}$ (blue circles) and $\{\hat{\beta}_{\text{Fin},w}\}$ (red triangles) from \eqref{eq:agm_event_study}, which regresses an indicator for whether a firm donated between January 15 and April 15, 2020, on the interaction between shareholder types and indicators for the firm’s AGM week. Blue circles report the coefficient for firms with at least one individual blockholder, while red triangles report the coefficient for firms with at least one financial institutional blockholder ($>5\%$). The gray vertical bar denotes the April 15 cutoff for the dependent variable. We include state and industry fixed effects. Ownership data is fixed as of December 2019 to ensure predetermined status. Vertical bars denote 95\% confidence intervals (CI). Standard errors are clustered by industry. The sample includes S\&P 500 firms; ownership data is from LSEG; AGM dates are from ISS; donations are hand-collected. The bottom panel shows the evolution of news (yellow) and web  (green) searches across weeks and compare them to the count of firms donating for covid relief (blue bars, right axis). Donations are capped at April 15. We consider only the first donations for each firm to avoid double counting across different news outlets. Donations are hand-collected; search data are from Google.
\par}
\end{minipage}
\end{figure}

The top panel of Figure~\ref{fig:agm_event_study} plots the estimated $\{\hat{\beta}_{\text{Ind},w}\}$ (blue circles) and $\{\hat{\beta}_{\text{Fin},w}\}$ (red triangles), while the bottom panel documents the evolution of the crisis in the media and the weekly count of firm donations. Note that the x-axis in the top panel represents the timing of the firm's AGM, whereas in the bottom panel it represents calendar time. Specifically, the bottom panel plots Google Trends searches for the term ``Coronavirus'' (a measure of pandemic salience) and the volume of donations (right axis) across weeks.

Three patterns emerge that support our conceptual framework. First, for firms with AGMs falling within the peak crisis window (late February to mid-April), the presence of an individual blockholder significantly increases the probability of donating. This effect grows as the pandemic intensifies, consistent with rising reputational gains available to prominent shareholders during periods of high public attention. Second, we find a symmetric but opposite effect for financial blockholders: these firms are significantly \emph{less} likely to donate when their AGM falls in the crisis window, consistent with type-$B$ shareholders opposing costs that yield limited visibility gains.

Third, the effects for both groups converge toward zero for firms with AGMs after April 15. Although this is a cross-sectional specification, this convergence acts as a placebo test analogous to a parallel-trends requirement. It suggests that the differential behavior of these firms is not a permanent trait but is specifically triggered by the alignment of the AGM ``spotlight'' with the onset of the crisis.

\subsubsection{Treatment Effect Estimates} 
We next estimate shareholder influence directly using the linear probability model in Equation~\eqref{eq:lin-prob-model-cross-section}, with the same dependent variable and fixed effects as in the event study \eqref{eq:agm_event_study}. We also control for firm size  as larger firms may find donations less expensive.\footnote{We use the log of total assets as of December 2019; results are robust to alternative size measures, as they are highly correlated. In Section~\ref{s:robustness} we extend the framework to \emph{panel} regressions with firm fixed effects and industry-by-time fixed effects, both to absorb unobserved, time-invariant firm characteristics and to net out industry-specific policy or market shocks (see also Appendix \ref{apndx:r1}).}

Table~\ref{tab:reg-results-block-static} reports the results. Across columns, \textit{Ownership}$_f$ measures the December 2019 equity share held by individual (Columns 1–3) or financial (Columns 4–6) blockholders, defined at different thresholds: above 10\%, 5\%, or below 2\%. To facilitate comparisons across columns, \textit{Ownership}$_f$ is standardized.

\input{Tables/block_groupped_inst.tex}

The interaction term $\hat{\beta}_{\text{treat}}$ is positive and significant in Columns 1 and 2: a one standard deviation increase in individual shareholder ownership raises the probability of donating by almost 9 percentage points (pp). By contrast, small individual shareholders exert no significant influence (Column 3). These patterns are consistent with the theoretical prediction that prominent individual investors, who face higher visibility gains, are more likely to push for prosocial behavior. Notably, $\hat{\beta}_1$ and $\hat{\beta}_2$ are insignificant across all columns, suggesting no selection on observables, further supporting the exogeneity of the AGM treatment.

Turning to financial investors, we find the opposite pattern. In Columns 4 and 5, firms with high financial ownership are significantly less likely to donate, with a one standard deviation increase in financial blockholding reducing donations by roughly 17pp. Smaller financial investors (Column 6) have no significant effect. Again, there is no evidence of selection bias, as $\hat{\beta}_1$ and $\hat{\beta}_2$ remain indistinguishable from zero.

Taken together, the estimates indicate sizable effects. In Column 1, a standard deviation increase in individual or family blockholding (10\%+ stake) makes a donation 18\% more likely. In contrast, firms with large financial blockholders are 37\% less likely to donate (Column 4).

\subsubsection{Extensions} 
Appendix Tables~\ref{tab:horse_race}--\ref{tab:covid_slack} summarize three
extensions. First, Table~\ref{tab:horse_race} shows that the main estimates are
unchanged when we include both shareholder types in the same specification and
add standard firm controls (cash-to-assets, market valuation, and ESG scores as of
December~2019). Results are qualitatively similar when using total individual ownership shares rather than blockholder indicators. This is expected, as the presence of individual blockholders is an important driver of aggregate individual ownership levels in our sample.

Second, Table~\ref{tab:covid_inst} broadens the institutional category beyond
financial institutions to all institutions (including foundations, pension funds,
private equity, and sovereign wealth funds). The estimates remain in line with
Table~\ref{tab:reg-results-block-static}, though less precise; in particular, the
effects are strongest among the very largest institutional blocks, consistent with
the largest diversified asset managers playing a distinct governance role.

Third, Table~\ref{tab:covid_slack} shows that the effects depend on firms’
financial slack. Individual-shareholder influence is concentrated among firms
with above-median slack, whereas institutional influence is concentrated among
firms with below-median slack, consistent with a reputational--cost trade-off:
reputational gains are easier to pursue when the firm can absorb the associated
costs, while opposition is strongest when those costs are hardest to bear.

In the next section, we uncover the underlying mechanism in greater detail.

\subsection{Mechanism} \label{s:mechanism}

\subsubsection{The Pass-Through of Image Gains}\label{s:image}
First, we test whether whether individual and institutional investors experience different levels of public visibility following a donation, as our conceptual framework in Section~\ref{s:conceptual} assumes. We proxy for public exposure using shareholder-level Google search activity around donation events. Specifically, we estimate the following event-study:
\begin{equation}\label{eq:Google}
\begin{aligned}
    y_{jft} &= \sum_{d=-9, \ d \neq -6}^{9}\gamma_d \, \textnormal{News}_{f(j),t+d} \times \textnormal{Individual}_{j} + \alpha_{jf,m(t)} + \tau_{ft} + \tau_{t,\kappa(j)}+ \varepsilon_{ift},
\end{aligned}
\end{equation}
where $y_{jft}$ is the cumulative Google Trends score for shareholder $j$ in firm $f$ over a five-day window starting on day $t$. The indicator $\textnormal{News}_{f,t+d}$ equals one on each day $d$ relative to a donation announcement by firm $f(j)$ in which $j$ is a shareholder. 

The interaction with $\textnormal{Individual}_j$ captures differential visibility for individual shareholders. Because we focus on a short window, the specification includes firm–shareholder-month fixed effects ($\alpha_{jf,m(t)}$), which absorb time-invariant effects such as whether shareholder $j$ holds currently or held in the past a C-level position or directorship at firm $f$. To identify differences across shareholders within the same firm-day, we include day-by-firm ($\tau_{ft}$) fixed effects. These absorb everything happening to firm $f$ on day $t$ that is common to all shareholders of that firm, including the donation announcement. Therefore, they control for the event itself and any firm-day shocks: firm news intensity, market moves, general attention to the firm that day, industry-wide news hitting that firm on that day, etc. Standard errors are clustered at the firm–shareholder and day levels.

\begin{figure}[!ht]
    \caption{Individual-institutional shareholder gap in Google Trend scores after a firm donates}
    \label{fig:Google_10}
        \centering
        \includegraphics[width=0.66\textwidth]{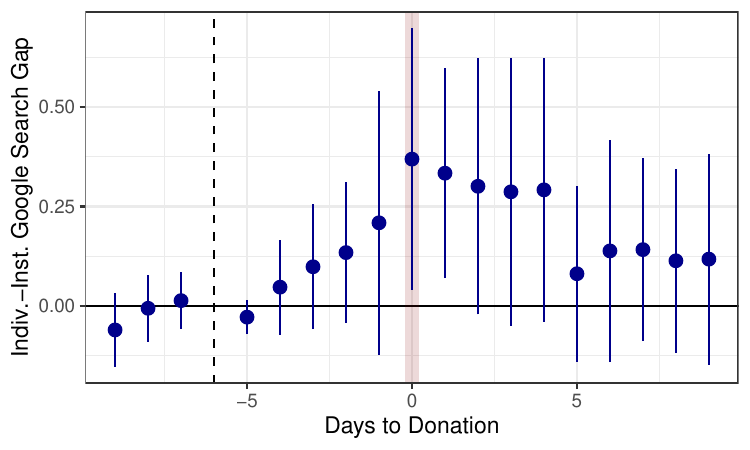} 
\begin{minipage}{1 \textwidth} 
{\footnotesize Note: This figure reports the estimated $\gamma_d$ from Equation \eqref{eq:Google} which regresses a standardized measure of a shareholder's Google Trends score (cumulated over five days) in day $t$ on the interaction between an indicator that the shareholder is an individual shareholder and time dummies indicating day $t$ is $d$ days away from the firm's Covid donation announcement. We include shareholder-firm-month fixed effects and firm-day fixed effects. Vertical bars represent 95\% confidence intervals, with standard errors clustered at the firm-shareholder and day level. Ownership data from LSEG; AGM dates from ISS; search data from Google; donations are hand-collected.
\par}
\end{minipage}
 \end{figure}

Figure~\ref{fig:Google_10} plots the estimated $\{\hat{\gamma}_d\}$ coefficients. Search activity is flat prior to the donation, indicating that individual and financial shareholders are equally searched before the event, but rises sharply afterward for individual shareholders, consistent with a reputational gain. The effect is economically large---an increase of 40\% of a standard deviation, effectively doubling the value taken by these shareholders' Google search index on average.

\subsubsection{Donate Directly or Through the Financial Footprint?}  \label{s:footprint}
Do financial investors oppose corporate donations during crises? To shed light on this question and better understand the trade-offs faced by institutional investors, we compare the donation behavior of financial corporations in the S\&P 500 (e.g., BlackRock, Bank of America) with the donation decisions of the firms in which they hold equity stakes. 

Specifically, we construct two vectors: one records whether each of the 37 financial firms in our sample made a donation by April 15, 2020 (binary indicator), and the other reports the share of portfolio firms (within the S\&P 500) that donated. A positive correlation between these two vectors suggests alignment between the donor behavior of financial investors and that of the firms they hold; a negative correlation indicates a divergence—i.e., financial firms may withhold donations while their portfolio companies contribute. Although we limit the analysis to donations of S\&P 500 firms due to data constraints, we expect the misalignment to be even more pronounced for smaller, more influenceable firms outside the index.\footnote{We focus on financial corporations within the S\&P 500 that hold shares in other S\&P 500 firms for consistency and data availability.}


\input{Tables/correlations}

Table \ref{tab:correlations} presents this correlation. Across rows, we vary the minimum ownership share for a firm to be considered in the portfolio of a financial firm. 

Column 2 weights donations by the investor’s ownership stake in each firm and shows weakly positive, but statistically insignificant, correlations—driven largely by a few large holdings. Column 3 switches to unweighted averages and finds consistently negative correlations when the ownership threshold exceeds 2\%, suggesting that financial investors may discourage donations by the firms they own. This pattern becomes more pronounced in Column~4, which restricts the analysis to firms with AGMs at the crisis onset. We find that focusing on variation plausibly driven by predetermined AGM timing, which helps interpret the correlation patterns, estimates are negative and statistically significant, approaching $-1$ at higher thresholds. 

Overall, financial institutions do not oppose \textit{all} donations. They donate themselves but tend to discourage donations of portfolio firms when they bear the financial cost without gaining visibility. These findings are consistent with the limited reputational upside shown in Figure~\ref{fig:Google_10} and support the idea that these donations have distributional consequences across shareholders, a point on which we will return in Section~\ref{s:distributional}.



\subsubsection{Robustness Checks}\label{s:robustness} 
Appendix \ref{apndx:rob} investigates potential confounding factors that may incentivize donations. We list the main results below.

\paragraph{Different AGM treatment windows} To test the sensitivity of our three-month window, we re-estimate Equation \eqref{eq:lin-prob-model-cross-section} over varying horizons ranging from one to four months (from January 15 to any date between February 15 and May 15). As shown in Appendix Figure \ref{windows_covid}, the estimated $\hat{\beta}_{\text{treat}}$ coefficients remain consistent with Table \ref{tab:reg-results-block-static}. These findings indicate that the observed effects are concentrated during the height of the pandemic’s salience, as documented in the bottom panel of Figure \ref{fig:agm_event_study}.

\paragraph{Validation using total 2020 donations} Appendix Table~\ref{tab:donations_2020} helps address three concerns: that treated firms merely shift donations earlier---the idea that control firms would catch up later---that Covid donations substitute for other philanthropy, and data quality. Using total 2020 donations scaled by revenue as outcome variable (data from LSEG), we still find effects only for firms with an early AGM and large individual blockholders: at treated firms, a one–standard-deviation increase in individual blockholding raises annual giving by about \$1 per \$1,000 of revenue.\footnote{Additional tests suggest that higher donations rather than lower revenues drive this estimate.} In contrast, we find no effect for institutional and financial ownership, suggesting their negative effect is specific to salient, early-crisis giving rather than overall annual philanthropy.

\paragraph{Different measures of  shareholder-firm association}
Large shareholding is a proxy for a firm's association with its most visible shareholders. To refine it, we analyze the correlation between Google Trends data for each firm and its individual shareholders over 2010–2019. We compute the Spearman correlation (and p-value) between every firm-shareholder pair, focusing on shareholders owning at least 1\% of the firm's equity. We then compute the t-statistic of that correlation, and average it by firm. That gives us a measure of a firm's association to its shareholders. 
Using this measure for $\textnormal{Ownership}_f$ in Equation~\eqref{eq:lin-prob-model-cross-section},  Table \ref{tab:other_association} shows that those firms more associated with shareholders are more likely to donate, confirming our findings from Table \ref{tab:reg-results-block-static}. 

Overall, we favor shareholding as our measure of association for two reasons. First, Google Trends–based measures correlate strongly with blockholding: the correlation between this measure and ownership share is positive and significant at the 1\% level for each shareholder. Second, the Google-based measure requires specifying a training period. Shorter periods increase the salience of current events but reduce statistical power, while longer periods place more weight on them and may qualitatively change results. Estimates using the Google-based measure are nonetheless consistent with those in Section~\ref{s:results}. For these reasons, we rely on shareholding in the rest of the paper.


\paragraph{Pandemic intensity} We next examine whether the severity of the pandemic shaped shareholder influence on corporate donations  in Appendix \ref{apndx:r1}. To do so, we extend specification \eqref{eq:lin-prob-model-cross-section} to a panel setting with firm and day fixed effects, and interact the AGM treatment with both Ownership$_f$ and local measures of pandemic intensity--cumulative COVID-19 deaths in the firm’s headquarters state. The results align with those in Table~\ref{tab:reg-results-block-static}: as the pandemic worsens, individual shareholders are more likely to push for donations, while financial shareholders are more likely to oppose them. This heterogeneity may help explain why donations are not always synchronized with AGM dates, but instead respond flexibly to the evolving public health emergency.

\paragraph{Financial motives of shareholders and managers} We find no evidence that abnormal stock returns drive COVID-related donations (see Appendix \ref{apndx:r2}). Using a Fama–French–Carhart four-factor model, we estimate firm-level cumulative abnormal returns (CARs) around donation announcements. The event study reveals short-lived negative CARs, consistent with \citet{kruger2015corporate}, who finds that markets often react negatively to CSR news. These results suggest that financial markets did not reward donations, supporting the interpretation that reputational motives rather than shareholder value maximization underpinned these actions. Furthermore, our estimates rule out the hypothesis that managers donated to inflate stock prices. 

In a related context, \citet{liang2025disaster} exploit proximity to the fiscal year-end to distinguish between strategic donations and managerial agency. Our empirical design, which relies on the interaction between AGM timing and the presence of individual versus institutional blockholders, isolates the joint effect of shareholder identity and the early-crisis window, thereby controlling for such agency effects. Specifically, while agency problems may differ between firms with and without individual blockholders, there should be no systematic difference between two firms with individual blockholders whose AGM periods happen to be pre-determined at different times of the year. Moreover, our results are robust to the inclusion of fixed effects for the month of the fiscal year-end and remain unchanged when restricted to the subsample of firms whose fiscal year ended in the three months immediately preceding the pandemic (Appendix Tables \ref{tab:covid_fyr} and \ref{tab:covid_fyr2}.)

\paragraph{Consumer pressure} We test whether consumer pressure influenced donations by exploiting variation in firms’ local COVID-19 exposure, measured using branch-level weights across U.S. states. Estimating a linear probability model, Appendix \ref{apndx:r3} interacts this exposure with the number of branches as a proxy for consumer-facing operations. The results show no significant relationship between local COVID-19 severity and donation behavior, suggesting that consumer pressure was not a key driver of prosocial actions.

\paragraph{Competition} Appendix \ref{apndx:r5} shows that donations do not follow previous donations by firms in the same industry.  These findings are consistent with prior evidence that managers prioritize financial performance around AGMs, and that large crisis-related donations may be perceived as financially costly \citep{dimitrov2011s}.

\paragraph{Defense Production Act}
\sloppy To rule out confounding effects from government-mandated procurement, we address the potential influence of the Defense Production Act (DPA), which led firms like \emph{3M} and \emph{GM} to fulfill government contracts for essential goods during the COVID-19 pandemic. First, note that, since the relevant executive orders were issued after our sample period (April 20, 2020), they should not affect our main findings. Additionally, Appendix \ref{apndx:r6} shows that our results still hold if we were to exclude firms subject to DPA orders in 2020, confirming that our results are not driven by actual or anticipated DPA procurement.

\section{Influence on Russian Invasion of Ukraine}
\label{s:russia}

We next examine whether the shareholder influence patterns documented during the
COVID-19 pandemic generalize to a distinct geopolitical setting. Specifically, we
study how shareholders shaped firms’ responses to Russia’s invasion of Ukraine in
early 2022. This episode provides a demanding test of external validity. Decisions
to exit the Russian market were highly salient and economically costly---often
involving asset write-downs, operational restructuring, and foregone revenues
totaling billions of dollars \citep{jack2022exit}. Importantly, during the first
month of the conflict, such exits were not legally required, making them well
suited to isolate reputational motives from regulatory compliance.

\paragraph{Data}
Our sample consists of U.S.-listed firms tracked by \citet{sonnenfeld2022business}
and the website \texttt{leave-russia.org}. We focus on firms that publicly announced
their response to the invasion by March 23, 2022—one month after hostilities began. 
Restricting attention to this initial window allows us to capture voluntary actions
taken under intense public scrutiny, while minimizing confounding effects from
government sanctions introduced later. We further limit the sample to U.S.-listed
firms, for which SEC rules governing AGM timing apply. Table \ref{tab:russia_exits} provides examples of such exits.

The initial sample includes 188 firms. We exclude firms with no reported revenues
from Russia in the previous fiscal year, using LSEG data, yielding a final
sample of 165 firms with active Russian exposure.

\input{Tables/sumstat_groups_russia}

Table~\ref{apndx:summary-rus} reports summary statistics. Panel~(i) shows that firms
in this sample are substantially larger than those in the COVID-19 analysis. For
example, average market capitalization is roughly three times that reported in
Table~\ref{tab:summary-meeting}. Crucially, however, we find no statistically
significant differences in firm size between treated and control firms. Nor do we
detect differences in exposure to Russia: the share of revenues derived from Russian
operations is balanced across groups (\textit{p}-value $=0.30$).

Panel~(ii) reports ownership structure. Because firms in this sample are larger on
average, ownership is more diffuse, as reflected in lower concentration indices
(HHI). As a consequence, individual blockholders with stakes exceeding 5\% are
relatively rare. We therefore focus on individual shareholders holding at least 1\%
of equity.\footnote{All results are robust to using a 5\% threshold, but only two
treated firms meet this criterion.} While this threshold yields a statistically
significant difference in individual ownership between treated and control firms,
we interpret any resulting treatment effect as a lower bound: the prevalence of zero
ownership among many firms mechanically attenuates estimated influence.

Panel~(iii) presents ESG characteristics from MSCI and shows no systematic
differences between treated and control firms. The remaining panels classify firms
according to their announced response to the invasion. Fifty firms—about 30\% of
the sample—fully exited the Russian market (Grade ``A''); treated firms are less likely to fully exit. 
The remaining categories capture progressively weaker responses, down to firms that explicitly refused to exit (Grade ``F''). Our results are robust to excluding this final category.

\paragraph{Main results}
We estimate Equation~\eqref{eq:lin-prob-model-cross-section}, paralleling the COVID
analysis. Results are reported in Table~\ref{tab:regression_russia2}. Across columns, we vary the shareholder category (individual versus institutional) and ownership thresholds. All specifications control for firm size using log total assets, ensuring comparability with Table~\ref{tab:reg-results-block-static}.

\input{Tables/regression_russia_2}

Across all specifications, AGM timing alone is uncorrelated with exit decisions,
confirming its exogeneity in this setting. Ownership composition, by contrast,
matters. A higher share of dispersed ownership is associated with fewer exits for
individual shareholders (Column 3) and more exits for institutional shareholders (Column 6). This pattern aligns with the model: shareholders with limited ability or willingness to engage in costly influence are less likely to press management toward high-cost, high-visibility actions. As the mass of such shareholders increases, the preferred outcome of the competing shareholder type is more likely to prevail.

We begin with individual shareholders in Columns~(1)–(3). Column~(1) focuses on
firms with larger individual shareholders. A one–standard deviation increase in
their ownership share raises the probability of exiting Russia by about 10
percentage points (\textit{p}-value $=0.06$). Column~(2) restricts attention to
firms with below-median revenue exposure to Russia---that is, firms for which exit
is financially less costly. In this subsample, the effect more than doubles,
reaching 21 percentage points per standard deviation. This amplification is
consistent with a trade-off between reputational gains and financial losses: when
the economic cost of exit is limited, visible individual shareholders are more
likely to favor actions that yield public recognition. Column~(3) considers small
individual shareholders (ownership below 1\%). The estimated interaction effect
is positive but imprecise, consistent with earlier evidence that small individual
shareholders exert little influence.

Columns~(4)–(6) turn to institutional shareholders. Columns~(4)
and~(5) focus on institutions with at least 5\% ownership. While the interaction
effect is small and statistically insignificant on average, it becomes negative
and significant when restricting the sample to firms with above-median exposure
to Russia. In other words, institutional shareholders become active precisely
when the financial stakes are large, consistent with their emphasis on preserving
firm value rather than capturing reputational rents. In contrast, Column~(4) shows that small institutional shareholders do not exert detectable influence on exit decisions. Larger institutional blockholders, however, intervene selectively consistent with a trade-off between the cost of influence and the financial loss from not doing so.

\begin{figure}[!ht]
    \caption{Individual-institutional shareholder gap in Google Trend scores after a firm exits Russia}
    \label{fig:google_russia}
        \centering
        \includegraphics[width=0.66\textwidth]{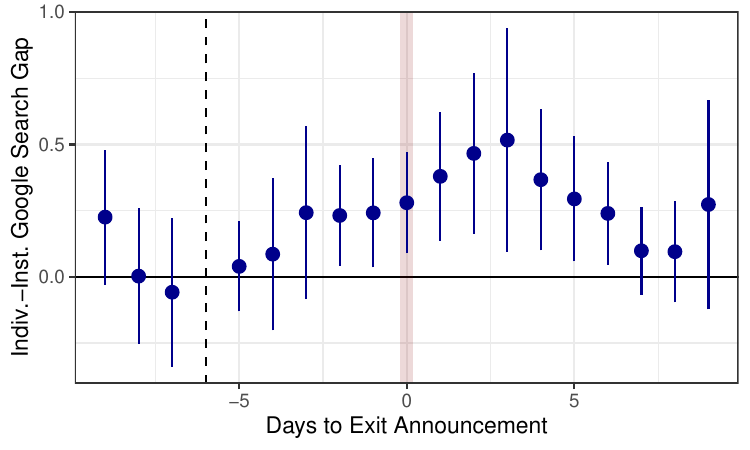} 
\begin{minipage}{1 \textwidth} 
{\footnotesize Note: This figure reports the estimated $\gamma_d$ from Equation \eqref{eq:Google} which regresses a standardized measure of a shareholder's Google Trends score (cumulated over five days) in day $t$ on the interaction between an indicator that the shareholder is an individual shareholder and time dummies indicating day $t$ is $d$ days away from the firm's exit from Russia announcement. We include shareholder-firm-month fixed effects and firm-day fixed effects. Vertical bars represent 95\% confidence intervals, with standard errors clustered at the firm-shareholder and day level. Ownership data from LSEG; AGM dates from ISS; exits are from \cite{sonnenfeld2022business} and \texttt{leave-russia.org}.
\par}
\end{minipage}
 \end{figure}
 
\paragraph{Mechanism} We estimate Equation \ref{eq:Google} using firms’ announcements to exit Russia as the focal event. We include as control group the shareholders of firms with exposure to Russia that do not appear on the Yale CELI list. Figure \ref{fig:google_russia} plots the estimated coefficients. We again observe a flat pre-trend: in the weeks leading up to the announcement, Google search activity does not differentially evolve for individual versus institutional shareholders. In contrast, around the announcement date, Google searches for individual shareholders rise by roughly 0.5 standard deviations relative to other shareholders.

\paragraph{Robustness checks} 

Estimates remain stable when varying the threshold for being considered a blockholder and when considering only financial investors (Appendix Table \ref{tab:regression_russia_fin}), when removing airplane carriers (Table \ref{tab:regression_russia_airlines}), an industry that received immediate government bans, or removing firms that took no action (Table \ref{tab:regression_russia_grade_f}). Appendix Figure \ref{fig:excess_returns_russia} shows that the decisions of exiting Russia  resulted in  similar cumulative abnormal returns as donating for COVID.

\section{The Shared Costs of Influence}\label{s:distributional} 

What are the implications for ``unheard'' shareholders when others influence managerial decisions? Consider a firm with revenues \( R \), operating costs \( C \), and operating income \( \pi = R - C \). The firm can reinvest \( \pi \) to support future income, distribute it as dividends, or yield to shareholder demands by spending \( \pi \) on non-productive activities, such as donations (Section \ref{s:covid}) or abrupt exits from markets like Russia (Section \ref{s:russia}). Such choices can reduce the discounted value of future incomes, imposing net losses on shareholders who do not benefit privately (e.g., via image returns). 

\paragraph{Measuring productivity} We leverage exogenous variation in AGM and crisis timing to measure productivity changes with two complementary metrics: \emph{operating income}, which reflects how effectively the firm converts costs into revenues given demand \citep[e.g.,][]{harberger1954monopoly}, and \emph{market valuation} (the market-to-book ratio), which captures shareholder losses through stock-price changes.\footnote{Dividends are not a good measure of rent extraction because firms often adopt dividend policies aimed at keeping payouts constant over time \citep[e.g.,][]{la2000agency}.}%

As productivity influences both \( R \) and \( C \), identifying productivity changes requires isolating demand fluctuations affecting \( R \). Consumer demand likely reacted to both COVID-19 and the invasion of Ukraine through lockdowns and protests. However, it is reasonable to assume that demand changes are unrelated to the share of individual or institutional shareholders and to the AGM treatment before a crises as shown in Section \ref{s:robustness}. 

\paragraph{Empirical design} To assess the differential influence of shareholder types, we implement a triple-differences (DDD) design. Treated firms are those holding AGMs at the onset of each crisis (Sections~\ref{s:results} and~\ref{s:russia}), while others serve as controls. We further split firms based on the presence of large individual shareholders just before the crisis. The DDD estimator combines two difference-in-differences (DD) comparisons—each interacting a post-crisis indicator with an AGM-timing indicator—but estimated separately for firms with and without individual blockholders. Identification relies on the standard DDD assumption that the \emph{difference in trends} between these two groups would have remained constant in the absence of treatment; that is, parallel trends need not hold within each group, but only in their difference. This relaxed condition allows for level differences or group-specific shocks, as long as they evolve similarly across the two groups \citep{olden2022triple}.\footnote{For example, if NGOs systematically target firms during AGMs \citep{fioretti2026ngo}, this would not threaten identification unless targeting varies with shareholder composition. Following \citet{olden2022triple}, we include control variables to account for compositional differences between groups, thereby addressing potential biases arising from observable characteristics that may influence treatment state or group assignment.}

\subsection{Data and Empirical Strategy} 
We perform our analysis on two distinct samples. The first sample examines the pandemic's effects, focusing on the 1,000 largest US-listed firms as of December 2019. For the Ukraine invasion case, the sample is restricted to firms with dealings in Russia prior to 2022, as identified by LSEG. The final sample for the Ukraine-Russia case consists of the largest 1,153 US-listed firms with Russian ties.\footnote{As a result, the first sample includes larger firms than the second, with average operating income and market capitalization approximately four times greater. The Ukraine-Russia sample is also more heterogeneous, exhibiting operating incomes nearly twice as skewed as those in the COVID-19 sample. For both samples, historical ownership and financial data are sourced from LSEG and Compustat, while AGM dates are from ISS.}

To assess the costs borne by unheard shareholders and the mechanism, we estimate the following event study in an intent-to-treat framework: 
\begin{equation}\label{eq:triple_EV}
    y_{ft} = \sum_{k \neq 0} \theta_k \; \mathds{1}_{\{t-\tau = k\}} \times \text{AGM}_{f} \times \text{Individual}_{f}  + \textbf{X}_{ft} \; \textbf{$\beta$} +\alpha_{f} + \iota_{i(f)t}  + \varepsilon_{ft},
\end{equation}
where AGM$_{f}$ is one for firms holding AGMs at the onset of a crisis, and Individual$_{f}$ is the standardized share held by individual shareholders with at least a 5\% equity stake at firm $f$ in the reference period.\footnote{To avoid outliers, we winsorize yearly variables at 1\% level and quarterly variables at 2.5\%.} $\tau$ is the reference period (end-2019 or end-2021), and $k$ indexes time relative to $\tau$ through the indicators $\mathds{1}_{\{t-\tau = k\}}$; we omit $k=0$ as the baseline. Thus, we interpret the coefficient vector $\{\theta_k\}$ as intent-to-treat. The controls in $\textbf{X}_{ft}$ include the interactions of time indicators with the AGM indicator and with shareholder composition ($\sum \gamma_k \; \mathds{1}_{\{t-\tau = k\}} \times \text{AGM}_{f}$ and $\sum \delta_k \; \mathds{1}_{\{t-\tau = k\}} \times \text{Individual}_{f}$) to control for group-specific time trends. $\alpha_{f}$ and $\iota_{i(f)t}$ denote firm and industry-by-year (or industry-by-quarter) fixed effects. 

On the left-hand side, $y_{ft}$ denotes either a yearly accounting variable such as operating income on assets or a quarterly market valuation measure such as the market-to-book ratio. All outcomes are standardized. As a result, the coefficients $\theta_k$ are expressed in standard deviation units of $y_{ft}$, allowing for comparability across specifications.\footnote{Standardizing $y_{ft}$ by its standard deviation improves robustness to additive and multiplicative transformations, mitigates sensitivity to outliers and specification searching, and avoids inflated significance estimates from arbitrary scale choices  \citep{mitton2024economic}. That said, the results are qualitatively unchanged when $y_f$ is not standardized. Quarterly data are used for market valuation outcomes to better capture high-frequency market reactions.}

Finally, the regression residuals are likely correlated within states (e.g., state policy) and across firms in the same industry (e.g., demand responses). Thus, we cluster the standard errors at these levels.


\subsection{Methodology and Results} 

\subsubsection{Crisis 1: Covid Pandemic} \label{s:dist_covid}
To examine the impact of the COVID-19 crisis, we estimate  \eqref{eq:triple_EV} using an indicator variable, $\text{AGM}_f$, which equals 1 if firm $f$ held an AGM within 90 days of the COVID-19 outbreak in the U.S. (defined as January 15, 2020). 

\begin{figure}[ht]
    \caption{Documenting rents: covid case} \label{fig:covid}
  \begin{subfigure}{7cm}
    \centering\includegraphics[width=7cm]{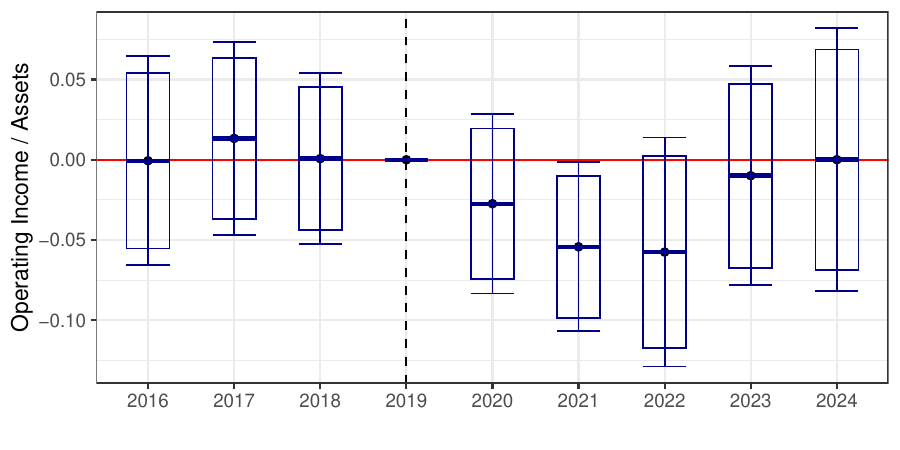}
    \caption{Operating income/Asset}
  \end{subfigure}\hfill
  \begin{subfigure}{7cm}
    \centering\includegraphics[width=7cm]{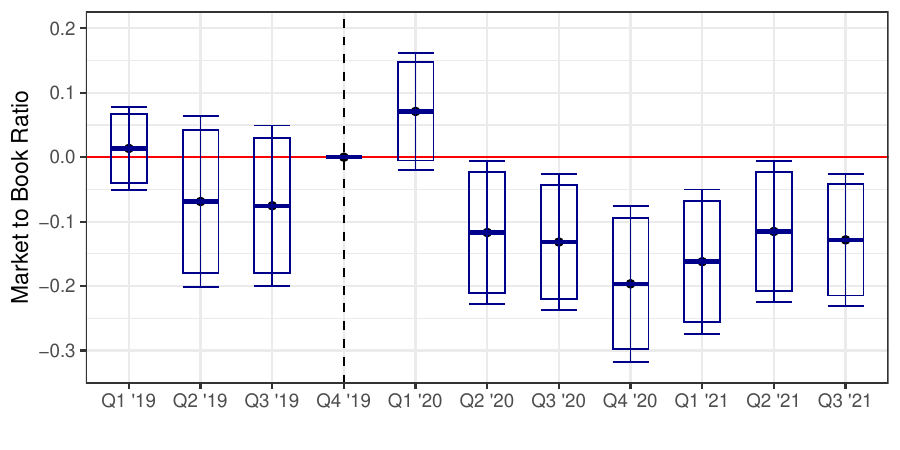}
    \caption{Market valuation}
  \end{subfigure}
  
  \begin{subfigure}{7cm}
    \centering\includegraphics[width=7cm]{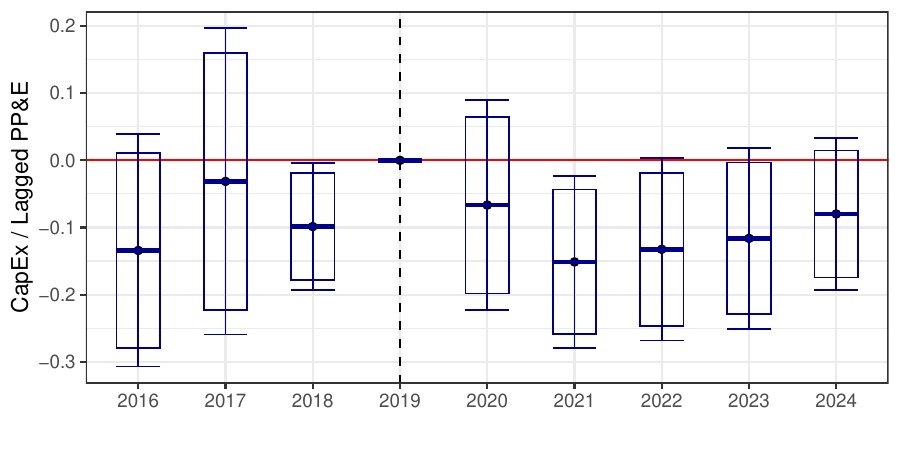}
    \caption{Investments}
  \end{subfigure}\hfill%
  \begin{subfigure}{7cm}
  \captionsetup{justification=centering}
    \centering\includegraphics[width=7cm]{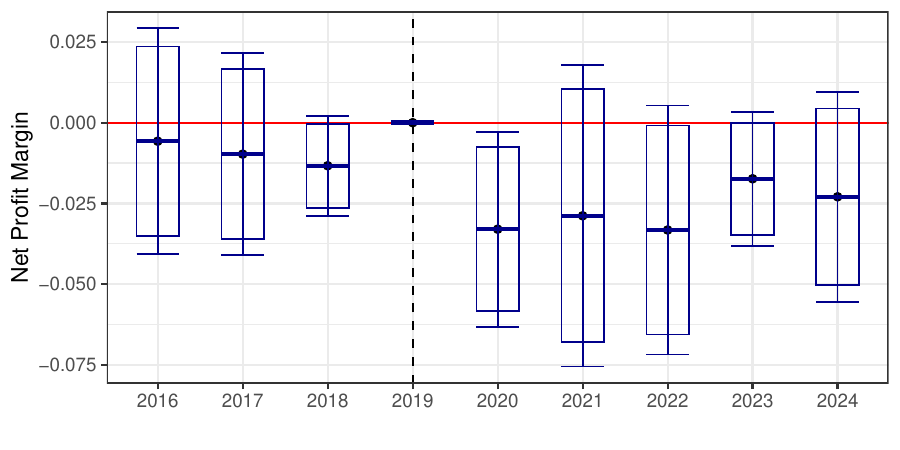}
    \caption{Net profit margin (Net income/Sales)}
  \end{subfigure}
    \begin{minipage}{1 \textwidth} 
    {\footnotesize Note: This figure plots the estimated coefficients from (\ref{eq:triple_EV}), where the coefficients of interest are the interaction between time indicators, an indicator equal to one if the firm holds  an AGM between January 15 and April 15, 2020, and the share of individual shareholders with at least a 5\% share. The dependent variables are standardized. The regressions include firm and time-by-industry fixed effects. Regressions reported in panels (c) and (d) also include log total assets to control for firm size. Standard errors are clustered at the industry and state levels. Error bars (boxes) report the 95\% (90\%) CI. Vertical dashed lines indicate the first covid case in the USA. Balance sheet data from Compustat; shareholding data from LSEG.
    \par}
\end{minipage}
\end{figure}

The top panels of Figure \ref{fig:covid} show the effect of a one--standard deviation increase in the share of individual blockholders ($>5\%$) at treated firms on operating income scaled by assets (Panel a) and market-to-book valuation (Panel b). Higher individual blockholding in firms with AGMs at the onset of the crisis led to a decline in operating income starting in 2020. The effect became economically meaningful in 2021--2022, corresponding to a 0.07--0.08 decline of a standard deviations of operating income per one standard deviation increase in blockholder share. This translates into a 1.16\% drop in operating income per 1pp increase in individual blockholder ownership. The effect is statistically significant ($p<0.05$) in 2021 and 2022, before reverting to pre-pandemic levels.

In contrast, the decline in market valuation manifested more rapidly: the market-to-book ratio dropped by over 10\% of a standard deviation as early as the second quarter of 2020 and fell to 20\% of a standard deviation in the fourth quarter of 2021. These drops correspond to approximately a 2\% decline in the unstandardized market-to-book ratio per 1pp increase in the ownership of individual blockholders.\footnote{In Q1 2020, the market-to-book ratio at treated firms rose by 0.07 (s.e. 0.13) standard deviations per one–standard deviation increase in blockholding (insignificant). Donations reduce retained earnings, lowering book equity and mechanically raising this ratio. As this is intent-to-treat and not all treated firms donated (Section \ref{s:covid}), the difference with control group is insignificant.
}

\paragraph{Mechanism: fewer investments at treated firms} Costly donations led to fewer investment in 2021: Panel (c) estimated a drop by 0.15 of a standard deviation relative to 2019, as standardized measure by capital expenditures scaled by lagged property, plants, and equipment \citep[as in][]{dessaint2019noisy}. In levels, this drop is 2\% per 1pp increase in the share of individual blockholders, which is similar in magnitude to the estimate in \citet{masulis2015agency}, who study donations to nonprofits as a form of managerial rent extraction.

Thus, the decline in operating income likely reflects higher costs rather than lower revenues. Panel (a) of Appendix Figure \ref{fig:es_sales_opincome} supports this claim by showing that sales relative to assets remained constant, consistent with Section \ref{s:robustness}, which indicates that COVID-related donations were not demand-driven. Panel (d) of Figure \ref{fig:covid} further shows that standardized net profit margin fell by 0.03 standard deviations per one-standard deviation increase in individual blockholding at treated firms since 2020 (equivalent to a 20\% decline in levels per 1pp increase in blockholder ownership). Unlike operating income in Panel (a), the net profit margin also incorporates non-operating costs, indicating an overall decline in profitability unrelated to sales.


\subsubsection{Crisis 2: Invasion of Ukraine} \label{s:dist_russia}

\begin{figure}[ht]
    \caption{Documenting rents: Russian invasion of Ukraine }\label{fig:russia}
  \begin{subfigure}{7cm}
    \centering\includegraphics[width=7cm]{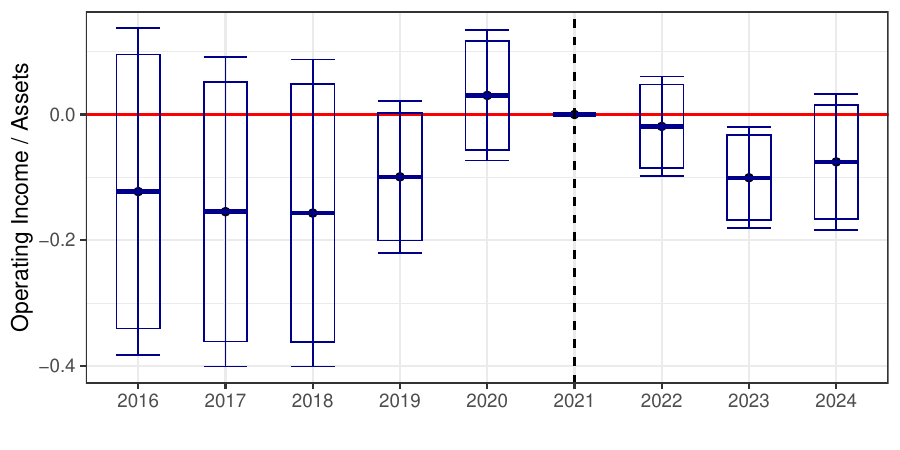}
    \caption{Operating income/Asset}
  \end{subfigure}\hfill
  \begin{subfigure}{7cm}
    \centering\includegraphics[width=7cm]{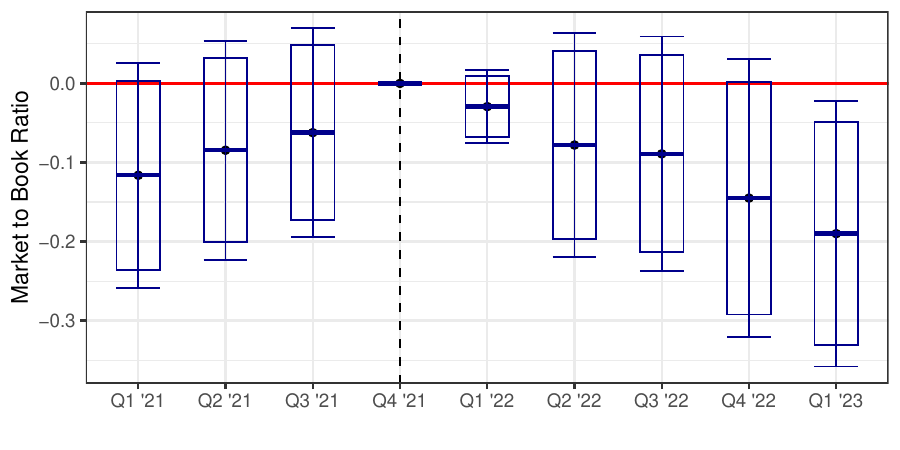}
    \caption{Market valuation}
  \end{subfigure}
  
  \begin{subfigure}{7cm}
  \captionsetup{justification=centering}
    \centering\includegraphics[width=7cm]{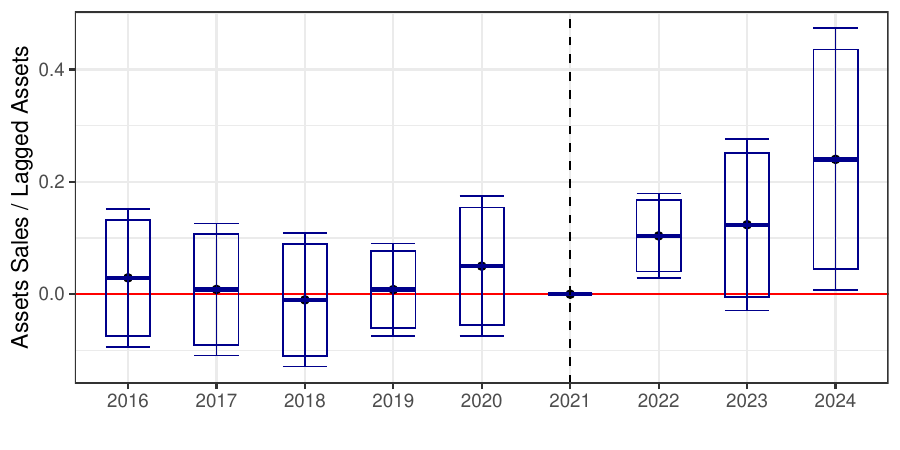}
    \caption{Sales of Investments and Property, \\[-1ex] Plants, and Equipments}
  \end{subfigure}\hfill%
  \begin{subfigure}{7cm}
    \centering\includegraphics[width=7cm]{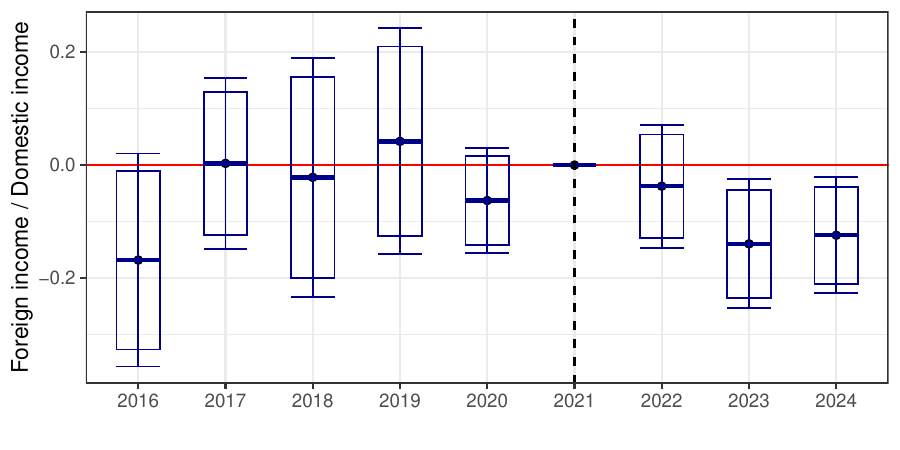}
    \caption{Foreign to domestic \\[-1ex] income ratio}
  \end{subfigure}
    \begin{minipage}{1 \textwidth} 
    {\footnotesize Note: This figure plots the estimated coefficients from (\ref{eq:triple_EV}), where the coefficients of interest are the interaction between time dummies, an indicator equal to one if the firm holds  an AGM between February 24 and May 24, 2022, and the share of individual shareholders with at least a 5\% share. The dependent variables are standardized. The regressions include firm and time-by-industry-fixed effects. Standard errors are clustered at the industry and state levels. Error bars (boxes) report the 95\% (90\%) CI. Vertical dashed lines indicate the start of the invasion of Ukraine. Balance sheet data from Compustat; shareholding data from LSEG.
    \par}
\end{minipage}
\end{figure}

Figure \ref{fig:russia} recasts the analysis in the context of the Ukrainian invasion, defining treated firms as those holding their AGM within \textit{one} month of the onset (February 22, 2022). A one-standard deviation increase in individual blockholding at treated firms is associated with a 0.12 standard deviation decline in operating income in 2023 (Panel a), equivalent to a 2\% drop in levels per 1pp increase in blockholder ownership. This effect persisted in 2024. Market valuation adjusted more slowly: the market-to-book ratio fell by 0.19 standard deviations in Q1 2023 (Panel b), corresponding to a 2\% decline in levels per 1pp increase in blockholder ownership, consistent with other evidence on the war's firm-level effects \citep{bougias2022valuation}.

\paragraph{Mechanism: restructuring costs} As before, we investigate whether the losses at treated firms stem from revenues or costs. Panel (b) of Appendix Figure \ref{fig:es_sales_opincome} shows that sales remained constant, suggesting that the costs of exiting Russia---such as finding new suppliers (see Section \ref{s:russia})---outweighed foregone revenues. Thus, rushed exits appear to have primarily raised costs rather than reduced demand.  

What were these costs? One possibility is that firms were forced to sell past investments at steep discounts or even at a loss. For example, carmaker Renault sold its plant for \$0.01. Panel (c) of Figure \ref{fig:russia} shows that treated firms were significantly more likely than controls to divest past investments, with the likelihood rising by more than 0.10 standard deviations per one--standard deviation increase in blockholding in 2022 (about a 4\% increase in levels). Consistent with market exit, Panel (d) shows that the ratio of foreign to domestic income at treated firms fell by 0.10 standard deviations in 2023--2024, equivalent to a 27\% increase in the domestic income share per one--standard deviation increase in blockholding.

\subsection{Discussion: The Costs for ``Unheard'' Shareholders}

We estimate that vocal, well-connected shareholders extracting visibility or reputational gains impose measurable costs on passive investors. We now turn to measuring these costs for ``silent'' shareholders.

\begin{figure}[!htbp]
 \caption{The distributional consequences of shareholders' voice} \label{fig:eps}
 \begin{subfigure}{7cm}
    \centering\includegraphics[width=7cm]{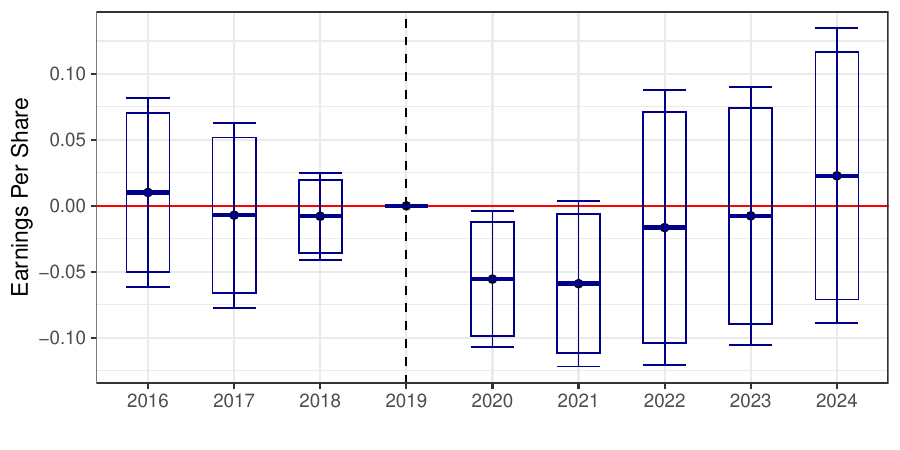}
    \caption{Covid: Earnings per share \\[-1ex] (EPS)} 
  \end{subfigure}\hfill
  \begin{subfigure}{7cm}
  \captionsetup{justification=centering}
    \centering\includegraphics[width=7cm]{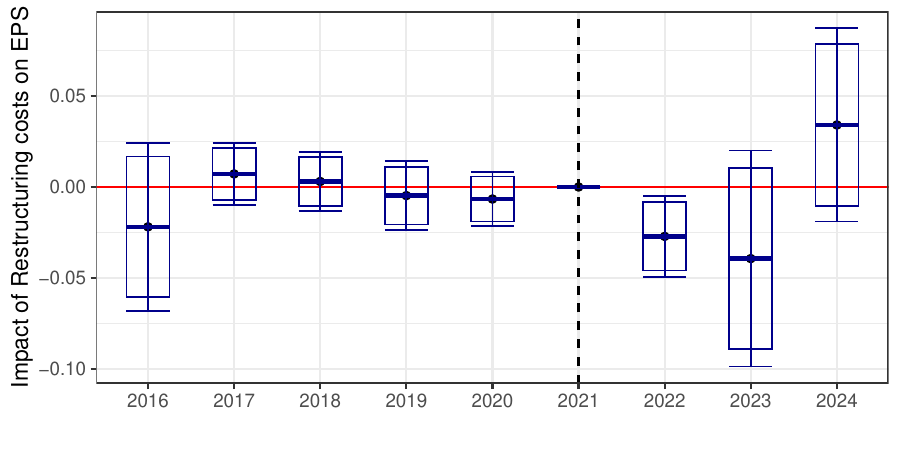}
    \caption{Ukraine: Proportional Impact of \\[-1ex] Restructuring Costs on EPS}
  \end{subfigure}
    \begin{minipage}{1 \textwidth} 
    {\footnotesize Note: The figure reports estimated coefficients from (\ref{eq:triple_EV}). The dependent variable is EPS, including extraordinary items (Panel a), and the change in EPS due to restructuring gains and costs (Panel b), which include Chapter 11 costs, workforce reductions, and relocation charges (incomes are coded as positive, and expenses as negative), scaled by current EPS. The dependent variables are standardized. Standard errors are clustered at the state and industry levels. Error bars (boxes) represent 95\% (90\%) CI. Vertical dashed lines indicate the event time. Balance sheet data from Compustat; shareholding data from LSEG.
   \par}
\end{minipage}
\end{figure}

In the COVID case, Section \ref{s:dist_covid} shows that donations raised operating costs without boosting revenues, reducing cash flows and future investment. Panel (a) of Figure \ref{fig:eps} shows that earnings per share (EPS) fell more at treated firms in 2020--2021, by 0.06 standard deviations (a 1.89\% decline in levels). These donations thus reduced both current and future shareholder payoffs.  

In the Ukraine--Russia case, costs stemmed from disorganized exits, such as relocating activities and selling assets quickly at discounted prices (Section \ref{s:dist_russia}). The higher rate of asset sales in Panel (c) of Figure \ref{fig:russia} translated into greater restructuring costs, eroding 0.02 standard deviations of EPS per one--standard deviation increase in blockholding, as shown in Panel (b) of Figure \ref{fig:eps}---three times more than in control firms.

\paragraph{Placebo analysis } As a falsification exercise, we replicate the analysis using firms where individual ownership is below 2\%, a threshold where shareholders have less incentive from reputational rents. Appendix Figure \ref{fig:es_small} shows no significant effect on operating income or market capitalization for this group. Consistently with this finding, Appendix Figure \ref{fig:placebo} confirms no change in EPS following either crisis for firms dominated by these small-stake shareholders. These results verify that our main findings are driven specifically by individual blockholders seeking reputational gains.

\paragraph{Back-of-the-envelope costs and benefits.}
We compare the private costs and gains to prominent individual shareholders from influencing management during COVID-19—the episode for which we can quantify both the cost (the EPS response) and the private benefit (changes in Google search activity).

On the cost side, Column 2 of Table \ref{tab:reg-results-block-static} indicates that the probability of donating increases by 0.088 per standard deviation in the share of individual shareholders holding stakes of at least 5\%. Figure \ref{fig:eps} shows that, in 2020, standardized EPS fell by 6\% of a standard deviation for the same variation in blockholding, which corresponds to a \$0.23 decline in non-standardized EPS per standard deviation. Taken together, these ITT estimates imply an expected EPS reduction of roughly $(\$0.23/0.088) \approx \$ 2.6$ cents per 1pp increase in the probability of donating.\footnote{Using \(\text{AGM}_f \times \text{Ownership}_f\)—with \(\text{Ownership}_f\) defined as in Column 2 of Table \ref{tab:reg-results-block-static}—as an instrument in a regression of EPS on COVID-related donations (sample as in Table \ref{tab:reg-results-block-static}) yields a similar LATE: a 4.5\% decline in EPS per standard deviation increase in individual blockholding. Estimating the same specification with EPS one and two years after the crisis yields null effects, consistent with Panel~(a) of Figure \ref{fig:eps}.} Given that a 5\% shareholder in this group holds roughly 7 million shares, this amounts to an earnings loss of about \$182{,}000, compared with average dollar-valued donations on the order of \$7 million (Panel~(iii) of Table \ref{tab:summary-meeting}). 

On the benefit side, Figure \ref{fig:Google_10} finds that a donation approximately doubles Google searches for prominent individual shareholders; hence, a 1pp increase in the probability of donating raises expected search activity by about 1\%. 

Therefore, tying our results in Sections \ref{s:covid} and \ref{s:distributional}, our back-of-the-envelope computation suggests that a prominent individual shareholder must value a 1\% increase in search activity at least \$182{,}000 to be privately optimal to push for donations through the associated firm. We document that this figure is comparable to the cost of elite PR campaigns in the public record—an otherwise opaque market.\footnote{Bell Pottinger, for example, was retained by the Gupta family on a three-month trial at about \$130{,}000 per month, plus costs, to mount a broad communications effort that advanced an ``economic-emancipation'' narrative and provided crisis support \citep{caesar2018-bellpottinger}. Likewise, Edelman’s CEO has noted that many large clients pay around \$100{,}000 per month in retainers \citep{ryan2013-prcosts}. These figures are of the same order as our \$182{,}000 benchmark.} Despite hiding large heterogeneity, this estimate highlights the large private gains that individual shareholders may achieve through activism at the expense of others.

While we document clear costs for other investors, their broader implications could differ. To the extent that these actions generate positive societal externalities—a dimension we cannot quantify—facilitating the pass-through of reputational gains from firms to connected shareholders may encourage prosocial behavior.

\paragraph{Discussion} Together, our analyses demonstrate that majority shareholders—motivated by private benefits—can impose both immediate and persistent losses on minority investors. Protecting “silent” shareholders thus requires more than board-composition rules:\footnote{Note that board composition is by construction uncorrelated with the AGM treatment.} it calls for mechanisms that make preference heterogeneity explicit, curb undue influence, and align managerial decisions with the broader shareholder base. One avenue is to facilitate coordination among like-minded shareholders, who may otherwise remain unaware of each other’s influence yet have stronger incentives to act collectively despite holding relatively small stakes.

Our stylized framework in Section \ref{s:conceptual} and the empirical evidence demonstrate that large shareholders do not act in isolation but continually monitor one another’s incentives before exerting influence on management. This \emph{intra}‑shareholder vigilance arises from diverse preferences over prosocial actions, the managerial costs of accommodating those demands, and the threat of reduced future earnings. Consequently, a firm’s governance must account not only for the presence of blockholders but for the heterogeneity of their objectives.

Prior work has emphasized how concentrated owners discipline managerial shirking \citep[e.g.,][]{shleifer1997survey}. Our findings build on this by highlighting the flipside: when blockholders share aligned preferences—whether to pursue high‑visibility donations or to execute rapid asset sales—they can coordinate to extract private rents at the expense of diffuse investors and other stakeholders. By contrast, heterogeneity in blockholder preferences serves as a natural check on such collusion, protecting minority interests, a conclusion that may extend beyond prosocial matters.

These insights highlight the limits of a narrow focus on board composition or voting thresholds. Because reputational incentives generate hidden preference heterogeneity, firms may benefit from formal mechanisms to aggregate and reconcile shareholder views—such as mandatory disclosures or structured investor–management dialogues—so that divergent preferences are visible before costly actions are undertaken. Governance mechanisms could also encourage broader consensus among large shareholders before implementing decisions with substantial distributional impacts. Finally, greater transparency around shareholder–manager interactions would allow passive investors to anticipate—and, where appropriate, contest—efforts at coordinated rent extraction.

From a policy perspective, our findings point to a blind spot in current sustainability rules. European disclosure laws, such as the Corporate Sustainability Reporting Directive and the Taxonomy Regulation, have improved information on firms’ environmental and social footprints, but they say little about who triggers high-profile “prosocial” actions or how the costs are shared across investors \citep{hummel2024overview}. Existing work on CSR and ESG reporting mostly studies firm-level outcomes and capital-market effects, not the conflicts that can arise among shareholders with different preferences \citep{frankel2025economics}. Policies could therefore add a simple process layer to reporting: firms would explain who proposed major prosocial decisions, what role large shareholders played, and how the views of other investors were taken into account. This would help regulators tell apart cases where reputation and virtue align from cases where prominent investors capture most of the reputational gains while diffuse shareholders bear most of the losses.

Early evidence from founding-family ownership indicates that these shareholders improve the performance of S\&P 500 firms when their ownership stakes are moderate, as they act as stewards vis-\'a-vis managers and other investors \citep{anderson2003founding}. However, when their stakes are large, performance deteriorates, likely because no counterweight constrains their influence. This pattern is consistent with our findings, since founding families are highly visible shareholders, and it reinforces the case for governance arrangements that balance influence across heterogeneous owners, even outside periods of crisis.

More broadly, the transition to a sustainable economy introduces new rent-seeking opportunities for shareholders, adding concerns such as private image gains to traditional profit motives. As \cite{grossman1979theory} emphasize, ``[...] it is the job of the manager of a firm [...] not only to [...] organize production, but also to learn about the \textit{preferences} of the firm's shareholders.''—a task that has grown more complex as managers must now navigate voices driven by diverse incentives and by agents with different access to managers, with consequences not only for other shareholders but also, though not the focus of this paper, for a broader set of stakeholders (e.g., governments, COVID-19 patients, potential joint ventures in the COVID case, and governments, Russian workers, and consumers in the context of the Russian invasion).

Looking ahead, our analytical framework opens several avenues for research. One important question is whether managers respond more to realized shareholder pressure (clear, observable demands) or to anticipated pressure (expectations about future activism). Microfoundations from behavioral models—such as warm‑glow utility \citep{andreoni1989giving} or reputational concerns \citep{bar2008seller}—could illuminate these channels. Empirical work exploiting variation in shareholder turnover or repeated shocks may help trace how influence wanes or intensifies over time offering identification power.

\section{Conclusion} \label{s:conclusion}
Stakeholder concerns shape firms’ strategic decisions. We develop a framework that leverages quasi-experimental variation to show how shareholders, as a central stakeholder group, influence corporate strategies through visible but costly prosocial actions. Applying it to COVID-19 donations and exits from Russia, we find that prominent individual investors promote such actions to enhance personal reputations, while financial investors resist them and donate privately instead, consistent with a smaller pass-through of reputational gains from firm actions to agents not immediately associated with a firm. This unequal distribution of reputational benefits creates misaligned preferences, resulting in lower investment, productivity, and profitability that persist for up to two years at the firms most exposed to shareholder rent-seeking.

A back-of-the-envelope calculation suggests that the cost of influence in terms of lost EPS is comparable to the expense of high-profile PR campaigns but far below typical donation amounts, illustrating how private image gains can drive costly firm behavior. Our results highlight shareholder preferences as a novel dimension of monitoring and competition, and open avenues for future research on how stakeholders influence the provision of social goods.

\bibliographystyle{ecca-mod}
\bibliography{covid_bibliography}	
\onehalfspacing

\newpage
\appendix
\onehalfspacing

    \pagenumbering{arabic}
    \setcounter{page}{1}

\setcounter{footnote}{0}




\section*{\Huge Online Appendix}
\setcounter{table}{0}
\renewcommand{\thetable}{A\arabic{table}}
\setcounter{figure}{0}
\renewcommand{\thefigure}{A\arabic{figure}}
\setcounter{equation}{0}
\renewcommand{\theequation}{A\arabic{equation}}
\section{Mathematical Derivations}\label{apndx:deriv} 
We begin from the first-order conditions for type $A$ and $B$ shareholders, which implicitly define their equilibrium effort levels $e_A$ and $e_B$:
\begin{align*}
F_A(e_A, e_B; v_A, D) &:= (v_A - D) \cdot \frac{\partial s(e_A, e_B)}{\partial e_A} - c'_A(e_A) = 0, \\
F_B(e_A, e_B; D) &:= (-D) \cdot \frac{\partial s(e_A, e_B)}{\partial e_B} - c'_B(e_B) = 0.
\end{align*}
We assume that $F_A$ and $F_B$ are continuously differentiable, and that a unique solution exists around a reference point $(\bar{e}_A, \bar{e}_B, \bar{v}_A, \bar{D})$. To derive a linear approximation, we apply the total differential to both conditions.

\paragraph{Total differentiation} We now take total differentials of the first-order conditions with respect to the endogenous variables \( e_A \) and \( e_B \), and the parameters \( v_A \) and \( D \). That is, we assume the functions \( F_A \) and \( F_B \) are continuously differentiable and expand around a reference point \( (\bar{e}_A, \bar{e}_B, \bar{v}_A, \bar{D}) \):

\begin{align*}
dF_A &= \frac{\partial F_A}{\partial e_A} \, de_A + \frac{\partial F_A}{\partial e_B} \, de_B + \frac{\partial F_A}{\partial v_A} \, dv_A + \frac{\partial F_A}{\partial D} \, dD = 0, \\
dF_B &= \frac{\partial F_B}{\partial e_A} \, de_A + \frac{\partial F_B}{\partial e_B} \, de_B + \frac{\partial F_B}{\partial D} \, dD = 0.
\end{align*}
\noindent We now define the following terms for notational convenience:

\begin{itemize}
  \item \( A_1 := \frac{\partial F_A}{\partial e_A} = \frac{\partial^2}{\partial e_A^2} \left[ (v_A - D) \cdot s(e_A, e_B) - c_A(e_A) \right] \): marginal sensitivity of type A’s FOC to their own effort;
  \item \( A_2 := \frac{\partial F_A}{\partial e_B} = \frac{\partial^2 s}{\partial e_A \partial e_B} \cdot (v_A - D) \): cross-sensitivity of A's payoff to B's effort;
  \item \( A_3 := \frac{\partial F_A}{\partial v_A} = \frac{\partial s}{\partial e_A} \): marginal reputational return for type A;
  \item \( A_4 := \frac{\partial F_A}{\partial D} = -\frac{\partial s}{\partial e_A} \): marginal cost effect for type A;
  \item \( B_1 := \frac{\partial F_B}{\partial e_A} = -D \cdot \frac{\partial^2 s}{\partial e_A \partial e_B} \): sensitivity of B’s FOC to A’s effort;
  \item \( B_2 := \frac{\partial F_B}{\partial e_B} = -D \cdot \frac{\partial^2 s}{\partial e_B^2} - c''_B(e_B) \): marginal sensitivity of B’s FOC to their own effort;
  \item \( B_3 := \frac{\partial F_B}{\partial D} = -\frac{\partial s}{\partial e_B} \): marginal cost sensitivity for type B.
\end{itemize}

Using this notation, the system becomes:
\begin{align*}
A_1 \, de_A + A_2 \, de_B &= -A_3 \, dv_A - A_4 \, dD, \\
B_1 \, de_A + B_2 \, de_B &= -B_3 \, dD.
\end{align*}
We solve this system for $de_A$ and $de_B$ using Cramer's rule. Let $\Delta = A_1 B_2 - A_2 B_1$ be the Jacobian determinant. Then:
\begin{align*}
de_A &= \frac{-A_3 B_2 \, dv_A - A_4 B_2 \, dD + A_2 B_3 \, dD}{\Delta}, \quad \text{and}\quad
de_B &= \frac{A_3 B_1 \, dv_A + A_4 B_1 \, dD - A_1 B_3 \, dD}{\Delta}.
\end{align*}

We can write this as:
\begin{equation} \label{eq:local_der}
\begin{aligned}
de_A &= \tilde{\alpha}_1 \, dv_A + \tilde{\alpha}_2 \, dD, \quad \text{and}\quad
de_B &= \tilde{\delta}_1 \, dv_A + \tilde{\delta}_2 \, dD,
\end{aligned}    
\end{equation}
where:
\begin{align*}
\tilde{\alpha}_1&= \frac{-A_3 B_2}{\Delta}, \quad
\tilde{\alpha}_2 = \frac{-A_4 B_2 + A_2 B_3}{\Delta},\quad
\tilde{\delta}_1 = \frac{A_3 B_1}{\Delta}, \quad\text{and}\quad 
\tilde{\delta}_2 = \frac{A_4 B_1 - A_1 B_3}{\Delta}.
\end{align*}

\paragraph{Integration to recover levels} The total differential Equations \eqref{eq:local_der} derived above describe how small changes in the parameters \( v_A \) and \( D \) affect the optimal effort levels \( e_A \) and \( e_B \). To recover approximate expressions for the levels of \( e_A \) and \( e_B \), we integrate the total differentials in \eqref{eq:local_der} around the reference point \( (\bar{v}_A, \bar{D}) \). Assuming that changes in \( v_A \) and \( D \) are small, we can use a first-order Taylor expansion of the effort functions \( e_A(v_A, D) \) and \( e_B(v_A, D) \):
\begin{align*}
e_A(v_A, D) &\approx e_A(\bar{v}_A, \bar{D}) 
+ \left.\frac{\partial e_A}{\partial v_A}\right|_{\bar{v}_A, \bar{D}} (v_A - \bar{v}_A) 
+ \left.\frac{\partial e_A}{\partial D}\right|_{\bar{v}_A, \bar{D}} (D - \bar{D}), \\
e_B(v_A, D) &\approx e_B(\bar{v}_A, \bar{D}) 
+ \left.\frac{\partial e_B}{\partial v_A}\right|_{\bar{v}_A, \bar{D}} (v_A - \bar{v}_A) 
+ \left.\frac{\partial e_B}{\partial D}\right|_{\bar{v}_A, \bar{D}} (D - \bar{D}).
\end{align*}

We denote the base levels \( \tilde{\alpha}_0 := e_A(\bar{v}_A, \bar{D}) \) and \( \tilde{\delta}_0 := e_B(\bar{v}_A, \bar{D}) \), and use the total differential coefficients as estimates for the partial derivatives. This yields the linear approximations:
\begin{align*}
e_A &\approx \tilde{\alpha}_0 + \tilde{\alpha}_1 \cdot (v_A - \bar{v}_A) + \tilde{\alpha}_2 \cdot (D - \bar{D}), \\
e_B &\approx \tilde{\delta}_0 + \tilde{\delta}_1 \cdot (v_A - \bar{v}_A) + \tilde{\delta}_2 \cdot (D - \bar{D}).
\end{align*}

Without loss of generality, and for notational simplicity, we re-center the data so that \( \bar{v}_A = \bar{D} = 0 \), yielding:
\begin{equation}
\begin{aligned}\label{eq:linear_efforts}
e_A \approx \tilde{\alpha}_0 + \tilde{\alpha}_1 v_A + \tilde{\alpha}_2 D, \qquad \text{and} \qquad 
e_B \approx \tilde{\delta}_0 + \tilde{\delta}_1 v_A + \tilde{\delta}_2 D.
\end{aligned}    
\end{equation}

\paragraph{Reduced-form equation}
To link the derived efforts to observable firm behavior, we assume that the firm undertakes a prosocial action when the net influence of shareholder pressure exceeds a threshold. Let its latent utility of prosocial behavior be:
\[
y_f^* = \tilde{\theta}_0 + \tilde{\theta}_1 e_A + \tilde{\theta}_2 e_B + \tilde{\epsilon}_f,
\]
where $\varepsilon_f$ follows a standard logistic distribution. The observed outcome is then:
\[
y_f = \begin{cases}
1 & \text{if } y_f^* > 0, \\
0 & \text{otherwise}.
\end{cases}
\]
Hence, the probability of observing a prosocial action is:
\[
\Pr(y_f = 1) = \Lambda\left( \theta_0 + \theta_1 e_A + \theta_2 e_B \right),
\]
where $\Lambda(z) = 1 / (1 + e^{-z})$ denotes the logistic CDF.

We now substitute the linear approximations for $e_A$ and $e_B$ in \eqref{eq:linear_efforts}. This yields:
\[
\Pr(y_f = 1) \approx \Lambda\left( \beta_0 + \beta_1 v_A + \beta_2 D \right),
\]
where:
\begin{align*}
\beta_0 &:= \tilde{\theta}_0 + \tilde{\theta}_1 \alpha_0 + \tilde{\theta}_2 \delta_0, \\
\beta_1 &:= \tilde{\theta}_1 \tilde{\alpha}_1 + \tilde{\theta}_2 \tilde{\delta}_1, \\
\beta_2 &:= \tilde{\theta}_1 \tilde{\alpha}_2 + \tilde{\theta}_2 \tilde{\delta}_2.
\end{align*}

\paragraph{Linear probability model} Under small variations in $v_A$ and $D$, as one might expect from our AGM shocks, the logistic function can be well-approximated by a linear function. This leads to the reduced-form equation used in our empirical analysis:
\[
y_f \approx \theta_0 + \theta_1 v_A + \theta_2 D + \epsilon_f,
\]
which interprets prosocial behavior as a linear function of reputational incentives and cost, even when underlying decisions are interdependent and non-linear.

\newpage\clearpage
\setcounter{table}{0}
\renewcommand{\thetable}{B\arabic{table}}
\setcounter{figure}{0}
\renewcommand{\thefigure}{B\arabic{figure}}

\section{Omitted Tables}\label{ap:add_tables}
\input{Tables/donations_top10}
\clearpage\newpage\FloatBarrier

\input{Tables/horse_race}
\clearpage\newpage\FloatBarrier

\input{Tables/covid_inst}
\clearpage\newpage\FloatBarrier

\input{Tables/covid_slack}
\clearpage\newpage\FloatBarrier

\input{Tables/donations_2020}
\clearpage\newpage\FloatBarrier

\input{Tables/other_association}
\clearpage\newpage\FloatBarrier

\input{Tables/covid_fyr}
\clearpage\newpage\FloatBarrier

\input{Tables/covid_fyr2}
\clearpage\newpage\FloatBarrier

\input{Tables/exits_top10}
\clearpage\newpage\FloatBarrier

\input{Tables/regression_russia_fin}
\clearpage\newpage\FloatBarrier

\input{Tables/regression_russia_airlines}
\clearpage\newpage\FloatBarrier

\input{Tables/regression_russia_grade_f}
\clearpage\newpage\FloatBarrier

\section{Omitted Figures}\label{apndx:add_figures}
\setcounter{table}{0}
\renewcommand{\thetable}{C\arabic{table}}
\setcounter{figure}{0}
\renewcommand{\thefigure}{C\arabic{figure}}


 \begin{figure}[H]
    \caption{Stability of the AGM month over time}
     \label{fig:meetings}
     \centering
    \begin{subfigure}{0.46\textwidth}
        \centering
    \includegraphics[width=1\linewidth]{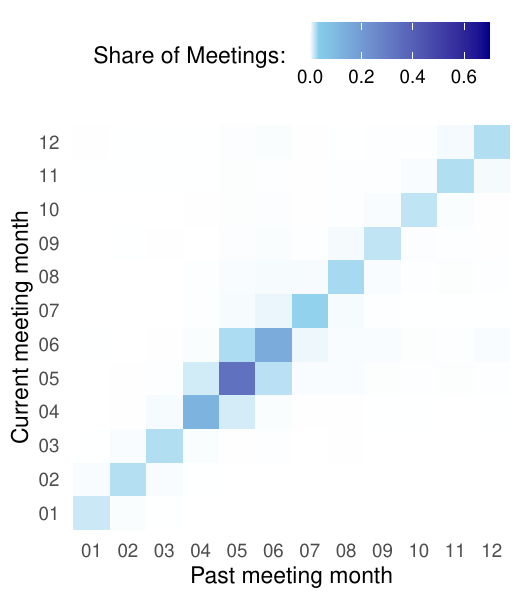}
    \caption{2012 to 2019}
    \end{subfigure}\hfill
    \begin{subfigure}{0.46\textwidth}
    \centering      
        \includegraphics[width=1\linewidth]{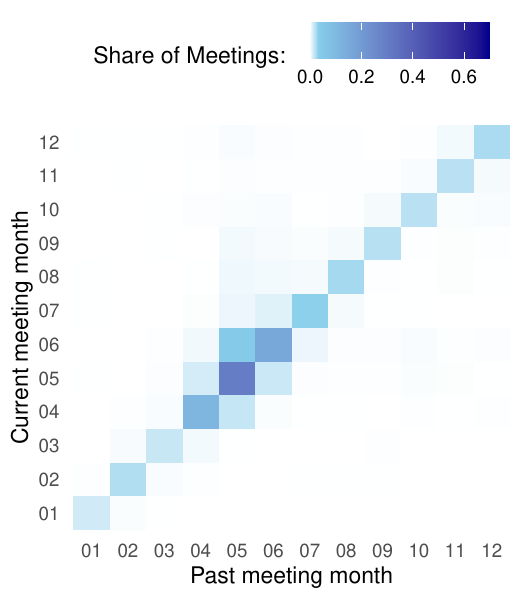}
        \caption{2020}  
    \end{subfigure}
		\medskip 
		
\begin{minipage}{1 \textwidth} 
{\footnotesize Note: Both panels count the occurrences of the AGM month over two adjacent years. A non-white cell in position (2,3) means that at least one firm with an AGM in March of year $t$ had an AGM in February of year $t-1$. The diagonal line indicate firms that did not change AGM month over time. Panel (a) focuses on data from 2012 to 2019, while Panel (b) zooms in on of 2020. Observations are at the firm-by-year level as firms have more than one AGM, in Panel (a). AGM dates from ISS .\par}
\end{minipage}
 \end{figure}

\clearpage\newpage\FloatBarrier
 \begin{figure}[H]
     \centering
         \caption{The distribution of ESG scores is invariant to AGM date \label{fig:agm_esg_score}}
     \includegraphics[width=0.5\linewidth]{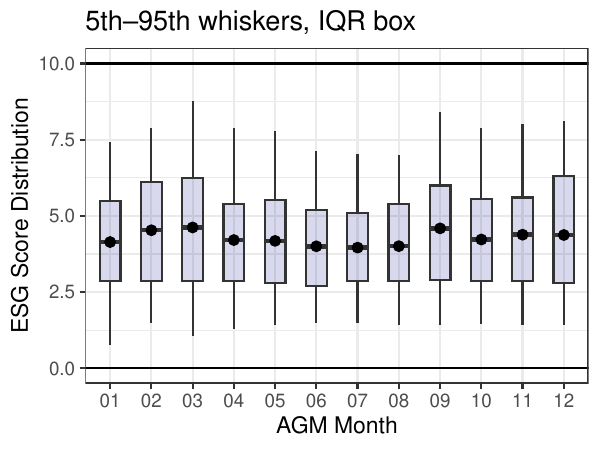}
     
\begin{minipage}{1 \textwidth} 
{\footnotesize Note: This figure plots the distribution of ESG scores of firms by AGM month between 2012 and 2020. The sample covers 2,715 US firms firms covered by ISS and MSCI. ESG scores are on a [0-10] scale. \par}
\end{minipage}
 \end{figure}
 \clearpage\newpage\FloatBarrier

\begin{figure}[!htbp]
\centering
\caption{Robustness to different AGM treatment windows\label{windows_covid}}

\includegraphics[width=\linewidth]{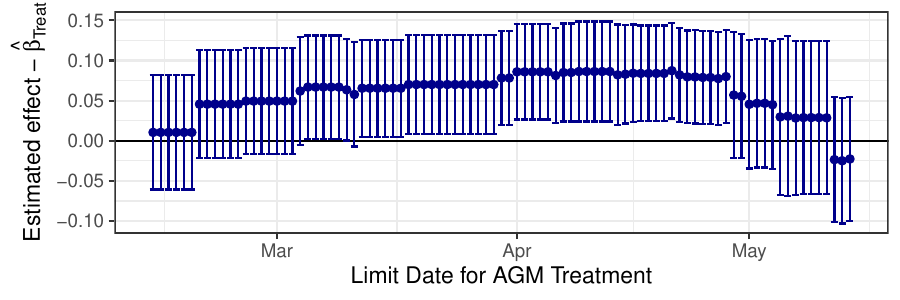}
\caption*{(a) Individual shareholders (5\% blockholders)}

\includegraphics[width=\linewidth]{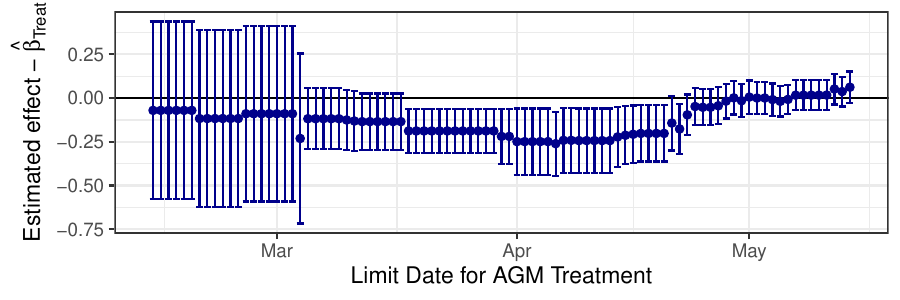}
\caption*{(b) Large financial shareholders (10\% blockholders)}

\begin{minipage}{\linewidth}
{\footnotesize \textit{Note:} This figure plots the estimated interaction coefficients $\hat{\beta}_{\text{treat}}$ from Eq.~\eqref{eq:lin-prob-model-cross-section}. The donation indicator equals one if firm $f$ announced a COVID-related donation by April 15, 2020. In Panel A, $Ownership$ is the fraction of equity held by individual shareholders who each own at least 5\% of firm $f$'s shares. In Panel B, $Ownership$ is defined for large financial shareholders who each own at least 10\% of firm $f$'s shares. For each point on the x-axis, we re-estimate the regression using an $AGM$ indicator equal to one if firm $f$'s 2020 AGM was scheduled between January 15, 2020 and the date shown on the x-axis. All specifications control for firm size (log total assets) and include state and subindustry fixed effects. Standard errors are clustered by industry. AGM dates from ISS; shareholding data from LSEG; donation data are hand-collected.\par}
\end{minipage}
\end{figure}
\clearpage\newpage\FloatBarrier

\begin{figure}[H]
    \caption{Sales or costs: What drove productivity down during crises?} \label{fig:es_sales_opincome}
  \begin{subfigure}{7cm}
    \centering\includegraphics[width=7cm]{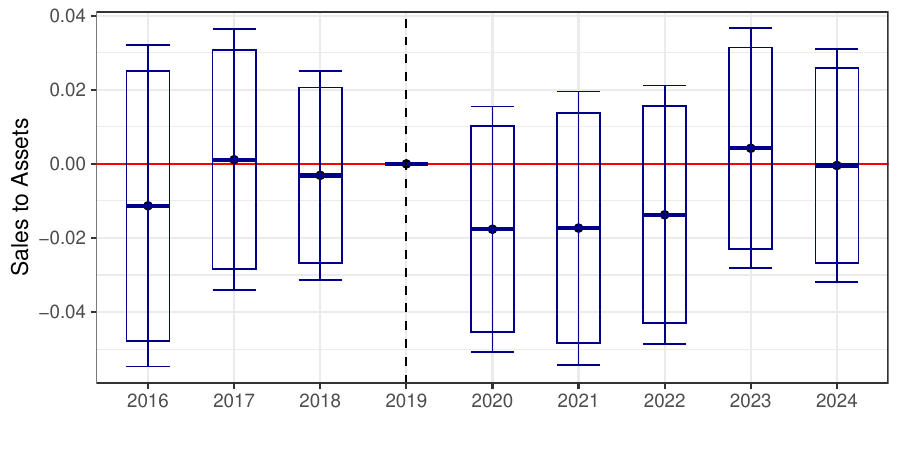}
    \caption{Sales to assets during Covid} 
  \end{subfigure}\hfill
  \begin{subfigure}{7cm}
    \centering\includegraphics[width=7cm]{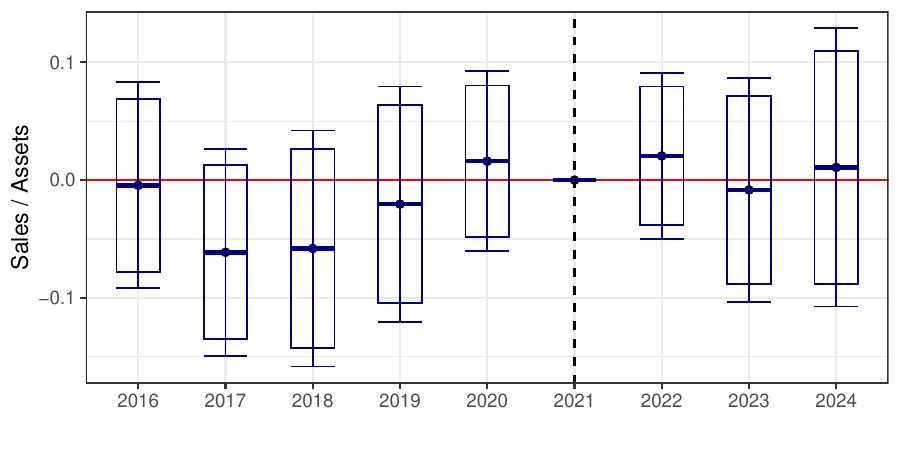}
    \caption{Sales to assets after the Russian invasion}
  \end{subfigure}
    \begin{minipage}{1 \textwidth} 
    {\footnotesize Note: The figure reports the estimated coefficients from an event study as in (\ref{eq:triple_EV}), with sales on total assets as the dependent variable. The dependent variables are standardized. We report only the estimated coefficients of interest -- the interaction between the time dummies and an indicator equal to one if the firm has an AGM in the 90 days after the onset of the emergency and the share of individual shareholders with more than 10\% equity shares. Each regression also includes firm and time-by-industry-fixed effects as described in Section \ref{s:distributional}. The dataset considers the largest 1,000 US-listed firms in Panel (a) and the largest 1,153 US-listed firms with exposure to Russia in Panel (b). Standard errors are clustered at the state and industry levels. Error bars (boxes) report the 95\% (90\%) CI. Vertical dashed lines indicate the time of the event. Balance sheet data from Compustat; shareholding data from LSEG.
    \par}
\end{minipage}
\end{figure}
\clearpage\newpage\FloatBarrier

\begin{figure}[ht]
    \caption{Documenting rents: small individual shareholders }\label{fig:es_small}
  \begin{subfigure}{7cm}
    \centering\includegraphics[width=7cm]{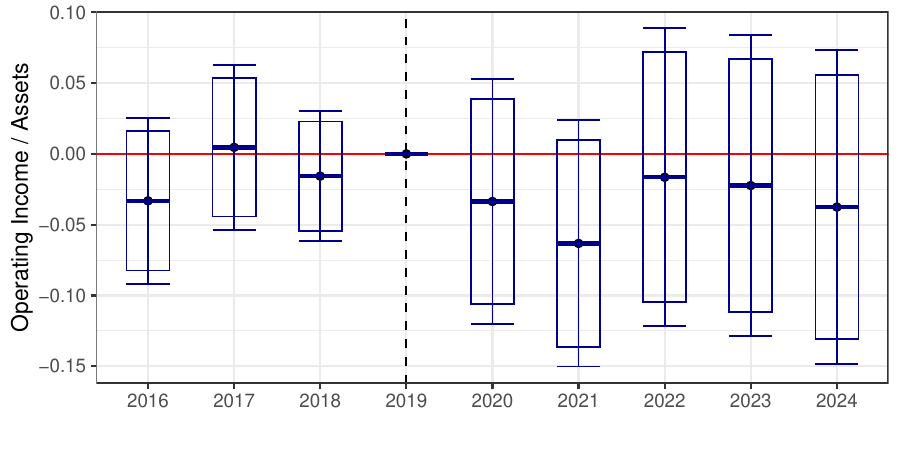}
    \caption{Operating income (Covid)}
  \end{subfigure}\hfill
  \begin{subfigure}{7cm}
    \centering\includegraphics[width=7cm]{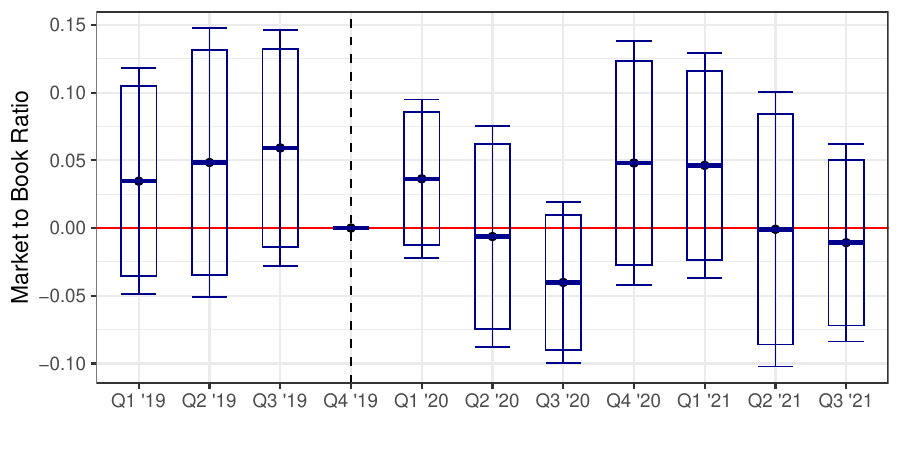}
    \caption{Market valuation (Covid)}
  \end{subfigure}
  
  \begin{subfigure}{7cm}
    \centering\includegraphics[width=7cm]{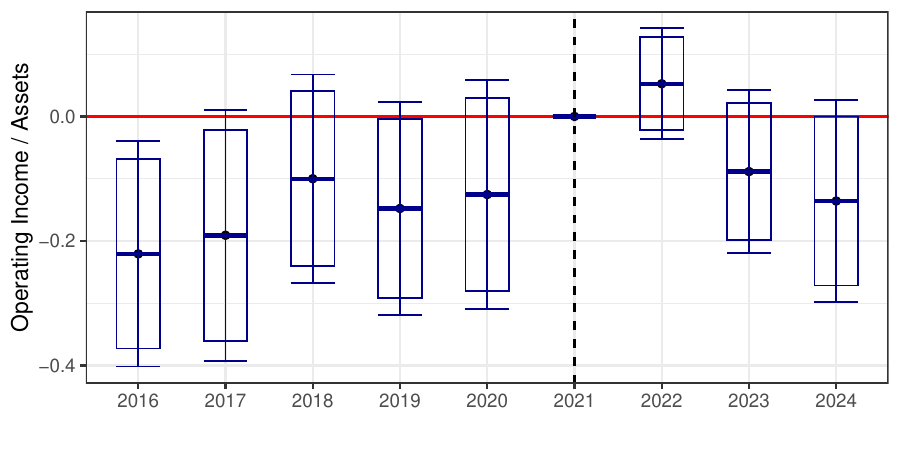}
    \caption{Operating income (Russia)}
  \end{subfigure}\hfill%
  \begin{subfigure}{7cm}
    \centering\includegraphics[width=7cm]{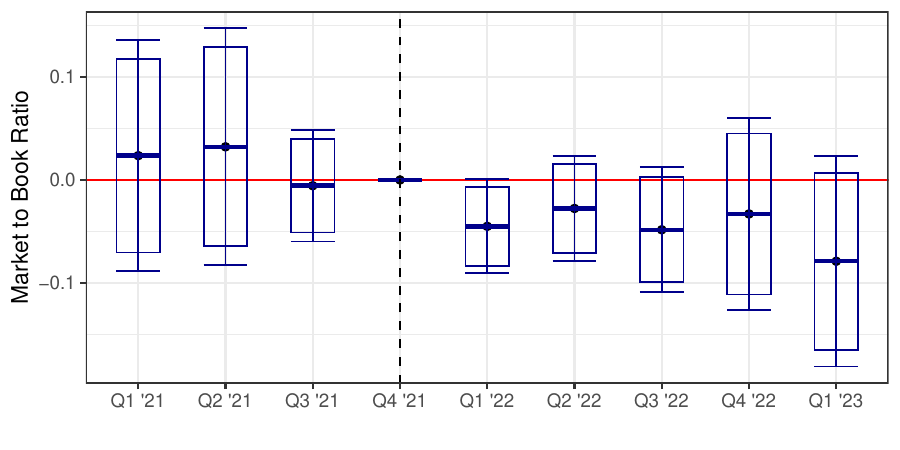}
    \caption{Market valuation (Russia)}
  \end{subfigure}
    \begin{minipage}{1 \textwidth} 
    {\footnotesize Note: This figure plots the estimated coefficients from (\ref{eq:triple_EV}), where the coefficients of interest are the interaction between time dummies, an indicator equal to one if the firm holds  an AGM between February 24 and May 24, 2022, and the share of individual shareholders with at less than a 2\% share. The regressions include firm and time-by-industry-fixed effects. The dependent variables are standardized. Standard errors are clustered at the industry and state levels. Error bars (boxes) report the 95\% (90\%) CI. Vertical dashed lines indicate the start of the invasion of Ukraine. Balance sheet data from Compustat; shareholding data from LSEG.
    \par}
\end{minipage}
\end{figure}
\clearpage\newpage\FloatBarrier

\begin{figure}[ht]
    \caption{Placebo Analysis: The distributional consequences of shareholders' voice}\label{fig:placebo}
  \begin{subfigure}{7cm}
    \centering\includegraphics[width=7cm]{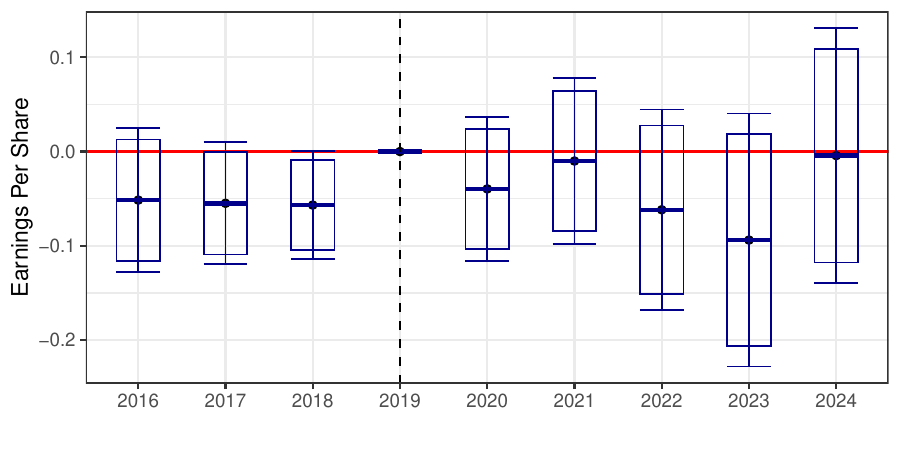}
    \caption{Covid: \\[-1ex] Market valuation}
  \end{subfigure}\hfill
  \begin{subfigure}{7cm}
    \centering\includegraphics[width=7cm]{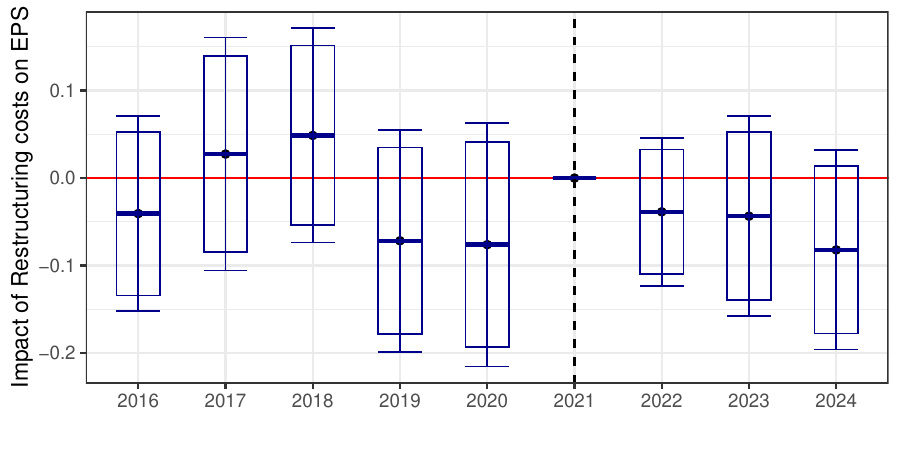}
    \caption{Ukraine: Proportional Impact \\[-1ex] of Restructuring Costs on EPS}
  \end{subfigure}
  
    \begin{minipage}{1 \textwidth} 
    {\footnotesize Note: This figure plots the estimated coefficients from (\ref{eq:triple_EV}), where the coefficients of interest are the interaction between time dummies, an indicator equal to one if the firm holds an AGM between January 15 and April 15 (panel a), or between February 24 and May 24, 2022 (panels b) and the share of individual shareholders with stakes smaller than 2\%. The dependent variables are standardized. The regressions include firm and time-by-industry-fixed effects. Standard errors are clustered at the industry and state levels. Error bars (boxes) report the 95\% (90\%) CI. Balance sheet data from Compustat; shareholding data from LSEG. \par}
\end{minipage}
\end{figure}

\clearpage\newpage\FloatBarrier

 \begin{figure}
     \caption{Cumulative abnormal returns around an exit from Russia announcement}
     \label{fig:excess_returns_russia}
     \centering
    \centering
    \includegraphics[width=0.75\linewidth]{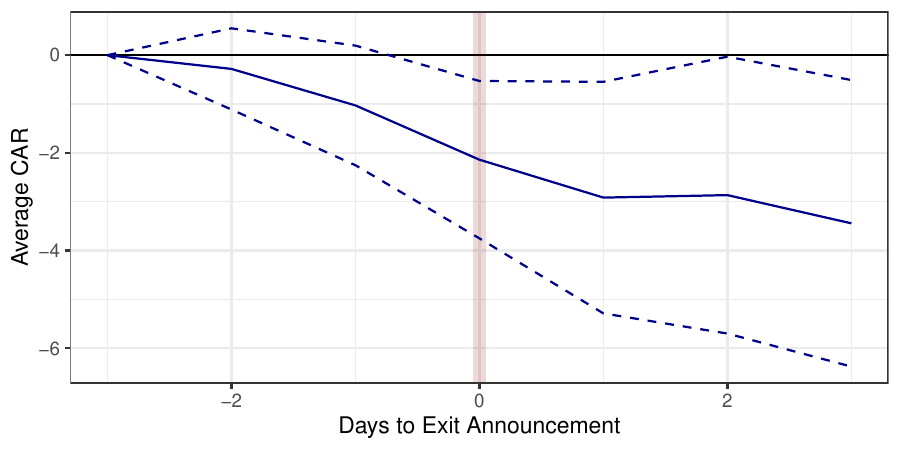}
		\medskip 
\begin{minipage}{1 \textwidth} 
{\footnotesize Note: This figure suggests that firms do not choose to donate to improve their performance on the stock market. The figure plots cumulative abnormal returns (CARs) computed using a similar methodology outlined in Section \ref{s:robustness}. Date 0 is the date when a firm publicly announced its decision to exit Russia after the invasion of Ukraine. CRISP data. 
\par}
\end{minipage}
 \end{figure}

\clearpage
\setcounter{table}{0}
\renewcommand{\thetable}{D\arabic{table}}
\setcounter{figure}{0}
\renewcommand{\thefigure}{D\arabic{figure}}
\setcounter{equation}{0}
\renewcommand{\theequation}{D\arabic{equation}}
\section{Robustness Checks}\label{apndx:rob}

\subsection{Intensity of the Pandemic} \label{apndx:r1} 
\begin{landscape}
 \begin{figure}[htbp]
    \caption{Covid cases, deaths, and corporate donations by US state and month}
     \label{fig:heatmap}
    \centering
    \includegraphics[width=1\linewidth]{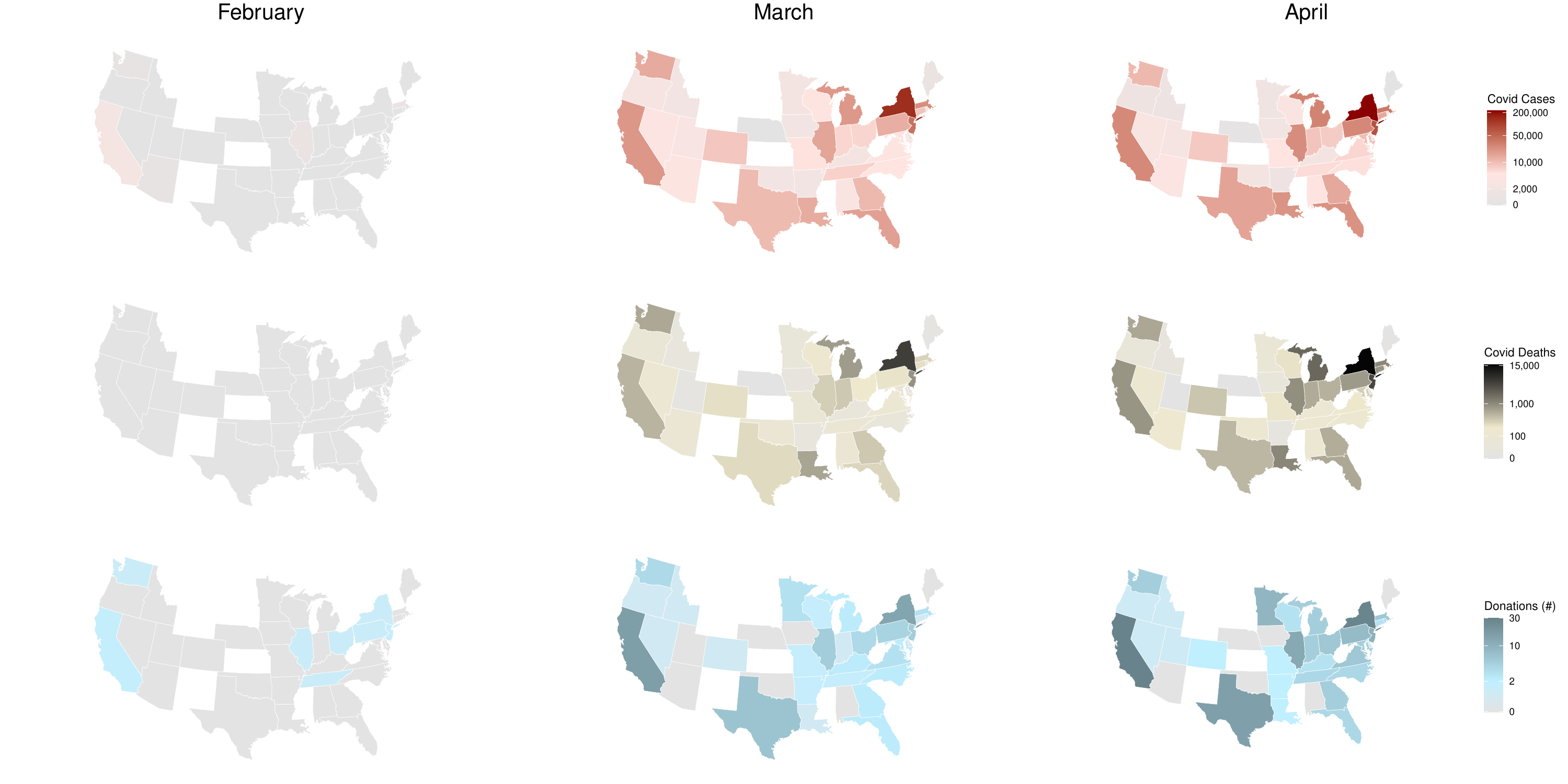}
    		\bigskip 
\begin{minipage}{1.5 \textwidth} 
{\footnotesize Note: The figure highlights the spatial and temporal correlation between the cumulative number of covid cases (first row), deaths (second row), and donations of S\&P 500 companies (third row). Each column reports the breakdown for each variable across US states on February 29 (Column 1), March 31 (Column 2), and April 15 (Column 3), when our sample ends. States in white do not house S\&P 500 firms. Covid rates come from Johns Hopkins University. Donation data are hand-collected using various online sources. Covid data from JHU; donation data are hand-collected. \par}
\end{minipage}
    \end{figure}
 \end{landscape}

Firms might be simply responding to the need for ventilators and similar items and our empirical approach could miss this if this need is correlated with the AGM treatment. We modify (\ref{eq:lin-prob-model-cross-section}) as follows to control for COVID cases and deaths occurring in a firm's headquarter state over time:
\vspace{-8pt}
\begin{equation} \label{eq:lin-prob-model2}
\begin{aligned}
y_{ft} &= \beta_1 \, \textnormal{Covid Rate}_{ft} + \beta_2 \, \textnormal{Covid Rate}_{ft} \times \textnormal{Ownership}_{f}  +\beta_3 \, \textnormal{Covid Rate}_{ft}   \\ 
        &\times \textnormal{AGM}_{f} + \beta_{\text{treat}} \, \textnormal{Covid Rate}_{ft} \times \textnormal{Ownership}_{f} \times \textnormal{AGM}_{f} + \alpha_f + \tau_{t} + \varepsilon_{ft},
\end{aligned}
\end{equation}
where the dependent variable, $y_{ft}$, is equal to one if firm $f$ has publicly committed to donating by day $t$ and zero otherwise. The main coefficient of interest is $\beta_{\text{treat}}$, which captures the interaction between the cumulative covid rate at firm $f$'s headquarters state, $\textnormal{Covid Rate}_{ft}$, the fraction of equity owned by the reference blockholder, $\textnormal{Ownership}_{f} $, which varies across specifications, and the AGM treatment, $\textnormal{AGM}_{f}$. We use headquarter-state covid rates because Appendix Figure \ref{fig:heatmap} shows a clear spatial pattern across these two variables, with firm $f$ being more likely to donate for covid relief as the pandemic heightens in $f$'s state. Finally, $\alpha_f$ and $\tau_{t}$ are firm- and day-fixed effects. 

\sloppy The first three columns of Appendix Table \ref{tab:reg-results-large-shareholders} present results using individual shareholders as the reference category, with $\textnormal{Ownership}_{f}$ and $\textnormal{Covid Rate}_{ft}$ standardized to facilitate comparisons across columns with different $x$-blockholding percentages, covid deaths. The only significant effect is that of the interaction between the AGM treatment and large individual blockholding. In contrast, treatment effect estimates for banks, mutual funds, and insurers (Columns 4-6) are negative, consistent with previous findings, while the unconditional effect of covid rates remains negligible.

\subsection{Financial Motives} \label{apndx:r2}
Abnormal returns do not explain covid-related donations. To compute abnormal returns, we predict daily stock returns using the Carhart four-factor model, namely, daily market returns ($R_{ft}^{MKT}$), daily returns  on a portfolio of ``small minus big stocks'' ($R_{ft}^{SMB}$), daily returns on a portfolio of stocks with ``high minus low'' book-to-market value ratios ($R_{ft}^{HML}$), and daily returns on a portfolio of ``winners minus losers'' ($R_{ft}^{UMD}$). All portfolio returns are from Kenneth French's website. We retrieve the stocks' betas ($\beta_f$) of those four portfolios for the stocks in our sample from WRDS' Beta Suite. Then, stock $f$'s abnormal return ($AR_{ft}$) on day $t$ is given by the difference between the actual excess return of the stock over the risk-free rate ($R_{ft}$) and the prediction of the 4-factor model, as 
    $AR_{ft} = R_{ft} - (\beta^{MKT}_f \cdot R^{MKT}_{ft} +\beta^{SMB}_f \cdot R^{SMB}_{ft} + \beta^{HML}_f \cdot R^{HML}_{ft} + \beta^{UMD}_f \cdot R^{UMD}_{ft}).$

\noindent where the left-hand side refers to either firm $f$'s abnormal return ($AR_{ft}$) or its cumulative abnormal return (CAR) on day $t$. The CAR estimates in Appendix Figure \ref{fig:excess_returns} drop on the day of the donation, of around 1\%.

\subsection{Consumer Pressure} \label{apndx:r3}
We exploit exogenous variation in a firm's exposure to COVID-19 through its branches to assess whether consumers pressured firms to donate. Using Orbis, we compute the weighted cumulative averages of covid cases and deaths using the number of branches a firm has in each state as weights. We denote the standardized versions of these two new variables by $\textnormal{Exposure at Branches}_{ft}$ and estimate the following linear probability model:
\begin{equation}\label{eq:lin-prob-model-exposure}
\begin{aligned}
y_{ft} &=  \beta_1 \, \textnormal{Exposure at Branches}_{ft} 
        + \beta_{\text{treat}} \, \textnormal{Exposure at Branches}_{ft} \\
        &\times \textnormal{Number of Branches}_{f} + \alpha_f + \tau_{t} + \varepsilon_{ft},
\end{aligned}
\end{equation}
where $\textnormal{Number of Branches}_{f}$ is the reported number of branches as of December 2019.\footnote{The distribution of the number of branches for S\&P 500 firms ranges from 0 to 13,582 with a median of 40 branches; since we do not know whether a branch is a shop or a factory, it is fair to assume that firms with more branches are the most exposed to final consumers.}  Appendix Table \ref{tab:exposure_industry_branches} shows no detectable effect of local COVID exposure at branch locations (and its interaction with branch counts) on donation behavior.


\subsection{Competition} \label{apndx:r5}
We also examine whether firms donated in response to past COVID-related donations by direct competitors by including in (\ref{eq:lin-prob-model2}) a variable equal to one if at least one S\&P 500 firm in the focal firm's industry donated in the previous week, and zero otherwise.\footnote{Competitors' donations in the past week help isolate a firm's response to peer donations. Results are robust to different lag specifications.} The results in Appendix Table \ref{tab:peer-pressure} indicate a limited role of competition, while the treatment effect coefficients remain consistent with those in Table \ref{tab:reg-results-large-shareholders}.

\subsection{Defense Production Act}\label{apndx:r6}
The Defense Production Act (DPA), originally enacted in 1950 during the Korean War, grants the U.S. President broad authority to direct industrial production for national defense purposes. In 2020, President Trump invoked the DPA to address critical supply shortages during the COVID-19 pandemic, requiring companies like 3M and General Motors to prioritize government contracts and produce essential goods such as masks and ventilators. 

While the use of the DPA could raise concerns about its potential impact on our results, we offer two arguments to mitigate this concern. First, the initial executive orders under the DPA relevant to our study were issued on April 20, 2020—after the end of our sample period—meaning they should not affect our main findings. Second, we conduct a robustness check by re-running our main analysis excluding firms that were specifically targeted by these orders. The results, presented in Appendix Table \ref{tab:dpac_regs}, show that the estimated coefficients remain stable relative to those reported in Table \ref{tab:reg-results-block-static}, suggesting that our conclusions are not driven by the (anticipation of) DPA interventions.



\newpage\clearpage

\input{Tables/block_capped_tic_regs_ALL_mod.tex}

\input{Tables/exposure_tic_regs_branches_mod.tex}


\input{Tables/peer_pressure_new}

\input{Tables/dpac_regs}


 \begin{figure}
     \caption{Cumulative abnormal returns around a COVID-related donation}
     \label{fig:excess_returns}
     \centering
    \centering
    \includegraphics[width=0.75\linewidth]{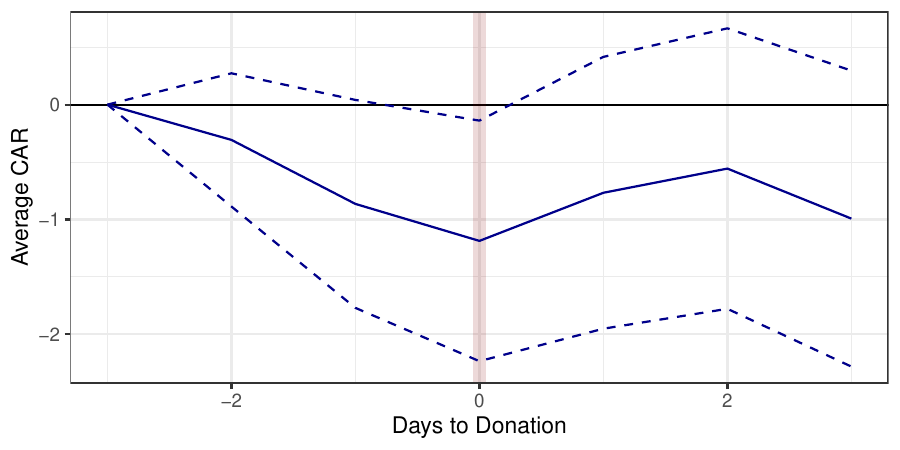}
		\medskip 
\begin{minipage}{1 \textwidth} 
{\footnotesize Note: This figure suggests that firms do not choose to donate to improve their performance on the stock market. The figure plots cumulative abnormal returns (CARs) computed using the methodology outlined in Section \ref{s:robustness}. Date 0 is the date when the donation was made public. CRISP data. 
\par}
\end{minipage}
 \end{figure}

\end{document}

%% file: Tables/summary_statistics_new.tex
\begin{table}[!t]
\centering
\caption{Firm and shareholder characteristics at treated and control groups} \label{tab:summary-meeting}
\centerline{\resizebox{1.2\textwidth}{!}{
\begin{tabular}{lrrrrrrrr}
  \toprule
&\multicolumn{3}{c}{Quantiles} &  \multicolumn{5}{c}{Average} \\
\cmidrule(lr){2-4} \cmidrule(lr){5-9}
 & \multicolumn{1}{c}{25\%} & \multicolumn{1}{c}{50\%} & \multicolumn{1}{c}{75\%} & \multicolumn{1}{c}{Overall} & \multicolumn{1}{c}{Treated} & \multicolumn{1}{c}{Control} & \multicolumn{1}{c}{Diff.} & \multicolumn{1}{c}{p-value} \\ 
     & \multicolumn{1}{c}{(1)} & \multicolumn{1}{c}{(2)} & \multicolumn{1}{c}{(3)} & \multicolumn{1}{c}{(4)} & \multicolumn{1}{c}{(5)} & \multicolumn{1}{c}{(6)} & \multicolumn{1}{c}{(5)-(6)} & \multicolumn{1}{c}{(8)} \\ 
\cmidrule(lr){2-4} \cmidrule(lr){5-5} \cmidrule(lr){6-9} 
 \multicolumn{9}{c}{\qquad \textbf{i. Finance \& Operations}} \\
  Market Capitalization (bn \$) & 13.43 & 24.24 & 51.48 & 52.78 & 65.05 & 52.03 & 13.03 & 0.40 \\
  Assets (bn \$) & 10.82 & 24.69 & 64.38 & 76.35 & 55.77 & 77.96 & -22.18 & 0.32 \\ 
  Revenue (bn \$) & 4.91 & 10.09 & 21.42 & 22.36 & 16.53 & 22.84 & -6.32 & 0.11 \\ 
  Cash to Total Assets & 0.01 & 0.04 & 0.10 & 0.07 & 0.07 & 0.07 & 0.01 & 0.71 \\ 
  Book-to-Market  & 0.17 & 0.34 & 0.58 & 0.42 & 0.40 & 0.41 & -0.02 & 0.83 \\ 
  Employees ('000s) & 9.00 & 21.00 & 56.93 & 51.14 & 58.50 & 50.49 & 8.01 & 0.58 \\ 
  Branches & 17.00 & 88.00 & 388.50 & 560.89 & 429.22 & 575.97 & -146.75 & 0.57 \\
  
  \multicolumn{9}{c}{\qquad \textbf{ii. Shareholding Composition}} \\
  Ownership Concentration (HHI) & 1.96 & 2.69 & 3.50 & 3.01 & 3.32 & 2.98 & 0.34 & 0.55 \\
 Individual Ownership & 0.00 & 0.00 & 0.00 & 0.63 & 1.66 & 0.54 & 1.13 & 0.29 \\ 
  Institutional Ownership (\%) & 76.70 & 86.53 & 93.98 & 84.03 & 81.31 & 84.28 & -2.97 & 0.12 \\ 
  Financial Ownership (\%) & 65.73 & 74.33 & 81.49 & 72.71 & 71.24 & 72.94 & -1.70 & 0.32 \\ 
  Individual Blockholding ($>5\%$) & 0.00 & 0.00 & 0.00 & 0.53 & 1.30 & 0.46 & 0.84 & 0.40 \\
 Institutional Blockholding ($>5\%$) & 14.11 & 20.99 & 28.37 & 21.55 & 18.15 & 21.88 & -3.73 & 0.00 \\ 
   Financial Blockholding ($>5\%$) & 14.01 & 20.55 & 27.35 & 20.96 & 18.15 & 21.24 & -3.09 & 0.02 \\ 
  
  \multicolumn{9}{c}{\qquad \textbf{iii. Prosocial Characteristics}} \\
  Composite ESG Score [0-10] & 4.00 & 5.50 & 6.80 & 5.36 & 4.89 & 5.41 & -0.52 & 0.11 \\ 
  Social Score [0-10] & 3.60 & 4.50 & 5.45 & 4.53 & 4.68 & 4.51 & 0.17 & 0.54 \\ 
 Governance Score [0-10] & 5.00 & 5.80 & 6.50 & 5.68 & 5.61 & 5.69 & -0.08 & 0.66 \\
  2019 Donations per 100\$ of Revenue & 0.04 & 0.10 & 0.21 & 0.32 & 0.43 & 0.31 & 0.13 & 0.63 \\ 
  {Covid donations before April 15:}  &  &  &  &  &  &  &  &  \\ 
  \quad - Indicator for donations (0/1) & 0.00 & 0.00 & 1.00 & 0.45 & 0.57 & 0.44 & 0.13 & 0.09 \\ 
  \quad - US\$ donation amount (m \$) & 1.00 & 2.00 & 5.00 & 7.36 & 7.35 & 7.50 & -0.15 & 0.97 \\ 
  \multicolumn{9}{c}{\qquad \textbf{iv. Definition of the Treatment Group}} \\
AGM date relative to April 15 & -& -& -& -& Before & After & -& -\\
Number of firms &-  &- &- & 482 & 47 & 435 & -&- \\ 
\bottomrule
\end{tabular} }}
\begin{tablenotes}
\footnotesize \vspace{1em}
\item\hspace{-1.2cm}{\parbox{1.16\textwidth}{Note: This table compares the firms in the treatment and control groups used in the empirical analysis of Covid-related corporate donation announcements. Firm financials from Compustat. Shareholding variables from LSEG. ESG variables are from MSCI. AGM dates are from ISS. Columns 5 focuses on firms with AGM within 3 months after the start of the pandemic and Columns 6 the other firms. }}  
\end{tablenotes}
\end{table}

%% file: Tables/block_groupped_inst.tex
\begin{table}[ht]
\center
\caption{Shareholders' influence on covid donations \label{tab:reg-results-block-static}}
\centerline{\resizebox{\textwidth}{!}{
\begin{tabular}{l*{6}{r @{} l}}
\toprule
& \multicolumn{12}{c}{Whether Firm $f$ has Donated (0/1)}       \\
 & \multicolumn{2}{c}{(1)}& \multicolumn{2}{c}{(2)}& \multicolumn{2}{c}{(3)}& \multicolumn{2}{c}{(4)}& \multicolumn{2}{c}{(5)}& \multicolumn{2}{c}{(6)} \\
 \cmidrule(r){2-7} \cmidrule(r){8-13} 
\textit{Ownership} ($\hat{\beta}_{1}$) & -0.001 &           & 0.018 &           & 0.006  &               & -0.032    &       & -0.048&$^{*}$    & 0.025&\\   
                      & (0.031)&           & (0.029)&          & (0.031)&               & (0.026)&          & (0.028)&         & (0.023)&\\   
AGM ($\hat{\beta}_{2}$)         & 0.109&             & 0.110&            & 0.073&                 & 0.077&            & 0.101&           & 0.139&\\   
                      & (0.091)&           & (0.092)&          & (0.099)&               & (0.089) &         & (0.081) &        & (0.092)&\\   
\textit{Ownership} $\times$ AGM ($\hat{\beta}_{treat}$)& 0.083&$^{***}$     & 0.088&$^{***}$    & 0.066 &                & -0.169&$^{**}$    & -0.163&$^{*}$    & -0.036&\\   
                      & (0.030)&           & (0.031)&          & (0.050)&               & (0.082)&          & (0.087)&         & (0.104)&\\   
\cmidrule(r){2-7} \cmidrule(r){8-13}
\textit{Ownership} is defined as & \multicolumn{6}{c}{Individuals} & \multicolumn{6}{c}{Financial} \\
\textit{Ownership} share: &\multicolumn{2}{c}{$>10\%$} & \multicolumn{2}{c}{$>5\%$} & \multicolumn{2}{c}{$(0\%, 2\%)$}&\multicolumn{2}{c}{$>10\%$} & \multicolumn{2}{c}{$>5\%$} & \multicolumn{2}{c}{$(0\%, 2\%)$}\\
\midrule
Control for size (log Assets) &  \checkmark &        &           \checkmark&        &           \checkmark&        &           \checkmark&        &           \checkmark&        &           \checkmark&        \\
Sub-Industry fixed effects &  \checkmark &        &           \checkmark&        &           \checkmark&        &           \checkmark&        &           \checkmark&        &           \checkmark&        \\
State fixed effects &  \checkmark &        &           \checkmark&        &           \checkmark&        &           \checkmark&        &           \checkmark&        &           \checkmark&        \\
$N$             & 482&                & 482&               & 482 &                   & 482 &              & 482 &             & 482&\\  
Adjusted R-squared & 0.55079 &           & 0.55282&           & 0.55073 &               & 0.55641&           & 0.55886&          & 0.54760&\\  
\bottomrule
\multicolumn{13}{l}{* -- $p < 0.1$; ** -- $p < 0.05$; *** -- $p < 0.01$.}\\
\end{tabular}}}
\begin{tablenotes}
\footnotesize 
\item \hspace{-0cm}{\parbox{1\textwidth}{Note: Estimated coefficient from (\ref{eq:lin-prob-model-cross-section}) using an indicator variable that is one if firm $f$ has donated by April 15, 2020. The variable \textit{Ownership} is the share of either individual (Columns 1-3) or financial investors (Columns 4-6) among all investors owning at least a share of total equity, as defined in the middle panel. The variable \textit{Ownership} is standardized. The AGM variable is an indicator variable taking value one if the firm has an AGM before April 15, 2020, and zero otherwise. Each regression controls for a firm's size as the $\log$ of total assets at December 2019. Standard errors are clustered by industry.}} 
\end{tablenotes}
\end{table}

%% file: Tables/correlations.tex
\begin{table}[ht!]
\begin{center}
\caption{The donations of financial shareholders' and of their portfolio firms} \label{tab:correlations}
\resizebox{.66\textwidth}{!}{
\begin{tabular}{cccc}
    \toprule
    & \multicolumn{3}{c}{Spearman Correlations} \\ \cmidrule(lr){2-4}
    Minimum  & Weighted  & Simple  & Weighted Avg. \\ 
    Equity Share  & Average  & Average &  $\times$ AGM \\ 
    Considered & [p-values]  & [p-values] &  [p-values] \\ 
    (1) & (2) & (3) & (4) \\ \cmidrule(lr){1-1} \cmidrule(lr){2-4}
 0\%  & -0.072& 0.097 & -0.260 \\ 
(Average $N$=377) & [0.704]  & [0.610] & [0.165] \\ 
 1\%  & -0.082 & 0.082 & -0.014 \\ 
(Average $N$=103) & [0.762] & [0.763] & [0.959] \\ 
 2\%   & -0.037 & -0.148 & -0.573$^{**}$ \\  
(Average $N$=90) & [0.900] & [0.613] & [0.032] \\  
 3\%   & -0.510 & -0.180 & -0.564$^{*}$ \\  
(Average $N$=103) & [0.109] & [0.597] & [0.071] \\
 4\% & -0.478 & 0.087 & -0.440 \\ 
 (Average $N$=106) & [0.193] & [0.825] & [0.235] \\
 5\% & -0.304 & 0.087 & -0.588$^{*}$ \\ 
(Average $N$=57) & [0.426] & [0.825] & [0.096] \\  
    \bottomrule
\end{tabular}}%
\end{center}
\begin{tablenotes}
\footnotesize \vspace{-1em}
\item Note: 
The table computes the Spearman correlation between whether a financial firm donates and the share of donations at the firms in its portfolio. In each row, we vary the minimum share ($x\%$) that a firm must have in another firm to be considered part of the portfolio of an investor. Column 1 also reports the average number of S\&P500 firms in the portfolio of a financial investor in parenthesis. Column 2 computes the total donations of the firms that financial firm $i$ has invested in using equity shares at Dec 2019 as weights (i.e.,  $\sum_{j} share_{ij} \times \mathbf{1}{[\text{firm }j \text{ donated}]} \times \mathbf{1}{[i\text{'s share in } j \text{ is greater than }x\% ]}$), Column 3 computes simple averages (i.e., $N_i^{-1} \times \sum_{j} \mathbf{1}{[\text{firm }j \text{ donated} ]} \times \mathbf{1}{[i\text{'s share in } j \text{ is greater than }x\% ]}$), and Column 4 considers only firms in $i$'s portfolio with an AGM before April 15 (i.e.,  $\sum_{j} share_{ij} \times \mathbf{1}{[\text{firm }j \text{ donated}]} \times \mathbf{1}{[i\text{'s share in } j \text{ is greater than }x\% ]} \times \mathbf{1}{[j \text{ has an AGM}]}$). p-values are in square brackets.
\end{tablenotes}
\end{table}

%% file: Tables/sumstat_groups_russia.tex
\begin{table}[!ht]
\centering
\begin{threeparttable}
\caption{Summary of the sample of listed firms exposed to the Russian economy by group \label{apndx:summary-rus}}

\small 
\setlength{\tabcolsep}{2pt}
\begin{tabular}{p{6cm}rrrrrrrr}
  \toprule
&\multicolumn{3}{c}{Quantiles} &  \multicolumn{5}{c}{Average} \\
\cmidrule(lr){2-4} \cmidrule(lr){5-9}
 & \multicolumn{1}{c}{25\%} & \multicolumn{1}{c}{50\%} & \multicolumn{1}{c}{75\%} & \multicolumn{1}{c}{Overall} & \multicolumn{1}{c}{Treated} & \multicolumn{1}{c}{Control} & \multicolumn{1}{c}{Diff.} & \multicolumn{1}{c}{p-value} \\ 
     & \multicolumn{1}{c}{(1)} & \multicolumn{1}{c}{(2)} & \multicolumn{1}{c}{(3)} & \multicolumn{1}{c}{(4)} & \multicolumn{1}{c}{(5)} & \multicolumn{1}{c}{(6)} & \multicolumn{1}{c}{(5)-(6)} & \multicolumn{1}{c}{(8)} \\ 
\cmidrule(lr){2-4} \cmidrule(lr){5-5} \cmidrule(lr){6-9} 
 \multicolumn{9}{c}{\quad\textbf{i. Firm characteristics}} \\
 Market Capitalization (bn \$) & 17.19 & 45.84 & 124.50 & 141.03 & 107.17 & 184.11 & -76.94 & 0.18 \\ 
Revenue (bn \$) & 13.17 & 39.50 & 113.74 & 128.37 & 103.72 & 157.23 & -53.51 & 0.29 \\ 
Assets (bn \$) & 7.42 & 22.32 & 72.46 & 65.58 & 78.28 & 50.71 & 27.57 & 0.12 \\ 
Revenue (bn \$)  & 3.63 & 10.09 & 29.45 & 24.10 & 27.68 & 20.04 & 7.63 & 0.15 \\
S\&P500 constituent & 0.00 & 1.00 & 1.00 & 0.74 & 0.79 & 0.68 & 0.10 & 0.14 \\  
Share of revenue from Russia (\%) & 0.80 & 1.29 & 1.74 & 1.41 & 1.48 & 1.32 & 0.16 & 0.30 \\  
 \multicolumn{9}{c}{\quad\textbf{ii. Shareholding composition}} \\
Ownership concentration & 1.64 & 2.26 & 3.47 & 2.98 & 2.84 & 3.13 & -0.29 & 0.52 \\ 
Individual shareholding (\%) & 0.00 & 0.00 & 0.04 & 1.01 & 0.20 & 1.96 & -1.76 & 0.03 \\ 
Institutional shareholding (\%) & 70.06 & 80.36 & 91.20 & 78.27 & 80.02 & 76.23 & 3.80 & 0.16 \\ 
Individual blockholding ($>$1\%) & 0.00 & 0.00 & 0.00 & 0.95 & 0.12 & 1.92 & -1.80 & 0.02 \\
Institutional blockholding ($>$5\%) & 9.14 & 16.98 & 26.24 & 19.70 & 20.25 & 19.07 & 1.17 & 0.55 \\
 \multicolumn{9}{c}{\quad\textbf{iii. ESG characteristics}} \\
Composite ESG Score [0-10] & 4.80 & 6.10 & 7.13 & 5.94 & 6.15 & 5.69 & 0.45 & 0.12 \\ 
Social Score [0-10] & 4.00 & 4.90 & 5.55 & 4.79 & 4.84 & 4.73 & 0.11 & 0.63 \\ 
Governance Score [0-10] & 4.25 & 5.00 & 5.60 & 4.83 & 4.93 & 4.71 & 0.23 & 0.20 \\ 
 \multicolumn{9}{c}{\quad\textbf{iv. Corporate Actions as of March 23, 2022} {\footnotesize (letters indicate the group grade)}} \\
A, Surgical Removal (50~firms) & 0.00 & 0.00 & 1.00 & 0.30 & 0.24 & 0.38 & -0.15 & 0.04 \\  
B, Keeping Options Open for Return (68~firms) & 0.00 & 0.00 & 1.00 & 0.41 & 0.40 & 0.42 & -0.02 & 0.83 \\
C, Reducing Current Operations (9~firms) & 0.00 & 0.00 & 0.00 & 0.05 & 0.09 & 0.01 & 0.08 & 0.02 \\ 
D, Holding Off New Investments/Development (25~firms) & 0.00 & 0.00 & 0.00 & 0.15 & 0.20 & 0.09 & 0.11 & 0.04 \\
F, Defying Demands for Exit (13~firms) & 0.00 & 0.00 & 0.00 & 0.08 & 0.07 & 0.09 & -0.02 & 0.56 \\
\multicolumn{9}{c}{\quad\textbf{v. Definition of the treatment group}} \\
AGM date relative to May 24 & -& -& -& -& Before & After & -& -\\
Number of firms &-  &- &- & 165 & 89 & 76 & -&- \\ 
\bottomrule
\end{tabular}

\begin{tablenotes}[flushleft]
    \footnotesize
    \item \textit{Note:} This table compares the firms in the treatment and control groups used in the empirical analysis of the corporate sanctions against the Russian economy executed in the first month of the 2022 war in Ukraine. Firm financials from Compustat. Shareholding variables and exposure to Russia come from LSEG. ESG variables are from MSCI. AGM dates are from ISS. This data was collected by Jeffrey Sonnenfeld and his team at Yale SOM, which we thank for sharing the data and explaining the data construction. Columns 5 focuses on firms with AGM within 3 months after the start of the war and Columns 6 the other firms. 
\end{tablenotes}
\end{threeparttable}
\end{table}

%% file: Tables/regression_russia_2.tex
\begin{table}[!ht]
\begin{center}
\caption{Shareholders' influence on the decision to exit the Russian market \label{tab:regression_russia2}}
\centerline{\resizebox{1.1\textwidth}{!}{
\begin{tabular}{l*{6}{r @{} l}}
\toprule
& \multicolumn{12}{c}{Exited Russia in the First Month of Conflict (0/1)} \\
\cmidrule(r){2-13}
& \multicolumn{2}{c}{(1)} &  \multicolumn{2}{c}{(2)} & \multicolumn{2}{c}{(3)} & \multicolumn{2}{c}{(4)} &  \multicolumn{2}{c}{(5)} & \multicolumn{2}{c}{(6)} \\
\cmidrule(r){2-7}\cmidrule(r){8-13}

AGM ($\hat{\beta}_1$)
& -0.102&        & -0.085&        & -0.106&        
& -0.099&        & -0.044&        & -0.121&        \\   
& (0.077)&       & (0.137)&       & (0.082)&       
& (0.080)&       & (0.082)&       & (0.078)&        \\  

\textit{Ownership} ($\hat{\beta}_2$)
& -0.044&        & -0.064&$^{**}$ & -0.108&$^{**}$ 
& -0.054&        & 0.020&         & 0.122&$^{**}$  \\   
& (0.031)&       & (0.028)&       & (0.041)&       
& (0.058)&       & (0.077)&       & (0.053)&        \\  

AGM $\times$ \textit{Ownership} ($\hat{\beta}_{treat}$)
& 0.110&$^{*}$   & 0.225&$^{***}$ & 0.115&         
& 0.029&         & -0.184&$^{*}$  & -0.055&        \\   
& (0.061) &      & (0.076)&S       & (0.090)&       
& (0.079)&       & (0.096)&       & (0.090)&        \\   

\cmidrule(r){2-7}\cmidrule(r){8-13}

\textit{Ownership} is defined as:
& \multicolumn{6}{c}{Individual} & \multicolumn{6}{c}{Institutional} \\

\textit{Ownership} share:
& \multicolumn{2}{c}{$>$1\%}
& \multicolumn{2}{c}{$>$1\%}
& \multicolumn{2}{c}{(0\%-1\%)}
& \multicolumn{2}{c}{$>$5\%}
& \multicolumn{2}{c}{$>$5\%}
& \multicolumn{2}{c}{(0\%-2\%)} \\ \cmidrule(lr){4-5} \cmidrule(lr){10-11}
\textit{Exposure to Russia (above / below median)}: & \multicolumn{2}{c}{} & \multicolumn{2}{c}{Below} & \multicolumn{4}{c}{} & \multicolumn{2}{c}{Above} & \multicolumn{2}{c}{} \\

\midrule
Control for size (log Assets)
& \checkmark && \checkmark && \checkmark
&& \checkmark && \checkmark && \checkmark \\

State fixed effects
& \checkmark && \checkmark && \checkmark
&& \checkmark && \checkmark && \checkmark \\

Sector fixed effects
& \checkmark && \checkmark && \checkmark
&& \checkmark && \checkmark && \checkmark \\

$N$
& 157& & 76& & 157&
& 157& & 77& & 157& \\  

Adjusted R-squared
& 0.36135&& 0.48358&       & 0.37860&
& 0.35885& & 0.53863& & 0.39441& \\ 

\bottomrule
 * -- $p < 0.1$; ** -- $p < 0.05$; *** -- $p < 0.01$.
\end{tabular}}}
\end{center}
\begin{tablenotes}
\footnotesize \vspace{-1em}
\item \hspace{-1.2cm}{\parbox{1.16\textwidth}{ 

\item Note: The table reports estimated coefficients from \eqref{eq:lin-prob-model-cross-section} using an indicator variable that is one if firm $f$ exited Russia by March 23, 2022, and zero otherwise. The variable \textit{Ownership} represents the share of either individual or institutional investors among all shareholders owning at least the threshold percentage of total equity indicated in the middle panel. Columns (2) and (5) subset the data to consider only firms with less than median exposure to Russia in terms of revenues or above median, respectively. Ownership categories and blockholding thresholds are reported directly below the main estimates; all \textit{Ownership} variables are standardized. The \textit{AGM} variable is an indicator taking the value one if the firm held an annual general meeting between February 24 and March 23, 2022, and zero otherwise. All regressions include controls for firm size ($\log$ of total assets), state fixed effects, and sector fixed effects. The sample shrinks from 165 (Table \ref{apndx:summary-rus}) to 157 because 8 firms have missing values for the variable asset. Standard errors are clustered by industry and reported in parentheses.
\par}}
\end{tablenotes}
\end{table}

%% file: Tables/donations_top10.tex
\begin{table}[htbp]
\begin{center}
    \caption{Examples of COVID-related corporate donations}
\label{tab:covid_contributions}
  \resizebox{\textwidth}{!}{
\begin{tabular}{l c c l}
\midrule
\textbf{Company} & \textbf{Date} & \textbf{Donation} & \textbf{Source} \\
\midrule
\noalign{\vskip 2mm}

\cmidrule(lr){1-4}
\multicolumn{4}{c}{\textit{Panel (a): Largest Initial Cash Donations}} \\
\cmidrule(lr){1-4}
Goldman Sachs & April 2 & \$300m & Finance Magnates \\
Cisco Systems & March 22 & \$225m & Company Website \\
Visa & April 6 & \$210m & Company Website \\
Automatic Data Processing & March 23 & \$51m & MarketWatch \\
BlackRock & March 23 & \$50m & The Guardian \\
Capital One Financial & March 27 & \$50m & Company Website \\
JPMorgan Chase \& Co. & March 18 & \$50m & Reuters \\
United Health Group & March 27 & \$50m & Fox News \\
Pfizer & April 6 & \$40m & Business Wire \\
Micron Technology & March 25 & \$35m & GlobeNewswire \\
\noalign{\vskip 2mm}

\cmidrule(lr){1-4}
\multicolumn{4}{c}{\textit{Panel (b): Examples of In-Kind Donations}} \\
\cmidrule(lr){1-4}

Apple &	March 21 &	Donation of 10 million masks &	Yahoo \\
Prudential Financial	& March 23 &	Donation of masks and ventilators & 	New Jersey News \\
Coty & March 25 & Production and donation of hand sanitizers & Company Website \\
Gap & March 25 & Production and donation of masks & Good Morning America \\
Disney & April 1 & Donation of masks & Company Website \\
Home Depot & April 1 & Donation of N95 respirator masks & USA Today \\
Oracle & April 1 & Donation of learning system tools & Business Insider \\
Gilead Science & April 5 & Medicine drug donation & Fox Business \\
Dupont & April 6 & Donation of N95 respirator masks & 	Company Website \\
Keysight & April 7 & Donation of cleaning supplies & CSR Wire \\

\bottomrule
\end{tabular} }
\end{center}
\vspace{-1.5mm}
\begin{tablenotes}
\footnotesize 
\item Note: This table reports examples of initial corporate donations either in cash (Panel (a)) or in-kind (Panel (b)) made by firms in the sample. The sample covers companies that were constituents of the S\&P 500 as of December 31, 2019. We manually identified the earliest publicly reported COVID-19–related donation by searching Google News and general Google results.
\end{tablenotes}
\end{table}

%% file: Tables/horse_race.tex
\begin{table}[ht]
\center
\caption{Shareholders' influence on covid donations: horse race and additional control variables \label{tab:horse_race}}
\begin{adjustbox}{width=.8\textheight,center}
\begin{threeparttable}
\begin{tabular}{l*{8}{r @{} l}}
\toprule
& \multicolumn{16}{c}{Whether Firm $f$ has Donated (0/1)}       \\
 & \multicolumn{2}{c}{(1)}& \multicolumn{2}{c}{(2)}& \multicolumn{2}{c}{(3)}& \multicolumn{2}{c}{(4)}& \multicolumn{2}{c}{(5)}& \multicolumn{2}{c}{(6)}& \multicolumn{2}{c}{(7)}& \multicolumn{2}{c}{(8)} \\
\cmidrule(lr){2-3} \cmidrule(lr){4-9} \cmidrule(lr){10-11} \cmidrule(lr){12-17}
\textit{Individual Ownership}               & 0.020 &        & -0.003&        & 0.013 &        & 0.006&         & 0.007&         & -0.016&        & 0.002&         & -0.0005&\\   
                      & (0.030)&       & (0.030)&       & (0.028)&       & (0.032) &      & (0.033) &      & (0.033)&       & (0.031)&       & (0.027)&\\
\textit{Financial Ownership}                & -0.020 &       & -0.026&        & -0.054&$^{*}$  & 0.021&         & -0.018&        & -0.021&        & -0.040 &       & 0.007&\\   
                      & (0.028) &      & (0.026)&       & (0.028)&       & (0.023)&       & (0.028)&       & (0.029)&       & (0.030) &      & (0.022)&\\   
   AGM                & 0.092&         & 0.032&         & 0.070 &        & 0.062&         & 0.042 &        & -0.019 &       & 0.021 &        & 0.0003&\\   
                      & (0.092)&       & (0.086)&       & (0.094)  &     & (0.090)&       & (0.095)&       & (0.090) &      & (0.101)&       & (0.094)&\\   
  \textit{Individual Ownership}  $\times$ AGM  & 0.106&$^{***}$ & 0.067&$^{**}$  & 0.062&$^{*}$   & 0.069&         & 0.122&$^{***}$ & 0.079&$^{**}$  & 0.083&$^{**}$  & 0.088&$^{*}$\\   
                      & (0.038)&       & (0.030)&       & (0.036) &      & (0.048)  &     & (0.038) &      & (0.032)  &     & (0.039) &      & (0.044)&\\   
   \textit{Financial Ownership}  $\times$ AGM  & 0.018&         & -0.198&$^{**}$ & -0.100&        & 0.034&         & 0.008&         & -0.209&$^{**}$ & -0.098&        & 0.046&\\   
                      & (0.106) &      & (0.086)&       & (0.091)&       & (0.119)&       & (0.104)&       & (0.089) &      & (0.091) &      & (0.120)&\\

   Cash to Assets               &&               &&               &&               & &              & -0.247&        & -0.293&        & -0.288&        & -0.196&\\   
                      &&               &&               & &              & &              & (0.301)&       & (0.289)&       & (0.288)&       & (0.295)&\\   
   Book to Market                &&              &&              & &              &&               & -0.190&$^{*}$  & -0.175&$^{*}$  & -0.157&       & -0.212&$^{**}$\\   
                      & &              &  &             &     &          &   &            & (0.102)&       & (0.103)  &     & (0.111)   &    & (0.092)&\\   
   Social Score      &  &             &     &          &    &           &    &           & 0.075&$^{***}$ & 0.072&$^{***}$ & 0.071&$^{***}$ & 0.075&$^{***}$\\   
                      &  &             &   &            &        &       &  &             & (0.025) &      & (0.022)   &    & (0.024)   &    & (0.024)&\\  
\cmidrule(lr){2-3}\cmidrule(lr){4-9} \cmidrule(lr){10-11}\cmidrule(lr){12-17} 
\textit{Ownership} share:&\multicolumn{2}{c}{Across all} &\multicolumn{6}{c}{Across shareholders with:}&\multicolumn{2}{c}{Across all} &\multicolumn{6}{c}{Across shareholders with:}\\
&\multicolumn{2}{c}{shareholders} &\multicolumn{2}{c}{$>10\%$} & \multicolumn{2}{c}{$>5\%$} & \multicolumn{2}{c}{$(0\%, 2\%)$}&\multicolumn{2}{c}{shareholders} &\multicolumn{2}{c}{$>10\%$} & \multicolumn{2}{c}{$>5\%$} & \multicolumn{2}{c}{$(0\%, 2\%)$}\\
\midrule
Control for firm size &  \checkmark &        &           \checkmark&        &           \checkmark& &  \checkmark &        &           \checkmark&        &           \checkmark&   &           \checkmark&        &           \checkmark&  \\
Sub-Industry fixed effects &  \checkmark &        &           \checkmark&        &           \checkmark& &  \checkmark &        &           \checkmark&        &           \checkmark&      &           \checkmark&        &           \checkmark&     \\
State fixed effects &  \checkmark &        &           \checkmark&        &           \checkmark& &  \checkmark &        &           \checkmark&        &           \checkmark&   &           \checkmark&        &           \checkmark&  \\
$N$            &          482&        &          482&        &          482& & 482 &&          482&        &          482&        &          482& & 482 &      \\
Adjusted R-squared& 0.55614&       & 0.56153&       & 0.56147 &      & 0.55198&       & 0.58370 &      & 0.59011  &     & 0.58607&       & 0.58081&\\  
\bottomrule
\multicolumn{7}{l}{* -- $p < 0.1$; ** -- $p < 0.05$; *** -- $p < 0.01$.}\\
\end{tabular}

\begin{tablenotes}
\footnotesize 
\item \hspace{0cm}{\parbox{1.51\textwidth}{Note: Estimated coefficient from Equation (\ref{eq:lin-prob-model-cross-section}) using an indicator variable that is one if firm $f$ has donated by April 15, 2020. The variable \textit{Ownership} is the ownership share of either individual or financial investors (Columns 1 and 5), the share owned by individual or financial investors owning at least 10\% (Columns 2 and 6), at least 5\% (Columns 3 and 7), or less than 2\% (Columns 4 and 8). The variable \textit{Ownership} is standardized. The AGM variable is an indicator variable taking value one if the firm has an AGM before April 15, 2020, and zero otherwise. All control variables are as of December 2019. Standard errors are clustered by industry.}} 
\end{tablenotes}
\end{threeparttable}
\end{adjustbox}
\end{table}

%% file: Tables/covid_inst.tex
\begin{table}[ht]
\center
\caption{Institutional Shareholders' influence on Covid donations \label{tab:covid_inst}}
\centerline{\resizebox{.8\textwidth}{!}{
\begin{tabular}{l*{3}{r @{} l}}
\toprule
& \multicolumn{6}{c}{Whether Firm $f$ has Donated (0/1)}       \\
 & \multicolumn{2}{c}{(1)}& \multicolumn{2}{c}{(2)}& \multicolumn{2}{c}{(3)} \\
 \cmidrule(r){2-7} 
\textit{Ownership} ($\hat{\beta}_{1}$) & -0.037&        & -0.057&$^{**}$ & 0.007&\\   
                      & (0.026) &      & (0.028) &      & (0.022)&\\  
AGM ($\hat{\beta}_{2}$)         & 0.039&         & 0.064&         & 0.127&\\   
                      & (0.089)&       & (0.096)&       & (0.094)&\\    
\textit{Ownership} $\times$ AGM ($\hat{\beta}_{treat}$) & -0.214&$^{**}$ & -0.139&        & 0.020&\\   
                      & (0.089) &      & (0.093)&       & (0.111)&\\  
 \cmidrule(r){2-7}
\textit{Ownership} share: &\multicolumn{2}{c}{$>10\%$} & \multicolumn{2}{c}{$>5\%$} & \multicolumn{2}{c}{$(0\%, 2\%)$}\\
\midrule
Control for size &  \checkmark &        &           \checkmark&        &           \checkmark& \\
Sub-Industry fixed effects &  \checkmark &        &           \checkmark&        &           \checkmark& \\
State fixed effects &  \checkmark &        &           \checkmark&        &           \checkmark& \\
$N$             & 482&                & 482&               & 482 & \\  
Adjusted R-squared & 0.55990 &       & 0.55787 &      & 0.54641 & \\  
\bottomrule
\multicolumn{7}{l}{* -- $p < 0.1$; ** -- $p < 0.05$; *** -- $p < 0.01$.}\\
\end{tabular}}}
\begin{tablenotes}
\footnotesize 
\item \hspace{-0cm}{\parbox{1\textwidth}{Note: Estimated coefficient from Equation (\ref{eq:lin-prob-model-cross-section}) using an indicator variable that is one if firm $f$ has donated by April 15, 2020. The variable \textit{Ownership} is the share of institutional shareholders---instead of only financial shareholders as in the main text analysis---owning at least a share of total equity, as defined in the middle panel. The variable \textit{Ownership} is standardized. The AGM variable is an indicator variable taking value one if the firm has an AGM before April 15, 2020, and zero otherwise. Each regression controls for a firm's size as the $\log$ of total assets at December 2019. Standard errors are clustered by industry.}} 
\end{tablenotes}
\end{table}

%% file: Tables/covid_slack.tex
\begin{table}[ht]
\center
\caption{Shareholders' influence on Covid donations -- By level of financial slack \label{tab:covid_slack}}
\centerline{\resizebox{\textwidth}{!}{
\begin{tabular}{l*{3}{r @{} l}}
\toprule
& \multicolumn{6}{c}{Whether Firm $f$ has Donated (0/1)}       \\
 & \multicolumn{2}{c}{(1)}& \multicolumn{2}{c}{(2)}& \multicolumn{2}{c}{(3)} \\
 \cmidrule(r){2-7} 
AGM        & 0.022&         & 0.174&$^{*}$   & -0.289&$^{**}$\\   
                      & (0.091) &      & (0.092)&       & (0.139)&\\   
                      
\textit{Individual Shareholding}  & 0.039&         & 0.028&         & 0.004&\\   
                      & (0.025) &      & (0.050) &      & (0.033)&\\  

\textit{Financial Blockholding} & -0.032 &       & -0.041 &       & -0.050&\\   
                      & (0.027)&       & (0.044)&       & (0.042)&\\   
                      
\textit{Individual Shareholding} $\times$ AGM & 0.330&$^{***}$ & 0.588&$^{***}$ & -0.552&\\   
                      & (0.116) &      & (0.141) &      & (0.499)&\\
\textit{Financial Blockholding} $\times$ AGM  & -0.215&$^{**}$ & -0.031&        & -0.495&$^{***}$\\   
                      & (0.092)&       & (0.066)&       & (0.171)&\\  
                      
 \cmidrule(r){2-7}
\textit{Financial slack (above / below median)} share: &\multicolumn{2}{c}{} & \multicolumn{2}{c}{Above} & \multicolumn{2}{c}{Below}\\
\midrule
Control for size &  \checkmark &        &           \checkmark&        &           \checkmark& \\
Sub-Industry fixed effects &  \checkmark &        &           \checkmark&        &           \checkmark& \\
State fixed effects &  \checkmark &        &           \checkmark&        &           \checkmark& \\
$N$            & 356&           & 152&           & 162&\\  
Adjusted R-squared & 0.54702&       & 0.61956&       & 0.70084&\\ 
\bottomrule
\multicolumn{7}{l}{* -- $p < 0.1$; ** -- $p < 0.05$; *** -- $p < 0.01$.}\\
\end{tabular}}}
\begin{tablenotes}
\footnotesize 
\item \hspace{-0cm}{\parbox{1\textwidth}{Note: Estimated coefficient from Equation (\ref{eq:lin-prob-model-cross-section}) using an indicator variable that is one if firm $f$ has donated by April 15, 2020. The AGM variable is an indicator variable taking value one if the firm has an AGM before April 15, 2020, and zero otherwise. \textit{Financial blockholding} is the share of equity owned by financial investors owning at least 10\% of firm $f$'s equity. Shareholding variables are standardized. Each regression controls for a firm's size as the $\log$ of total assets at December 2019. Standard errors are clustered by industry.}} 
\end{tablenotes}
\end{table}

%% file: Tables/donations_2020.tex
\begin{table}[ht]
\center
\caption{Shareholders' influence on 2020 total donations \label{tab:donations_2020}}
\centerline{\resizebox{1.1\textwidth}{!}{
\begin{tabular}{l*{6}{r @{} l}}
\toprule
& \multicolumn{12}{c}{2020 Donations to Revenue (\$ per \$ k of Revenue)}       \\
 & \multicolumn{2}{c}{(1)}& \multicolumn{2}{c}{(2)}& \multicolumn{2}{c}{(3)}& \multicolumn{2}{c}{(4)}& \multicolumn{2}{c}{(5)}& \multicolumn{2}{c}{(6)} \\
 \cmidrule(r){2-7} \cmidrule(r){8-13} 
\textit{Ownership} ($\hat{\beta}_{1}$) 
& -0.157&       & -0.092&        & -0.054&
& -0.116&        & -0.060&       & -0.117&\\   
& (0.111)&      & (0.228)&       & (0.150)&
& (0.128)&       & (0.139)&      & (0.129)&\\

AGM ($\hat{\beta}_{2}$) & -0.511&$^{*}$ & -0.595&$^{**}$ & -0.552&
& -0.741&$^{**}$ & -0.575&$^{*}$ & -0.763&$^{**}$\\   
& (0.256)&      & (0.285)&       & (0.333)&
& (0.288)&       & (0.311)&      & (0.334)&\\

\textit{Ownership} $\times$ AGM ($\hat{\beta}_{treat}$) & 1.19&$^{***}$ & -0.056&        & 0.452&       
& 0.488&         & 0.436&        & 0.424&\\   
& (0.123)&      & (0.184)&       & (0.564)&     
& (0.404)&       & (0.535)&      & (0.505)&\\

\cmidrule(lr){2-5} \cmidrule(lr){6-9} \cmidrule(lr){10-13}
\textit{Ownership} is defined as & \multicolumn{4}{c}{Individuals} & \multicolumn{4}{c}{Institutional} & \multicolumn{4}{c}{Financial} \\
\textit{Ownership} share: &\multicolumn{2}{c}{$>2\%$} & \multicolumn{2}{c}{$(0\%, 2\%)$} & \multicolumn{2}{c}{$>5\%$}&\multicolumn{2}{c}{$(0\%, 2\%)$} & \multicolumn{2}{c}{$>5\%$} & \multicolumn{2}{c}{$(0\%, 2\%)$}\\
\midrule
Control for size (log Assets) &  \checkmark &        &           \checkmark&        &           \checkmark&        &           \checkmark&        &           \checkmark&        &           \checkmark&        \\
Sub-Industry fixed effects &  \checkmark &        &           \checkmark&        &           \checkmark&        &           \checkmark&        &           \checkmark&        &           \checkmark&        \\
State fixed effects &  \checkmark &        &           \checkmark&        &           \checkmark&        &           \checkmark&        &           \checkmark&        &           \checkmark&        \\
$N$             & 242&          & 242&     & 242&     & 242&     & 242&     & 242&\\   
Adjusted R-squared & 0.61693&      & 0.61175& & 0.62825& & 0.61230& & 0.62825& & 0.61230&\\
\bottomrule
\multicolumn{13}{l}{* -- $p < 0.1$; ** -- $p < 0.05$; *** -- $p < 0.01$.}\\
\end{tabular}}}
\begin{tablenotes}
\footnotesize 
\item \hspace{-0cm}{\parbox{1\textwidth}{Note: Estimated coefficient from Equation (\ref{eq:lin-prob-model-cross-section}) using firm $f$'s 2020 amount donated per thousand dollars of revenue. The variable \textit{Ownership} is the share of either individual (Columns 1-2), institutional (Columns 3-4) or financial investors (Columns 5-6) among all investors owning at least a share of total equity, as defined in the middle panel. The variable \textit{Ownership} is standardized. The AGM variable is an indicator variable taking value one if the firm has an AGM before April 15, 2020, and zero otherwise. Each regression controls for a firm's size as the $\log$ of total assets at December 2019. Standard errors are clustered by industry.}} 
\end{tablenotes}
\end{table}

%% file: Tables/other_association.tex
\begin{table}[ht]
\center
\caption{Other measure of shareholder-firm association \label{tab:other_association}}
\centerline{\resizebox{0.6\textwidth}{!}{
\begin{tabular}{l*{2}{r @{} l}}
\toprule
& \multicolumn{4}{c}{Whether Firm $f$}       \\
& \multicolumn{4}{c}{has donated (0/1)}       \\
 & \multicolumn{2}{c}{(1)} & \multicolumn{2}{c}{(2)} \\
 \cmidrule(r){2-5}

   AGM                                      & -0.066&        & 0.128&\\   
                                            & (0.143)&       & (0.079)&\\   
   Firm Size                                & 0.130&$^{***}$ & 0.129&$^{***}$\\   
                                            & (0.028)&       & (0.029)&\\   
   Ownership concentration                  & -0.588 &       & -0.041&\\   
                                            & (1.18)&        & (1.25)&\\ 
Association $>0$ (0/1)                    & 0.046&         &   \\   
                                            & (0.058)&       &   \\ 
   Association $>0$ (0/1)    $\times$ AGM       & 0.334&$^{*}$   &&   \\   
                                            & (0.186)&       &&   \\   
   Association               & &              & 0.031&\\   
                                            &  &             & (0.031)&\\   
   Association$\times$ AGM  &  &             & 0.162&$^{**}$\\   
                                            &   &            & (0.079)&\\   
   \midrule
   State fixed effects                                    & \checkmark &           & \checkmark&\\  
   Sub-Industry fixed effects                                  & \checkmark&           & \checkmark&\\  
   $N$                             & 432&           & 432&\\  
   R$^2$                                    & 0.59665 &      & 0.59681&\\  
\bottomrule
\multicolumn{5}{l}{* -- $p < 0.1$; ** -- $p < 0.05$; *** -- $p < 0.01$.}\\
\end{tabular}}}
\begin{tablenotes}
\footnotesize 
\item \hspace{-0cm}{\parbox{1\textwidth}{Note: Estimated coefficient from Equation (\ref{eq:lin-prob-model-cross-section}) using an indicator variable that is one if firm $f$ has donated by April 15, 2020. The variable \textit{Association} is the average t-statistic of the correlation between firm $f$'s Google Trends score and each of its individual shareholders' Google Trends score. \textit{Association} is standardized in Column 2. The AGM variable is an indicator variable taking value one if the firm has an AGM before April 15, 2020, and zero otherwise.  Standard errors are clustered by industry.}} 
\end{tablenotes}
\end{table}

%% file: Tables/covid_fyr.tex
\begin{table}[ht]
\center
\caption{Shareholders' influence on covid donations -- Controlling for the distance to fiscal year-end \label{tab:covid_fyr}}
\centerline{\resizebox{1.05\textwidth}{!}{
\begin{tabular}{l*{6}{r @{} l}}
\toprule
& \multicolumn{12}{c}{Whether Firm $f$ has Donated (0/1)}       \\
 & \multicolumn{2}{c}{(1)}& \multicolumn{2}{c}{(2)}& \multicolumn{2}{c}{(3)}& \multicolumn{2}{c}{(4)}& \multicolumn{2}{c}{(5)}& \multicolumn{2}{c}{(6)} \\
 \cmidrule(r){2-7} \cmidrule(r){8-13} 
\textit{Ownership} ($\hat{\beta}_{1}$) 
& 0.005&         & 0.024&         & -0.003& 
& -0.028&         & -0.051&$^{*}$  & 0.026&\\
& (0.036) &      & (0.032)&       & (0.032)&
& (0.028)&        & (0.028)&       & (0.022)&\\  

AGM ($\hat{\beta}_{2}$) 
& 0.162&         & 0.170&         & 0.135&
& 0.080 &         & 0.111&         & 0.188&\\   
& (0.115) &      & (0.115) &      & (0.114) &
& (0.132) &       & (0.124)  &     & (0.116)& \\ 

\textit{Ownership} $\times$ AGM ($\hat{\beta}_{treat}$)
& 0.078&$^{**}$  & 0.082&$^{**}$  & 0.065&
& -0.208&$^{***}$ & -0.144&        & -0.036&\\   
& (0.030)&       & (0.031) &      & (0.048)&
& (0.075)  &      & (0.091)&       & (0.102)& \\   

\cmidrule(r){2-7} \cmidrule(r){8-13}
\textit{Ownership} is defined as & \multicolumn{6}{c}{Individuals} & \multicolumn{6}{c}{Financial} \\
\textit{Ownership} share: &\multicolumn{2}{c}{$>10\%$} & \multicolumn{2}{c}{$>5\%$} & \multicolumn{2}{c}{$(0\%, 2\%)$}&\multicolumn{2}{c}{$>10\%$} & \multicolumn{2}{c}{$>5\%$} & \multicolumn{2}{c}{$(0\%, 2\%)$}\\
\midrule
Control for size (log Assets) &  \checkmark &        &           \checkmark&        &           \checkmark&        &           \checkmark&        &           \checkmark&        &           \checkmark&        \\
Sub-Industry fixed effects &  \checkmark &        &           \checkmark&        &           \checkmark&        &           \checkmark&        &           \checkmark&        &           \checkmark&        \\
State fixed effects &  \checkmark &        &           \checkmark&        &           \checkmark&        &           \checkmark&        &           \checkmark&        &           \checkmark&        \\
Fiscal-year end month fixed effects &  \checkmark &        &           \checkmark&        &           \checkmark&        &           \checkmark&        &           \checkmark&        &           \checkmark&        \\
$N$             & 442&                & 442&               & 442 &                   & 442 &              & 442 &             & 442&\\  
Adjusted R-squared & 0.52618&       & 0.52852&       & 0.52477 &      & 0.53411 &       & 0.53241&       & 0.52322&\\
\bottomrule
\multicolumn{13}{l}{* -- $p < 0.1$; ** -- $p < 0.05$; *** -- $p < 0.01$.}\\
\end{tabular}}}
\begin{tablenotes}
\footnotesize 
\item \hspace{-0cm}{\parbox{1\textwidth}{Note: Estimated coefficient from Equation (\ref{eq:lin-prob-model-cross-section}) using an indicator variable that is one if firm $f$ has donated by April 15, 2020. The variable \textit{Ownership} is the share of either individual (Columns 1-3) or financial investors (Columns 4-6) among all investors owning at least a share of total equity, as defined in the middle panel. The variable \textit{Ownership} is standardized. The AGM variable is an indicator variable taking value one if the firm has an AGM before April 15, 2020, and zero otherwise. Each regression controls for a firm's size as the $\log$ of total assets at December 2019. Standard errors are clustered by industry.}} 
\end{tablenotes}
\end{table}

%% file: Tables/covid_fyr2.tex
\begin{table}[ht]
\center
\caption{Shareholders' influence on covid donations -- Among firms ending their fiscal year after September \label{tab:covid_fyr2}}
\centerline{\resizebox{1.05\textwidth}{!}{
\begin{tabular}{l*{6}{r @{} l}}
\toprule
& \multicolumn{12}{c}{Whether Firm $f$ has Donated (0/1)}       \\
 & \multicolumn{2}{c}{(1)}& \multicolumn{2}{c}{(2)}& \multicolumn{2}{c}{(3)}& \multicolumn{2}{c}{(4)}& \multicolumn{2}{c}{(5)}& \multicolumn{2}{c}{(6)} \\
 \cmidrule(r){2-7} \cmidrule(r){8-13} 
\textit{Ownership} ($\hat{\beta}_{1}$) 
& -0.045&$^{**}$ & -0.018&        & -0.006&
& -0.021&        & -0.030&        & 0.016&\\   
& (0.022)&       & (0.028)&       & (0.038)&
& (0.030)&       & (0.028)&       & (0.022)&\\

AGM ($\hat{\beta}_{2}$) 
& 0.100&         & 0.099&         & 0.065&
& 0.072&         & 0.060&         & 0.131&\\   
& (0.087) &      & (0.088)&       & (0.088)&
& (0.089)&       & (0.092)&       & (0.091)&\\ 

\textit{Ownership} $\times$ AGM ($\hat{\beta}_{treat}$)
& 0.121&$^{***}$ & 0.120&$^{***}$ & 0.088&
& -0.216&$^{**}$ & -0.196&$^{**}$ & -0.065&\\   
& (0.022)&       & (0.030)&       & (0.059)&
& (0.092)&       & (0.096)&       & (0.110)&\\

\cmidrule(r){2-7} \cmidrule(r){8-13}
\textit{Ownership} is defined as & \multicolumn{6}{c}{Individuals} & \multicolumn{6}{c}{Financial} \\
\textit{Ownership} share: &\multicolumn{2}{c}{$>10\%$} & \multicolumn{2}{c}{$>5\%$} & \multicolumn{2}{c}{$(0\%, 2\%)$}&\multicolumn{2}{c}{$>10\%$} & \multicolumn{2}{c}{$>5\%$} & \multicolumn{2}{c}{$(0\%, 2\%)$}\\
\midrule
Control for size (log Assets) &  \checkmark &        &           \checkmark&        &           \checkmark&        &           \checkmark&        &           \checkmark&        &           \checkmark&        \\
Sub-Industry fixed effects &  \checkmark &        &           \checkmark&        &           \checkmark&        &           \checkmark&        &           \checkmark&        &           \checkmark&        \\
State fixed effects &  \checkmark &        &           \checkmark&        &           \checkmark&        &           \checkmark&        &           \checkmark&        &           \checkmark&        \\
$N$             & 367&                & 367&               & 367 &                   & 367 &              & 367 &             & 367&\\  
Adjusted R-squared & 0.52567&       & 0.52537&       & 0.52466& 
& 0.53136 &      & 0.53055 &      & 0.51884& \\  
\bottomrule
\multicolumn{13}{l}{* -- $p < 0.1$; ** -- $p < 0.05$; *** -- $p < 0.01$.}\\
\end{tabular}}}
\begin{tablenotes}
\footnotesize 
\item \hspace{-0cm}{\parbox{1\textwidth}{Note: Estimated coefficient from Equation (\ref{eq:lin-prob-model-cross-section}) using an indicator variable that is one if firm $f$ has donated by April 15, 2020. The variable \textit{Ownership} is the share of either individual (Columns 1-3) or financial investors (Columns 4-6) among all investors owning at least a share of total equity, as defined in the middle panel. The variable \textit{Ownership} is standardized. The AGM variable is an indicator variable taking value one if the firm has an AGM before April 15, 2020, and zero otherwise. Each regression controls for a firm's size as the $\log$ of total assets at December 2019. Standard errors are clustered by industry.}} 
\end{tablenotes}
\end{table}

%% file: Tables/exits_top10.tex

\begin{table}[htbp]
\centering
\caption{Top 10 Earliest Exits from Russia Announcements (2022)}
\label{tab:russia_exits}

\begin{threeparttable}
\begin{tabular}{l c c l}
\toprule
{Company} & {Date} & Source &  {Original Source} \\[1ex]
\midrule
MSCI & February 24 & Yale List & Yahoo Finance \\[1ex]
Delta Air Lines & February 25 & Yale List & Company Website \\[1ex]
Global Foundries & February 25 & Leave-Russia.org & Washington Post \\[1ex]
Uber & February 27 &Yale List & The Moscow Times \\[1ex]
BlackRock & February 28 &Yale List & Company LinkedIn \\[1ex]
Netflix & February 28 &Yale List & Variety \\[1ex]
Roku & February 28 & Yale List &Reuters \\[1ex]
Alaska Airlines & March 1 & Yale List &Company Website \\[1ex]
ExxonMobil & March 1 & Yale List &Company Website \\[1ex]
Alcoa & March 2 &Yale List & Company Website \\[1ex]
\bottomrule
\end{tabular}

\begin{tablenotes}[flushleft]
\footnotesize
\item Note: This table reports the ten earliest public announcements by firms suspending or terminating business with Russia following the invasion of Ukraine on February 24, 2022. We combine the lists compiled by Yale’s CELI \citep{sonnenfeld2022business} and Leave-Russia.org. We retrieve the original sources from these websites.
\end{tablenotes}
\end{threeparttable}

\end{table}

%% file: Tables/regression_russia_fin.tex
\begin{table}[!ht]
\begin{center}
\caption{Shareholders' influence on the decision to exit the Russian market -- Alternative thresholds and financial investors \label{tab:regression_russia_fin}}
\centerline{\resizebox{1.2\textwidth}{!}{
\begin{tabular}{l*{6}{r @{} l}}
\toprule
& \multicolumn{12}{c}{Exited Russia in the First Month of Conflict (0/1)} \\
\cmidrule(r){2-13}
& \multicolumn{2}{c}{(1)} &  \multicolumn{2}{c}{(2)} & \multicolumn{2}{c}{(3)} & \multicolumn{2}{c}{(4)} &  \multicolumn{2}{c}{(5)} & \multicolumn{2}{c}{(6)} \\
\cmidrule(r){2-7}\cmidrule(r){8-13}

AGM ($\hat{\beta}_1$)
& -0.101&        & -0.094&        & -0.093&
& -0.101&        & -0.034&        & -0.120&\\   
& (0.079) &      & (0.138)&       & (0.081)&
& (0.080)&       & (0.082)&       & (0.080)&\\        

\textit{Ownership} ($\hat{\beta}_2$)
& -0.028 &       & -0.058&$^{*}$  & -0.130&$^{***}$
& -0.043&        & 0.029&         & 0.118&$^{**}$\\   
& (0.024)&       & (0.030)&       & (0.046)&
& (0.055) &      & (0.079)&       & (0.058)&\\     

AGM $\times$ \textit{Ownership} ($\hat{\beta}_{treat}$)
& 0.094&$^{*}$   & 0.199&$^{***}$ & 0.134&$^{*}$
& 0.018&         & -0.172&$^{*}$  & -0.085&\\   
& (0.053)&       & (0.071) &      & (0.078)&
& (0.074)&       & (0.093) &      & (0.094)&\\   

\cmidrule(r){2-7}\cmidrule(r){8-13}

\textit{Ownership} is defined as:
& \multicolumn{6}{c}{Individual} & \multicolumn{6}{c}{Financial} \\

\textit{Ownership} share:
& \multicolumn{2}{c}{$>$5\%}
& \multicolumn{2}{c}{$>$5\%}
& \multicolumn{2}{c}{(0\%-2\%)}
& \multicolumn{2}{c}{$>$5\%}
& \multicolumn{2}{c}{$>$5\%}
& \multicolumn{2}{c}{(0\%-2\%)} \\ \cmidrule(lr){4-5} \cmidrule(lr){10-11}
\textit{Exposure to Russia (above / below median)}: & \multicolumn{2}{c}{} & \multicolumn{2}{c}{Below} & \multicolumn{4}{c}{} & \multicolumn{2}{c}{Above} & \multicolumn{2}{c}{} \\

\midrule
Control for size (log Assets)
& \checkmark && \checkmark && \checkmark
&& \checkmark && \checkmark && \checkmark \\

State fixed effects
& \checkmark && \checkmark && \checkmark
&& \checkmark && \checkmark && \checkmark \\

Sector fixed effects
& \checkmark && \checkmark && \checkmark
&& \checkmark && \checkmark && \checkmark \\

$N$
& 157&           & 76&            & 157&
& 157 &          & 77 &           & 157&\\       

Adjusted R-squared
& 0.35729&       & 0.47862 &      & 0.37185&
& 0.35751 &      & 0.52909 &      & 0.38564&\\  

\bottomrule
 * -- $p < 0.1$; ** -- $p < 0.05$; *** -- $p < 0.01$.
\end{tabular}}}
\end{center}
\begin{tablenotes}
\footnotesize \vspace{-1em}
\item \hspace{-1.2cm}{\parbox{1.16\textwidth}{ 

\item Note: The table reports estimated coefficients from Equation \eqref{eq:lin-prob-model-cross-section} using an indicator variable that is one if firm $f$ exited Russia by March 23, 2022, and zero otherwise. The variable \textit{Ownership} represents the share of either individual or financial investors among all shareholders owning at least the threshold percentage of total equity indicated in the middle panel. Columns (2) and (5) subset the data to consider only firms with less than median exposure to Russia in terms of revenues or above median, respectively. Ownership categories and blockholding thresholds are reported directly below the main estimates; all \textit{Ownership} variables are standardized. The \textit{AGM} variable is an indicator taking the value one if the firm held an annual general meeting between February 24 and March 23, 2022, and zero otherwise. All regressions include controls for firm size ($\log$ of total assets), state fixed effects, and sector fixed effects. The sample size decreases from 165 (Table~\ref{apndx:summary-rus}) to 157 due to the exclusion of fixed-effect singletons. Standard errors are clustered by industry and reported in parentheses.
\par}}
\end{tablenotes}
\end{table}

%% file: Tables/regression_russia_airlines.tex
\begin{table}[!ht]
\begin{center}
\caption{Shareholders' influence on the decision to exit the Russian market -- Excluding airline companies\label{tab:regression_russia_airlines}}
\centerline{\resizebox{1.2\textwidth}{!}{
\begin{tabular}{l*{6}{r @{} l}}
\toprule
& \multicolumn{12}{c}{Exited Russia in the First Month of Conflict (0/1)} \\
\cmidrule(r){2-13}
& \multicolumn{2}{c}{(1)} &  \multicolumn{2}{c}{(2)} & \multicolumn{2}{c}{(3)} & \multicolumn{2}{c}{(4)} &  \multicolumn{2}{c}{(5)} & \multicolumn{2}{c}{(6)} \\
\cmidrule(r){2-7}\cmidrule(r){8-13}

AGM ($\hat{\beta}_1$)
& -0.054&         & -0.040&       & -0.058&
& -0.048&         & -0.011&       & -0.064&\\   
 & (0.069)&        & (0.129)&      & (0.075)&
 & (0.071)&        & (0.096)&       & (0.068)&\\     

\textit{Ownership} ($\hat{\beta}_2$)
& -0.038&         & -0.053&$^{*}$ & -0.096&$^{**}$  
& -0.051&        & 0.006&          & 0.151&$^{***}$\\   
& (0.030)&        & (0.026)&      & (0.044)&
& (0.058)&        & (0.081)&        & (0.048)&\\  

AGM $\times$ \textit{Ownership} ($\hat{\beta}_{treat}$)
& 0.091&          & 0.179&$^{**}$ & 0.105&          
& 0.012&          & -0.197&$^{*}$   & -0.096&\\   
& (0.059)&        & (0.080)&      & (0.091)&
& (0.081)&        & (0.098)&        & (0.085)&\\     

\cmidrule(r){2-7}\cmidrule(r){8-13}

\textit{Ownership} is defined as:
& \multicolumn{6}{c}{Individual} & \multicolumn{6}{c}{Institutional} \\

\textit{Ownership} share:
& \multicolumn{2}{c}{$>$1\%}
& \multicolumn{2}{c}{$>$1\%}
& \multicolumn{2}{c}{(0\%-1\%)}
& \multicolumn{2}{c}{$>$5\%}
& \multicolumn{2}{c}{$>$5\%}
& \multicolumn{2}{c}{(0\%-2\%)} \\ \cmidrule(lr){4-5} \cmidrule(lr){10-11}
\textit{Exposure to Russia (above / below median)}: & \multicolumn{2}{c}{} & \multicolumn{2}{c}{Below} & \multicolumn{4}{c}{} & \multicolumn{2}{c}{Above} & \multicolumn{2}{c}{} \\

\midrule
Control for size (log Assets)
& \checkmark && \checkmark && \checkmark
&& \checkmark && \checkmark && \checkmark \\

State fixed effects
& \checkmark && \checkmark && \checkmark
&& \checkmark && \checkmark && \checkmark \\

Sector fixed effects
& \checkmark && \checkmark && \checkmark
&& \checkmark && \checkmark && \checkmark \\

$N$
& 154&            & 74&           & 154&
& 154&            & 74&             & 154&\\   

Adjusted R-squared &
 0.38150&        & 0.46596&      & 0.39637&
 & 0.38213&        & 0.54214&        & 0.43446&\\ 

\bottomrule
 * -- $p < 0.1$; ** -- $p < 0.05$; *** -- $p < 0.01$.
\end{tabular}}}
\end{center}
\begin{tablenotes}
\footnotesize \vspace{-1em}
\item \hspace{-1.2cm}{\parbox{1.16\textwidth}{ 

\item Note: The table reports estimated coefficients from Equation \eqref{eq:lin-prob-model-cross-section} using an indicator variable that is one if firm $f$ exited Russia by March 23, 2022, and zero otherwise. This table excludes airline companies. The variable \textit{Ownership} represents the share of either individual or institutional investors among all shareholders owning at least the threshold percentage of total equity indicated in the middle panel. Columns (2) and (5) subset the data to consider only firms with less than median exposure to Russia in terms of revenues or above median, respectively. Ownership categories and blockholding thresholds are reported directly below the main estimates; all \textit{Ownership} variables are standardized. The \textit{AGM} variable is an indicator taking the value one if the firm held an annual general meeting between February 24 and March 23, 2022, and zero otherwise. All regressions include controls for firm size ($\log$ of total assets), state fixed effects, and sector fixed effects. The sample size decreases from 165 (Table~\ref{apndx:summary-rus}) to 157 due to the exclusion of fixed-effect singletons. Standard errors are clustered by industry and reported in parentheses.
\par}}
\end{tablenotes}
\end{table}

%% file: Tables/regression_russia_grade_f.tex
\begin{table}[!ht]
\begin{center}
\caption{Shareholders' influence on the decision to exit the Russian market -- Excluding grade F firms \label{tab:regression_russia_grade_f}}
\centerline{\resizebox{1.2\textwidth}{!}{
\begin{tabular}{l*{6}{r @{} l}}
\toprule
& \multicolumn{12}{c}{Exited Russia in the First Month of Conflict (0/1)} \\
\cmidrule(r){2-13}
& \multicolumn{2}{c}{(1)} &  \multicolumn{2}{c}{(2)} & \multicolumn{2}{c}{(3)} & \multicolumn{2}{c}{(4)} &  \multicolumn{2}{c}{(5)} & \multicolumn{2}{c}{(6)} \\
\cmidrule(r){2-7}\cmidrule(r){8-13}

AGM ($\hat{\beta}_1$)
& -0.122&         & -0.114&         & -0.116&
& -0.117&         & -0.080&         & -0.137&$^{*}$\\   
& (0.078)&        & (0.145)&        & (0.083)&
& (0.082)&        & (0.086)&        & (0.081)&\\      

\textit{Ownership} ($\hat{\beta}_2$)
& -0.063&$^{*}$   & -0.084&$^{***}$ & -0.101&$^{*}$
& -0.061&         & -0.026&         & 0.107&$^{*}$\\   
& (0.033)&        & (0.021)&        & (0.052)&
& (0.066)&        & (0.089)&        & (0.058)&\\  

AGM $\times$ \textit{Ownership} ($\hat{\beta}_{treat}$)
& 0.130&$^{*}$    & 0.216&$^{**}$   & 0.131&
& 0.041&          & -0.193&         & -0.050&\\   
& (0.073)&        & (0.082)&        & (0.096)&
& (0.095)&        & (0.124)&        & (0.103)&\\

\cmidrule(r){2-7}\cmidrule(r){8-13}

\textit{Ownership} is defined as:
& \multicolumn{6}{c}{Individual} & \multicolumn{6}{c}{Institutional} \\

\textit{Ownership} share:
& \multicolumn{2}{c}{$>$1\%}
& \multicolumn{2}{c}{$>$1\%}
& \multicolumn{2}{c}{(0\%-1\%)}
& \multicolumn{2}{c}{$>$5\%}
& \multicolumn{2}{c}{$>$5\%}
& \multicolumn{2}{c}{(0\%-2\%)} \\ \cmidrule(lr){4-5} \cmidrule(lr){10-11}
\textit{Exposure to Russia (above / below median)}: & \multicolumn{2}{c}{} & \multicolumn{2}{c}{Below} & \multicolumn{4}{c}{} & \multicolumn{2}{c}{Above} & \multicolumn{2}{c}{} \\

\midrule
Control for size (log Assets)
& \checkmark && \checkmark && \checkmark
&& \checkmark && \checkmark && \checkmark \\

State fixed effects
& \checkmark && \checkmark && \checkmark
&& \checkmark && \checkmark && \checkmark \\

Sector fixed effects
& \checkmark && \checkmark && \checkmark
&& \checkmark && \checkmark && \checkmark \\

$N$
& 142&            & 67&             & 142&
& 142&            & 70 &            & 142&\\     

Adjusted R-squared
& 0.41492&        & 0.49757&        & 0.41722&
& 0.40621&        & 0.62160&        & 0.42994&\\

\bottomrule
 * -- $p < 0.1$; ** -- $p < 0.05$; *** -- $p < 0.01$.
\end{tabular}}}
\end{center}
\begin{tablenotes}
\footnotesize \vspace{-1em}
\item \hspace{-1.2cm}{\parbox{1.16\textwidth}{ 

\item Note: The table reports estimated coefficients from Equation \eqref{eq:lin-prob-model-cross-section} using an indicator variable that is one if firm $f$ exited Russia by March 23, 2022, and zero otherwise. This table excludes firms that claimed that they would not leave Russia (Grade F firms) as they may be different compared to other firms fearing retaliation from the public.  The variable \textit{Ownership} represents the share of either individual or institutional investors among all shareholders owning at least the threshold percentage of total equity indicated in the middle panel. Columns (2) and (5) subset the data to consider only firms with less than median exposure to Russia in terms of revenues or above median, respectively. Ownership categories and blockholding thresholds are reported directly below the main estimates; all \textit{Ownership} variables are standardized. The \textit{AGM} variable is an indicator taking the value one if the firm held an annual general meeting between February 24 and March 23, 2022, and zero otherwise. All regressions include controls for firm size ($\log$ of total assets), state fixed effects, and sector fixed effects. The sample size decreases from 165 (Table~\ref{apndx:summary-rus}) to 157 due to the exclusion of fixed-effect singletons. Standard errors are clustered by industry and reported in parentheses.
\par}}
\end{tablenotes}
\end{table}

%% file: Tables/block_capped_tic_regs_ALL_mod.tex
	\begin{table}[htbp]
    \vspace{0cm}
	\begin{center}
		\caption{The impact blockholders on covid donations through pandemic exposure \label{tab:reg-results-large-shareholders}}
		\centerline{\resizebox{1.1\textwidth}{!}{
			\begin{tabular}{l*{6}{r @{} l}}
\toprule
& \multicolumn{12}{c}{Whether Firm $f$ has Donated (0/1)}       \\
 & \multicolumn{2}{c}{(1)}& \multicolumn{2}{c}{(2)}& \multicolumn{2}{c}{(3)}& \multicolumn{2}{c}{(4)}& \multicolumn{2}{c}{(5)}& \multicolumn{2}{c}{(6)} \\
 \cmidrule(r){2-7} \cmidrule(r){8-13} 
\textit{Cum. Covid Deaths}                               & -0.0013&       & -0.0017&        & -0.0016&  & -0.0027&         & -0.0027&         & -0.0012&\\   
                                            & (0.0082)&      & (0.0082)&       & (0.0083)& & (0.0079)&        & (0.0079)&        & (0.0082)&\\   
   \textit{Cum. Covid Deaths}  $\times$ \textit{Ownership}                 & -0.0039&       & -0.0072&        & -0.0023&  & -0.0333&$^{***}$ & -0.0307&$^{***}$ & 0.0002&\\   
                                            & (0.0065)&      & (0.0077)&       & (0.0104)& & (0.0059)&        & (0.0055)&        & (0.0071)&\\   
   \textit{Cum. Covid Deaths} $\times$ AGM                  & 0.0296&        & 0.0291&         & 0.0213&   & 0.0200&          & 0.0275&          & 0.0359&$^{*}$\\   
                                            & (0.0197)&      & (0.0197)&       & (0.0221)& & (0.0175)&        & (0.0182)&        & (0.0191)&\\   
   \textit{Cum. Covid Deaths} $\times$ \textit{Ownership} $\times$ AGM    & 0.0168&$^{**}$ & 0.0231&$^{***}$ & 0.0155&   & -0.0158&         & -0.0064&         & -0.0076&\\   
                                            & (0.0070)&      & (0.0084)&       & (0.0121)& & (0.0148)&        & (0.0169)&        & (0.0190)&\\   
\cmidrule(r){2-7} \cmidrule(r){8-13}
\textit{Ownership} is defined as & \multicolumn{6}{c}{Individuals} & \multicolumn{6}{c}{Financial} \\
\textit{Ownership} share: &\multicolumn{2}{c}{$>10\%$} & \multicolumn{2}{c}{$>5\%$} & \multicolumn{2}{c}{$(0\%, 2\%)$}&\multicolumn{2}{c}{$>10\%$} & \multicolumn{2}{c}{$>5\%$} & \multicolumn{2}{c}{$(0\%, 2\%)$}\\
\midrule
Day fixed effects &  \checkmark &        &           \checkmark&        &           \checkmark&        &           \checkmark&        &           \checkmark&        &           \checkmark&        \\
State fixed effects &  \checkmark &        &           \checkmark&        &           \checkmark&        &           \checkmark&        &           \checkmark&        &           \checkmark&        \\
$N$             & 120,380&                & 120,380&               & 120,380 &                   & 120,380 &              & 120,380 &             & 120,380&\\  
Adjusted R-squared & 0.77799&       & 0.77813&        & 0.77801&  & 0.78217&         & 0.78139&         & 0.77773&\\ 
\bottomrule
\multicolumn{13}{l}{* -- $p < 0.1$; ** -- $p < 0.05$; *** -- $p < 0.01$.}\\
\end{tabular}}}
	\end{center}
	\begin{tablenotes}
		\footnotesize \vspace{-1em}
		\item \hspace{0cm}{\parbox{1\textwidth}{ Note: The table reports the OLS regressions based on Equation \ref{eq:lin-prob-model2} where the dependent variable is an indicator that is one if firm $f$ has donated by day $t$ and zero otherwise on covariates. The variable \textit{Ownership} varies across columns based as the share of a certain class of investors owning at least a given share of total equity (greater than 10\%, greater than 5\%, or between 0 and 2\%). The AGM variable is an indicator variable taking value one if the firm has an AGM before April 15, 2020, and zero otherwise.  All columns include day- and firm-fixed effects. The interaction \textit{Ownership} $\times$ AGM and the direct effect of the variables \textit{Ownership} and AGM are accounted for by the firm-fixed effects. The dataset starts on January 15, 2020, and ends on April 15, 2020. Standard errors are clustered by firm and presented in parenthesis.} }
	\end{tablenotes}
\end{table}	

%% file: Tables/exposure_tic_regs_branches_mod.tex
\begin{table}[H]\caption{The impact of covid exposure at branches on donations} \label{tab:exposure_industry_branches}
	\begin{center}
		\centerline{\resizebox{1.1\textwidth}{!}{
			\begin{tabular}{l*{4}{r @{} l}}
				\toprule
				&\multicolumn{8}{c}{Whether Firm $f$ has Donated by Time $t$ (0/1)}\\
				&\multicolumn{2}{c}{(1)} &\multicolumn{2}{c}{(2)}&\multicolumn{2}{c}{(3)}&\multicolumn{2}{c}{(4)}  \\
				\cmidrule(r){2-9}
\textit{Exposure at branches}& -0.0260&$^{**}$ & -0.0201&$^{*}$ & -0.0192&  & -0.0190&\\   
                                    & (0.0129)&       & (0.0119)&      & (0.0118)& & (0.0118)&\\   

\textit{Exposure at branches} $\times$ Number of branches ($\ln$)& 0.0040&         &&               &&          &&   \\   
                                    & (0.0030) &      & &              &&         & &  \\
\textit{Exposure at branches} $\times$ \textit{More than $x$ branches}&&                & 0.0176&        & 0.0266 &  & 0.0072&\\   
                                    & &               & (0.0150)&      & (0.0181)& & (0.0261)&\\   
\cmidrule(r){2-9}                
\textit{More than $x$ branches} (0/1) is 1 if the firm has  &                  \multicolumn{2}{c}{ } & \multicolumn{2}{c}{ }	& \multicolumn{2}{c}{ }		& \multicolumn{2}{c}{ } \\
more branches than the $x$ quantile:  &                  \multicolumn{2}{c}{ } & \multicolumn{2}{c}{50\%}	& \multicolumn{2}{c}{75\%}		& \multicolumn{2}{c}{90\%}\\
				\midrule
				Time fixed effects &       $\checkmark$&        &       $\checkmark$&        &       $\checkmark$&        &       $\checkmark$&    \\
				Firm fixed effects &       $\checkmark$&        &       $\checkmark$&        &       $\checkmark$&        &       $\checkmark$&  \\
$N$             & 110,500&        & 110,500&       & 110,500&  & 110,500&\\
Adjusted R-squared& 0.77873&        & 0.77868 &      & 0.77882&  & 0.77844&\\
				\bottomrule
				\multicolumn{9}{l}{* -- $p < 0.1$; ** -- $p < 0.05$; *** -- $p < 0.01$.}\\
		\end{tabular}}}
	\end{center}
	\begin{tablenotes}
		\footnotesize \vspace{-1.em}
		\item \hspace{0cm}{\parbox{1\textwidth}{ Note: This table finds that consumers were not a driver of covid-related donations because covid exposure in the states where firms had their branches does not correlate with donations rates across firms.  The table presents the coefficients from OLS regressions based on Equation \ref{eq:lin-prob-model-exposure} based on an indicator variable that is one if firm $f$ has donated by day $t$ and zero otherwise on covariates. Covid exposure is measured by cumulative covid deaths.  The variable \textit{Number of branches} in Columns 1 is in log. The remaining columns use a dummy variable (\textit{More than $x$ branches}) for whether the focal firm has more than the $x$-percentile than the distribution of branches: this value is 88 branches in Column 2, 391 branches in Column 3, and 1,415 branches in Column 4. Orbis data do not report branches for 15 firms, which are therefore omitted from the analysis. All columns include day- and firm-fixed effects. For this reason, the table does not report the direct effect of the \textit{Number of Branches}, which does not vary over time and, thus, is captured by the firm-fixed effects. The dataset starts on January 15, 2020 and ends on April 15, 2020.  Standard errors are clustered by firm and presented in parenthesis. }}
	\end{tablenotes}
\end{table}

%% file: Tables/peer_pressure_new.tex
\begin{table}[htbp]
	\begin{center}
		\caption{The role of peer pressure in driving Covid donations \label{tab:peer-pressure}}
		\centerline{\resizebox{1.1\textwidth}{!}{
			\begin{tabular}{l*{6}{r @{} l}}
\toprule
& \multicolumn{12}{c}{Whether Firm $f$ has Donated (0/1)}       \\
 & \multicolumn{2}{c}{(1)}& \multicolumn{2}{c}{(2)}& \multicolumn{2}{c}{(3)}& \multicolumn{2}{c}{(4)}& \multicolumn{2}{c}{(5)}& \multicolumn{2}{c}{(6)} \\
 \cmidrule(r){2-7} \cmidrule(r){8-13} 
Competitor Donating (0/1) & 0.013&        & 0.013&        & 0.012&   & 0.016&          & 0.014&          & 0.012&\\   
                                            & (0.010)&      & (0.010)&      & (0.010)& & (0.010) &       & (0.010)  &      & (0.010)&\\   
\textit{Cum. Covid Deaths}     & -0.001&       & -0.001&        & -0.001&       & -0.002&         & -0.002&         & -0.001&\\   
                                            & (0.008)&       & (0.008) &      & (0.008)&       & (0.008)&        & (0.008)&        & (0.008)&\\   
   \textit{Cum. Covid Deaths}  $\times$ \textit{Ownership}             & -0.002&        & -0.004&        & 0.001&        & -0.035&$^{***}$ & -0.032&$^{***}$ & 0.002&\\   
                                            & (0.006)&       & (0.007)&       & (0.010)&       & (0.006)&        & (0.006)&        & (0.007)&\\     
   \textit{Cum. Covid Deaths} $\times$ AGM                 & 0.034&$^{*}$   & 0.034&$^{*}$   & 0.028&         & 0.025&          & 0.033&$^{*}$    & 0.040&$^{**}$\\   
                                            & (0.019) &      & (0.019)&       & (0.021)&       & (0.016)&        & (0.017)&        & (0.019)&\\   
   \textit{Cum. Covid Deaths} $\times$ \textit{Ownership} $\times$ AGM    & 0.017&$^{**}$ & 0.020&$^{**}$ & 0.011&   & -0.021&         & -0.012&        & -0.006&\\   
                                            & (0.007)&      & (0.010)&      & (0.015)& & (0.015)&        & (0.016) &       & (0.019)&\\   
\cmidrule(r){2-7} \cmidrule(r){8-13}
\textit{Ownership} is defined as & \multicolumn{6}{c}{Individuals} & \multicolumn{6}{c}{Financial} \\
\textit{Ownership} share: &\multicolumn{2}{c}{$>10\%$} & \multicolumn{2}{c}{$>5\%$} & \multicolumn{2}{c}{$(0\%, 2\%)$}&\multicolumn{2}{c}{$>10\%$} & \multicolumn{2}{c}{$>5\%$} & \multicolumn{2}{c}{$(0\%, 2\%)$}\\
\midrule
Day fixed effects &  \checkmark &        &           \checkmark&        &           \checkmark&        &           \checkmark&        &           \checkmark&        &           \checkmark&        \\
State fixed effects &  \checkmark &        &           \checkmark&        &           \checkmark&        &           \checkmark&        &           \checkmark&        &           \checkmark&        \\
$N$            & 120,380&       & 120,380 &      & 120,380&       & 120,380&        & 120,380&        & 120,380&\\
Adjusted R-squared & 0.78551 &      & 0.78556 &      & 0.78539 &      & 0.78997 &       & 0.78916&        & 0.78520&\\ 
\bottomrule
\multicolumn{13}{l}{* -- $p < 0.1$; ** -- $p < 0.05$; *** -- $p < 0.01$.}\\
\end{tabular}}}
	\end{center}
	\begin{tablenotes}
		\footnotesize \vspace{-1em}
		\item \hspace{0cm}{\parbox{1\textwidth}{ Note: The table reports the OLS regressions based on Equation \ref{eq:lin-prob-model2} where the dependent variable is an indicator that is one if firm $f$ has donated by day $t$ and zero otherwise on covariates. The variable \textit{Ownership} varies across columns based as the share of a certain class of investors owning at least a given share of total equity (greater than 10\%, greater than 5\%, or between 0 and 2\%). The AGM variable is an indicator variable taking value one if the firm has an AGM before April 15, 2020, and zero otherwise.  All columns include day- and firm-fixed effects. The interaction \textit{Ownership} $\times$ AGM and the direct effect of the variables \textit{Ownership} and AGM are accounted for by the firm-fixed effects. The dataset starts on January 15, 2020, and ends on April 15, 2020. Standard errors are clustered by firm and presented in parenthesis.} }
	\end{tablenotes}
\end{table}

%% file: Tables/dpac_regs.tex
\begin{table}[ht]
\center
\caption{Shareholders' influence on covid donations, excluding firms targeted by the DPA}
\label{tab:dpac_regs}
\centerline{\resizebox{1.1\textwidth}{!}{
\begin{tabular}{l*{6}{r @{} l}}
\toprule
& \multicolumn{12}{c}{Whether Firm $f$ has Donated (0/1)}       \\
 & \multicolumn{2}{c}{(1)}& \multicolumn{2}{c}{(2)}& \multicolumn{2}{c}{(3)}& \multicolumn{2}{c}{(4)}& \multicolumn{2}{c}{(5)}& \multicolumn{2}{c}{(6)} \\
 \cmidrule(r){2-7} \cmidrule(r){8-13} 
\textit{Ownership} ($\hat{\beta}_{1}$) & -0.001&            & 0.023&             & 0.004&                  & -0.026&            & -0.058&$^{**}$    & 0.018&\\   
                      & (0.031)&            & (0.029)&           & (0.032)&                & (0.026)&           & (0.027)&          & (0.023)&\\ 
AGM ($\hat{\beta}_{2}$)           & 0.106&              & 0.107&             & 0.074&                 & 0.049 &            & 0.082 &           & 0.130&\\   
                      & (0.090)&            & (0.091)&           & (0.098)&                & (0.089)&           & (0.094)&          & (0.093)&\\
\textit{Ownership} $\times$ AGM ($\hat{\beta}_{treat}$)& 0.083&$^{***}$      & 0.081&$^{**}$      & 0.063&                  & -0.210&$^{**}$     & -0.125&           & 0.013&\\   
                      & (0.029)&            & (0.032)&           & (0.049) &               & (0.085) &          & (0.084) &         & (0.116)&\\   
\cmidrule(r){2-7} \cmidrule(r){8-13}
\textit{Ownership} is defined as & \multicolumn{6}{c}{Individuals} & \multicolumn{6}{c}{Institutional} \\
\textit{Ownership} is the share of investors owning: &\multicolumn{2}{c}{$>10\%$} & \multicolumn{2}{c}{$>5\%$} & \multicolumn{2}{c}{$(0\%, 2\%)$}&\multicolumn{2}{c}{$>10\%$} & \multicolumn{2}{c}{$>5\%$} & \multicolumn{2}{c}{$(0\%, 2\%)$}\\
\midrule
Industry fixed effects &  \checkmark &        &           \checkmark&        &           \checkmark&        &           \checkmark&        &           \checkmark&        &           \checkmark&        \\
State fixed effects &  \checkmark &        &           \checkmark&        &           \checkmark&        &           \checkmark&        &           \checkmark&        &           \checkmark&        \\
$N$           & 474&                & 474&               & 474 &                   & 474&               & 474&              & 474&\\  
Adjusted R-squared& 0.55535&            & 0.55756&           & 0.55486&                & 0.56304&           & 0.56264&          & 0.55165&\\ 
\bottomrule
\multicolumn{13}{l}{* -- $p < 0.1$; ** -- $p < 0.05$; *** -- $p < 0.01$.}\\
\end{tabular}}}
\begin{tablenotes}
\footnotesize 
\item \hspace{0cm}{\parbox{1\textwidth}{Note: Estimated coefficient from (\ref{eq:lin-prob-model-cross-section}) using an indicator variable that is one if firm $f$ has donated by April 15, 2020, and excluding firms targeted by the \textit{Defense Production Act}. The variable \textit{Ownership} is the share of either individual (Columns 1-3) or institutional investors (Columns 4-6) among all investors owning at least a share of total equity, as defined in the middle panel. The variable \textit{Ownership} is standardized. The AGM variable is an indicator variable taking value one if the firm has an AGM before April 15, 2020, and zero otherwise.  Standard errors are clustered by industry.}} 
\end{tablenotes}
\end{table}